\documentclass{aa}
\usepackage{amsmath,graphics,amssymb}

\usepackage{natbib}
\bibpunct{(}{)}{,}{a}{}{,}
\defcitealias{nltei}{Paper I}

\newcommand{\Teff}{\ensuremath{T_\mathrm{eff}}}
\newcommand{\zav}[1]{\left(#1\right)}
\newcommand{\ms}{\ensuremath{\text{M}_{\odot}}}
\newcommand{\kms}{\ensuremath{\mathrm{km}\,\mathrm{s}^{-1}}}
\newcommand{\msr}{\ensuremath{\ms\,\text{yr}^{-1}}} 
\newcommand{\vel}{{v}}
\newcommand{\vinfty}{\ensuremath{\vel_\infty}}
\newcommand{\hzav}[1]{\left[#1\right]}

\begin{document}

\title{Global hot-star wind models for stars from Magellanic Clouds}

\author{J.  Krti\v{c}ka\inst{1} \and J. Kub\'at\inst{2}}
\authorrunning{J. Krti\v{c}ka and J. Kub\'at}

\institute{\'Ustav teoretick\'e fyziky a astrofyziky P\v{r}F MU,
            CZ-611 37 Brno, Czech Republic, \email{krticka@physics.muni.cz}
           \and
           Astronomick\'y \'ustav, Akademie v\v{e}d \v{C}esk\'e
           republiky, CZ-251 65 Ond\v{r}ejov, Czech Republic}

\date{}

\abstract{We provide mass-loss rate predictions for O stars from Large and Small
Magellanic Clouds. We calculate global (unified, hydrodynamic) model atmospheres
of main sequence, giant, and supergiant stars for chemical composition
corresponding to Magellanic Clouds. The models solve radiative transfer equation
in comoving frame, kinetic equilibrium equations (also known as NLTE equations),
and hydrodynamical equations from (quasi-)hydrostatic atmosphere to expanding
stellar wind. The models allow us to predict wind density, velocity, and
temperature (consequently also the terminal wind velocity and the mass-loss
rate) just from basic global stellar parameters. As a result of their lower
metallicity, the line radiative driving is weaker leading to lower wind
mass-loss rates with respect to the Galactic stars. We provide a formula that
fits the mass-loss rate predicted by our models as a function of stellar
luminosity and metallicity. On average, the mass-loss rate scales with
metallicity as $ \dot M\sim Z^{0.59}$. The predicted mass-loss rates are lower
than mass-loss rates derived from H$\alpha$ diagnostics and can be reconciled
with observational results assuming clumping factor $C_\text{c}=9$. On the other
hand, the predicted mass-loss rates either agree or are slightly higher than the
mass-loss rates derived from ultraviolet wind line profiles. The calculated
\ion{P}{v} ionization fractions also agree with values derived from observations
for LMC stars with $T_\text{eff}\leq40\,000\,$K. Taken together, our theoretical
predictions provide reasonable models with consistent mass-loss rate
determination, which can be used for quantitative study of stars from Magellanic
Clouds.}

\keywords{stars: winds, outflows -- stars:   mass-loss  -- stars:  early-type --
Magellanic Clouds -- hydrodynamics -- radiative transfer}

\maketitle

\section{Introduction}

The radiative force influences various types of astrophysical objects on
different spatial scales. Because the radiative force acts selectively on
individual species, it depends on their abundances, and consequently also on
metallicity. The metallicity dependence of radiatively driven hot-star winds
leads to the absence of line-driven outflows in metal-free Pop~III stars
\citep{bezvi}, while Galactic stars lose a significant part of their mass due to
the winds \citep[e.g.,][]{kostel}.

Besides metallicity, hot-star wind mass-loss rates depend also on other basic
stellar parameters (luminosity, mass, and radius). While dependence of hot-star
wind on most of the stellar parameters can be observationally studied using
a local stellar sample \citep[e.g.,][]{pulchuch}, similar study of metallicity
dependence is more complicated. Besides ultraviolet satellites, observational
study of mass loss at a fraction of the solar metallicity in the Magellanic
Clouds requires large telescopes. It was possible to extend the range of studied
metallicities using spectroscopic observations of stars residing in galaxies of
the Local Group \citep{tramp,hergagalic}. However, hot stars with metallicity
below about one tenth of the solar metallicity are still observationally
unattainable for a detailed wind study.

The most complete picture of the dependence of wind properties on the
metallicity can be therefore obtained from theoretical models. Such models are
able to predict the dependence of basic wind properties on metallicity. The most
important wind parameter is the mass-loss rate $ \dot M$, that is the amount of
mass lost by the star per unit of time. \citet{vikolamet} predicted that for a
broad range of metallicities $Z$ (given by the mass fraction of heavier
elements) the mass-loss rate varies as $\dot M \sim Z^{0.69}$ in O and early B
stars. A more complex dependence of the mass-loss rate on metallicity was found
by \citet{grahamz} for late-type WN stars. \citet{peta} calculated metallicity-dependent wind mass-loss rates from the balance between the radiative energy
deposited in the wind and the energy required to lift up the wind material. The
latter calculations based on the CMFGEN code predicted that the metallicity
dependence of the mass-loss rate in luminous B stars is a complicated function
of stellar effective temperature and is the strongest around
$\Teff=17\,500-20\,000\,$K.

Study of stellar winds at very low metallicities is important for understanding
stellar evolution and stellar feedback in the early Universe and in dwarf
galaxies from the Local Group. The gravitational-wave source GW150914 is also
expected to originate from low-metallicity environment \citep{jiniabbotti}.
Because observations of stars in such low-metallicity environments are not
always possible, the numerical models may provide missing information about the
wind physics. To reach this goal, the current models have to be tested against
observations for as broad a range of metallicities as possible.

However, testing of wind models at low metallicities is complicated by the
mismatch between individual mass-loss rate determinations. Even at Galactic
metallicity the estimates based on the X-ray spectroscopy \citep{cohcar,rahen},
combined optical and UV analysis \citep{bouhil,clres2}, and near-infrared line
spectroscopy \citep{najaro} seem to point to lower mass-loss rates than
predicted by \citet{vikolamet}. On the other hand, the results based purely on
the H$\alpha$ emission lines \citep{mokz} can be reconciled with theoretical
predictions of \citet{vikolamet}. Moreover, a multiwavelength analysis of
\citet{shendelori} shows good agreement with theoretical models. Consequently,
the existence of the discrepancy between empirical and theoretical mass-loss
rate determinations is not unanimously accepted. The discrepancy may be even
nonmonotonic, because massive stars at the top of the main sequence show
enhanced mass-loss rates with respect to theoretical models \citep{nejlepsi}.

Part of these discrepancies may be connected with the influence of
inhomogeneities on the observational mass-loss rate diagnostics
\citep{chuchcar,sund,clres1,clres2} and on theoretical predictions
\citep{muij,supujo}. On the other hand, wind blocking in global (unified) wind
models leads to reduction of predicted mass-loss rates in agreement with some
observational estimates \citep{cmfkont}. These mass-loss rates taken from global
wind models are typically lower than \citet{vikolamet} and
\citet{cinskasmrt} mass-loss rates by a factor of 2--5, however they are consistent with predictions
based on CMFGEN \citep{bouhil,vesnice,peta} and PoWR codes
\citep{gratub,powrdyn}. Here we apply our global wind models \citep{cmfkont} to
stars from Magellanic Clouds.

\section{Description of global wind models}

The wind models used here were calculated using the METUJE code \citep{cmf1,cmfkont},
which provides global (unified) models of the stellar photosphere and wind. The
code solves the radiative transfer equation, the kinetic (statistical)
equilibrium equations, and hydrodynamic equations (equations of continuity,
momentum, and energy) both in the photosphere and in the wind. The code assumes
that the flow is stationary (time-independent) and spherically symmetric.

The radiative transfer equation is solved in the comoving frame (CMF) following
the method proposed by \citet{mikuh}. We include line and continuum transitions
relevant in atmospheres of hot stars in the radiative transfer equation. The
inner boundary condition for the radiative transfer equation is derived from the
diffusion approximation, and we assume no infalling radiation at the outer
boundary.

The ionization and excitation state of considered elements was calculated from
the kinetic equilibrium equations (also called NLTE equations). These equations
account for the radiative and collisional excitation, deexcitation, ionization,
and recombination. Part of the models of ions was adopted from the TLUSTY model
stellar atmosphere input data \citep{ostar2003,bstar2006}. To prepare the
remaining ionic models listed in \citet{nlteiii} and not included in TLUSTY
input data we use the same strategy as in TLUSTY, that is, the data are based on
the Opacity and Iron Project calculations \citep{topt,zel0} and corrected by the
observational data available in the NIST database \citep{nist}. For phosphorus
the ionic model was prepared using data described by \citet{pahole}. The
low-lying levels of ions are included explicitly in the calculations, while the
higher levels are merged into superlevels \citep[see][for
details]{ostar2003,bstar2006}. The bound-free radiative rates are consistently
calculated from the CMF mean intensity, while for the bound-bound rates we still
use the Sobolev approximation.

\begin{figure*}[t]
\centering
\resizebox{\hsize}{!}{\includegraphics{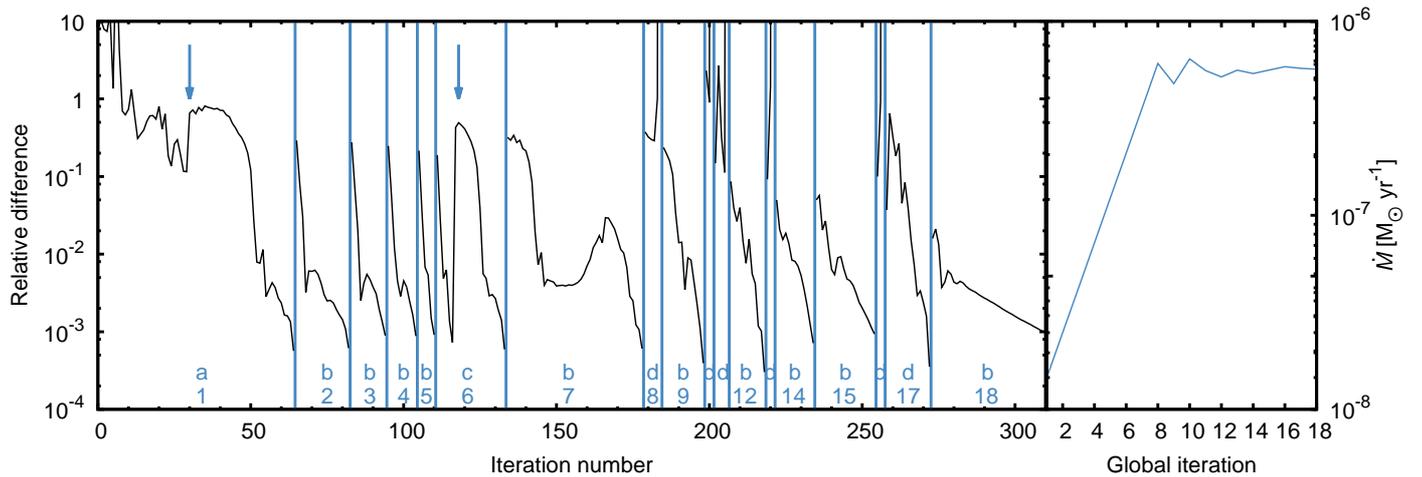}}
\caption{Convergence of the hydrodynamical structure of the LMC 400-3 model.
{\em Left plot}: Plot of the maximum relative difference between hydrodynamical
variables in subsequent iteration steps as a function of total number of
hydrodynamical iterations. We note that kinetic equilibrium equations are
iterated for each step of hydrodynamical iterations. Blue vertical strips denote
a new step of a global iteration (numbered in the bottom of the plot) when the
inner boundary velocity (and the mass-loss rate) is changed. Labels a-d in the
bottom of the plot indicate (see also \citealt{cmfkont}, Sect. 2): a)
Calculation of the initial model for subcritical velocities. In the first few steps
the CMF radiative transfer is not solved. Afterwards, the CMF calculations are
switched on (indicated by a blue arrow). b) Iteration of wind model with
mass-loss rate lower than or equal to the final one. Each model denoted as `b' has
a higher mass-loss rate than the previous one. c) Iteration of wind model with
mass-loss rate lower than the final one and inclusion of model for supercritical
velocities. Here the blue arrow denotes inclusion of model for supercritical
velocities. d) Iteration of wind model with mass-loss rate higher than the final
one. After this step the model mass-loss rate decreases. The plot shows that the
convergence of hydrodynamical structure in each step is stable and relatively
fast. {\em Right plot}: Mass-loss rate in individual global iteration steps
(numbers on horizontal axis are the same as in blue labels in the left plot).
The models converge to a final mass-loss rate given in Table~\ref{ohvezpar}.}
\label{konver}
\end{figure*}

Three different methods are needed to solve the energy equation. We use
a differential form of the transfer equation deep in the photosphere, while the
integral form of this equation is advantageous in the upper layers of the
photosphere \citep{kubii}. The electron thermal balance method \citep{kpp} is
applied in the wind. The individual terms in the energy equation are taken from
the CMF radiative field. The hydrodynamical equations, that is, the continuity
equation, equation of motion, and the energy equation, are solved iteratively to
obtain the wind density, velocity, and temperature structure. The iterations of
hydrodynamical equations are performed using the Newton-Raphson method (see
Fig.~\ref{konver} for convergence properties). The derivatives of the CMF
radiative force with respect to the flow variables within the Newton-Raphson
iteration step are approximated from the line force in the Sobolev approximation
corrected by the CMF line force \citep{cmf1}. We use the ``shooting method'' to
derive an estimate of the mass-loss rate. We calculate a series of wind models with
variable base velocity and search for the base velocity which provides a smooth
transonic solution with maximum mass-loss rate \citep{cmfkont}. The radiative
force due to line and continuum transitions is calculated in the CMF. The line
data for the calculation of the line force were taken from the VALD database
(Piskunov et al. \citeyear{vald1}, Kupka et al. \citeyear{vald2}) with some
updates using the NIST data \citep{nist}.

We utilize the TLUSTY model stellar atmospheres \citep{ostar2003,bstar2006} to
derive the initial guess of the solution in the photosphere. The TLUSTY models
are calculated for the same effective temperature, surface gravity, and chemical
composition as the wind models.

\begin{table*}
\caption{Stellar parameters of the model grid with
derived values of the terminal velocity $v_\infty$ and mass-loss rate $\dot M$}
\centering
\label{ohvezpar}
\begin{tabular}{ccrccccc}
\hline
\hline
Model &$\Teff$ & $R_{*}$ & $M$& \multicolumn{2}{c}{$\vinfty$} &
\multicolumn{2}{c}{$\dot M$} \\
& $[\text{K}]$ & $[\text{R}_{\odot}]$ &
$[\text{M}_{\odot}]$ &\multicolumn{2}{c}{$[\kms]$} 
&\multicolumn{2}{c}{$[\text{M}_{\odot}\,\text{yr}^{-1}$]}\\
&&&&$0.5Z_\odot$&$0.2Z_\odot$&$0.5Z_\odot$&$0.2Z_\odot$\\
\hline \multicolumn{8}{c}{main sequence stars}\\
300-5 & 30000 & 6.6 & 12.9  & 2560 & 2360 & $ 6.4\times10^{-9 }$ & $ 2.7\times10^{-9 }$ \\
325-5 & 32500 & 7.4 & 16.4  & 2110 & 2200 & $ 7.1\times10^{-9 }$ & $ 4.0\times10^{-9 }$ \\
350-5 & 35000 & 8.3 & 20.9  & 1070 & 1500 & $ 2.7\times10^{-8 }$ & $ 1.5\times10^{-8 }$ \\
375-5 & 37500 & 9.4 & 26.8  & 2070 & 1940 & $ 8.1\times10^{-8 }$ & $ 5.2\times10^{-8 }$ \\
400-5 & 40000 & 10.7 & 34.6 & 2380 & 2180 & $ 1.5\times10^{-7 }$ & $ 1.0\times10^{-7 }$ \\
425-5 & 42500 & 12.2 & 45.0 & 2170 & 2180 & $ 3.2\times10^{-7 }$ & $ 2.2\times10^{-7 }$ \\
450-5 & 45000 & 13.9 & 58.6 & 1890 & 2060 & $ 7.1\times10^{-7 }$ & $ 5.1\times10^{-7 }$ \\
\hline \multicolumn{8}{c}{giants}\\
300-3 & 30000 & 13.1 & 19.3 & 2020 & 2020 &  $ 4.7\times10^{-8 }$ & $ 2.5\times10^{-8 }$ \\
325-3 & 32500 & 13.4 & 22.8 & 1290 & 1750 &  $ 7.2\times10^{-8 }$ & $ 2.4\times10^{-8 }$ \\
350-3 & 35000 & 13.9 & 27.2 & 1170 & 1320 &  $ 1.7\times10^{-7 }$ & $ 1.2\times10^{-7 }$ \\
375-3 & 37500 & 14.4 & 32.5 & 1670 & 1730 &  $ 3.0\times10^{-7 }$ & $ 2.0\times10^{-7 }$ \\
400-3 & 40000 & 15.0 & 39.2 & 1360 & 1800 &  $ 6.5\times10^{-7 }$ & $ 3.1\times10^{-7 }$ \\
425-3 & 42500 & 15.6 & 47.4 & 1340 & 1600 &  $ 9.0\times10^{-7 }$ & $ 6.1\times10^{-7 }$ \\
450-3 & 45000 & 16.3 & 57.7 & 1400 & 1690 &  $ 1.3\times10^{-6 }$ & $ 1.0\times10^{-6 }$ \\
\hline \multicolumn{8}{c}{supergiants}\\
300-1 & 30000 & 22.4 & 28.8 & 1250 & 1280 &  $ 3.0\times10^{-7 }$ & $ 1.7\times10^{-7 }$ \\
325-1 & 32500 & 21.4 & 34.0 & 1010 & 860  &  $ 3.8\times10^{-7 }$ & $ 2.2\times10^{-7 }$ \\
350-1 & 35000 & 20.5 & 40.4 & 1120 & 1350 &  $ 6.0\times10^{-7 }$ & $ 3.6\times10^{-7 }$ \\
375-1 & 37500 & 19.8 & 48.3 & 1340 & 1650 &  $ 9.0\times10^{-7 }$ & $ 4.6\times10^{-7 }$ \\
400-1 & 40000 & 19.1 & 58.1 & 1440 & 1890 &  $ 1.1\times10^{-6 }$ & $ 5.4\times10^{-7 }$ \\
425-1 & 42500 & 18.5 & 70.3 & 1550 & 1960 &  $ 1.0\times10^{-6 }$ & $ 7.4\times10^{-7 }$ \\
450-1 & 45000 & 18.0 & 85.4 & 1930 & 2090 &  $ 1.3\times10^{-6 }$ & $ 9.5\times10^{-7 }$ \\
\hline
\end{tabular}
\end{table*}

\section{Calculated global wind models}

We calculated a grid of global wind models for metallicities corresponding to
Magellanic Cloud O stars in the effective temperature range
$30\,000-45\,000\,\text{K}$. Stellar masses and radii given in
Table~\ref{ohvezpar} were calculated using relations of \citet{okali} for main
sequence stars, giants, and supergiants. Although these relations were derived
for Galactic stars, they fairly describe also the parameters of stars from
Magellanic Clouds \citep[c.f.,][]{mamrac2,mokv}. We assumed solar chemical
composition \citep{asp09} scaled for elements heavier than helium by a factor of
0.5 and 0.2 corresponding to typical chemical composition of stars in Large and
Small Magellanic Clouds \citep[e.g.,][]{kopec,ven,rolles,bourak}, respectively.

\begin{figure*}[tp]
\centering
\resizebox{0.310\hsize}{!}{\includegraphics{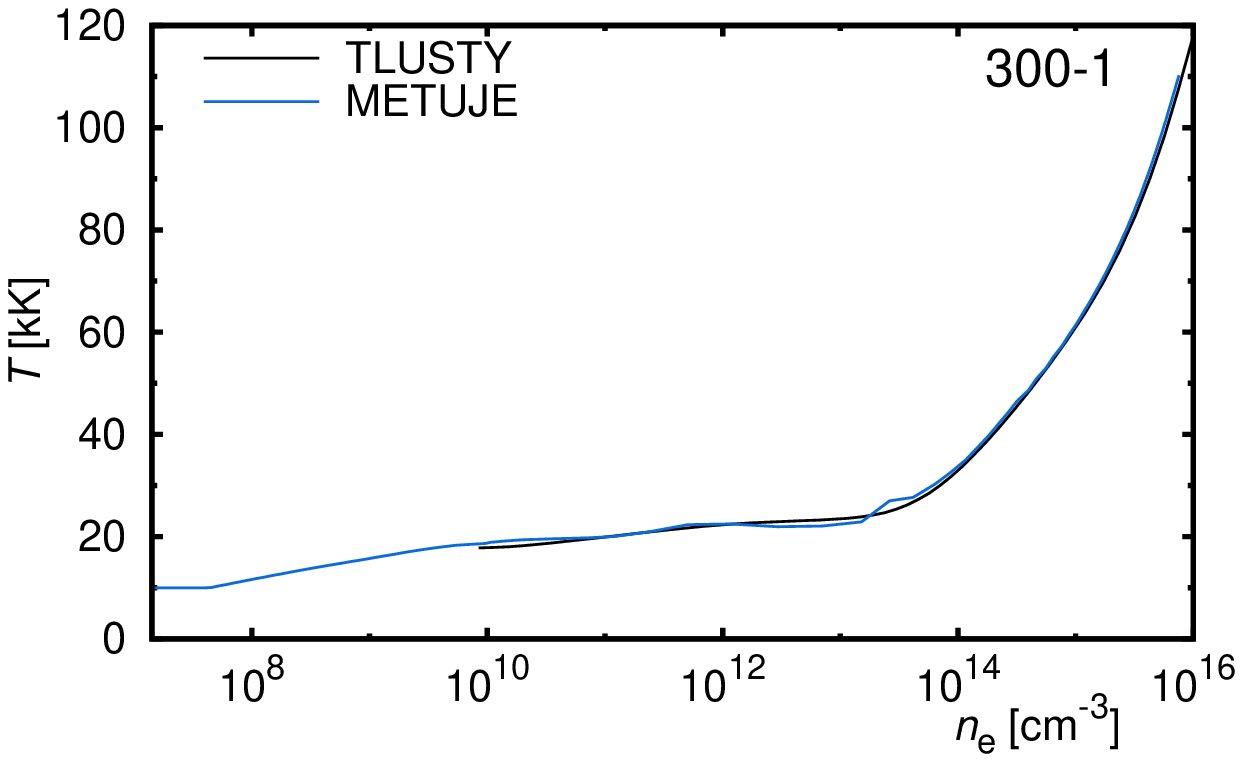}}
\resizebox{0.310\hsize}{!}{\includegraphics{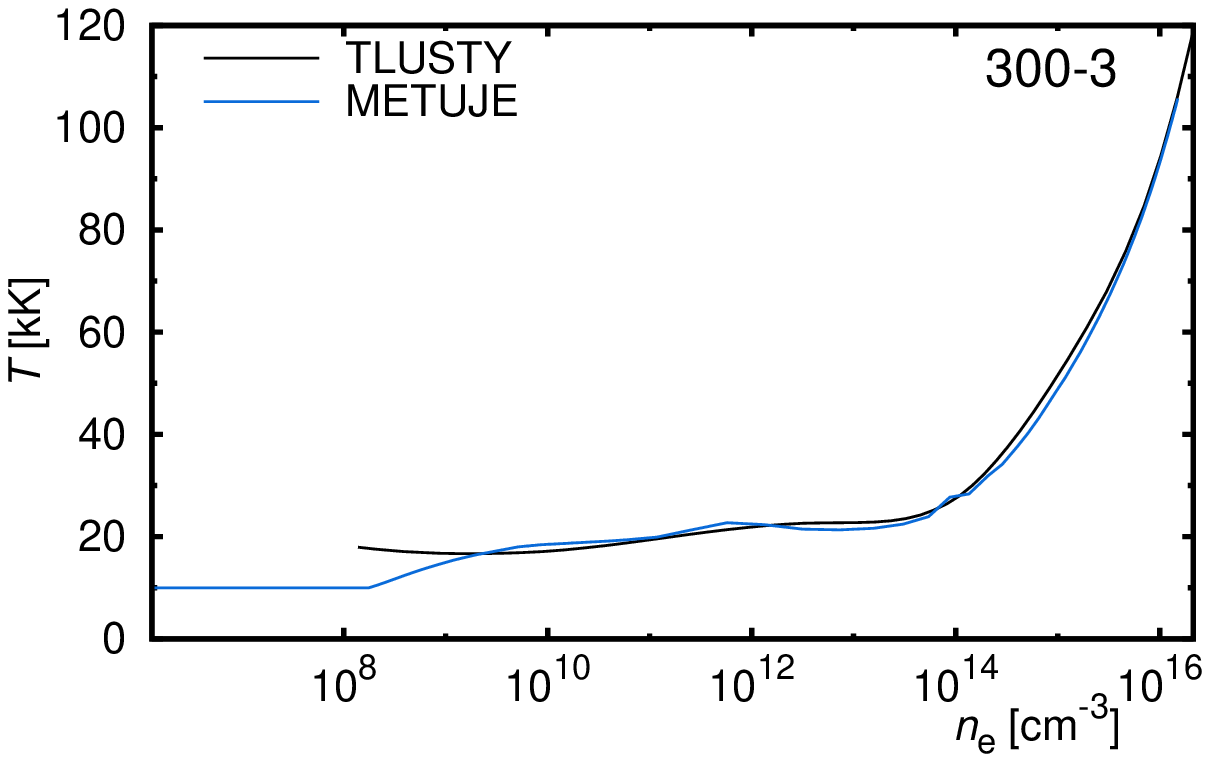}}
\resizebox{0.310\hsize}{!}{\includegraphics{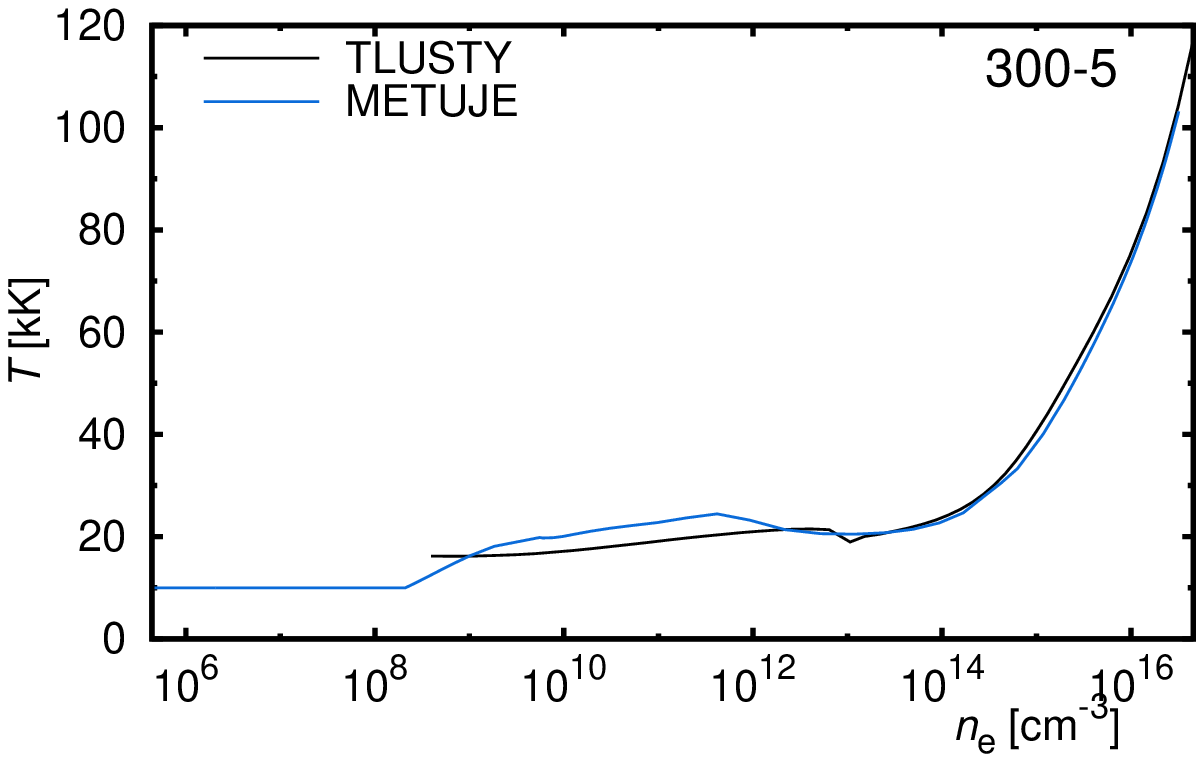}}\\
\resizebox{0.310\hsize}{!}{\includegraphics{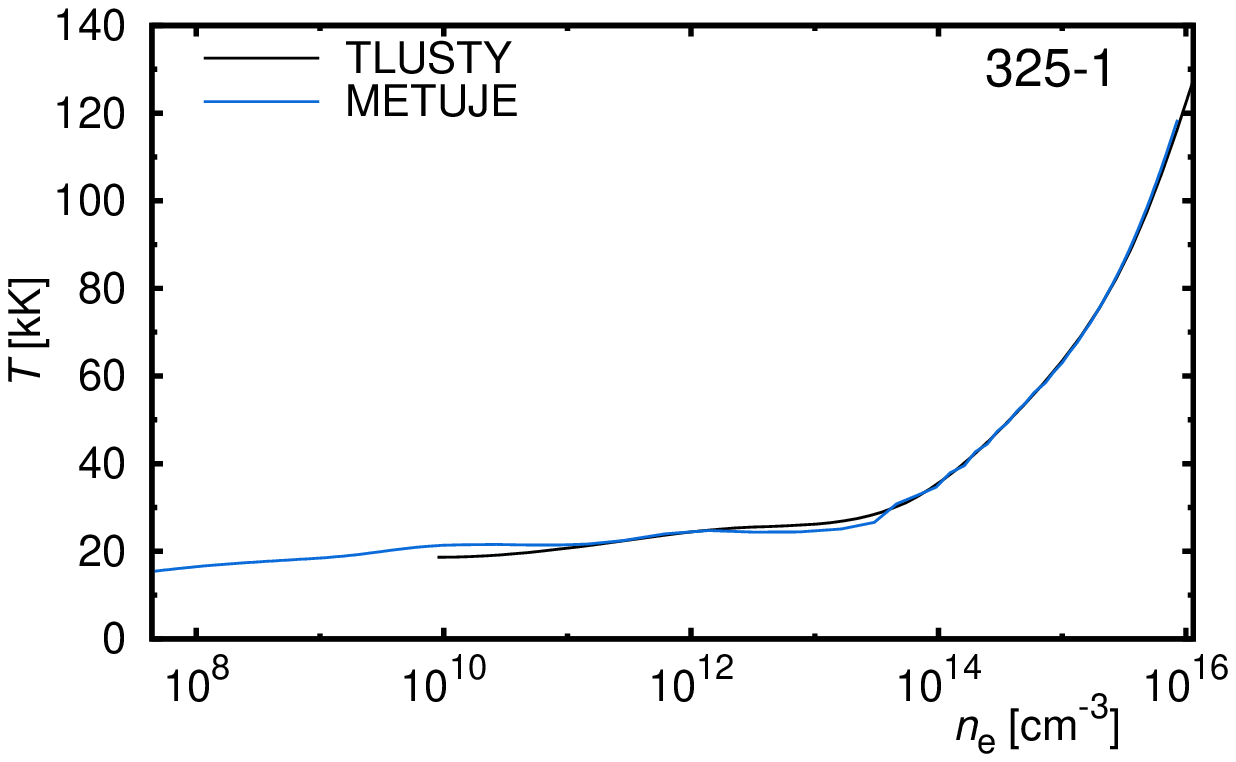}}
\resizebox{0.310\hsize}{!}{\includegraphics{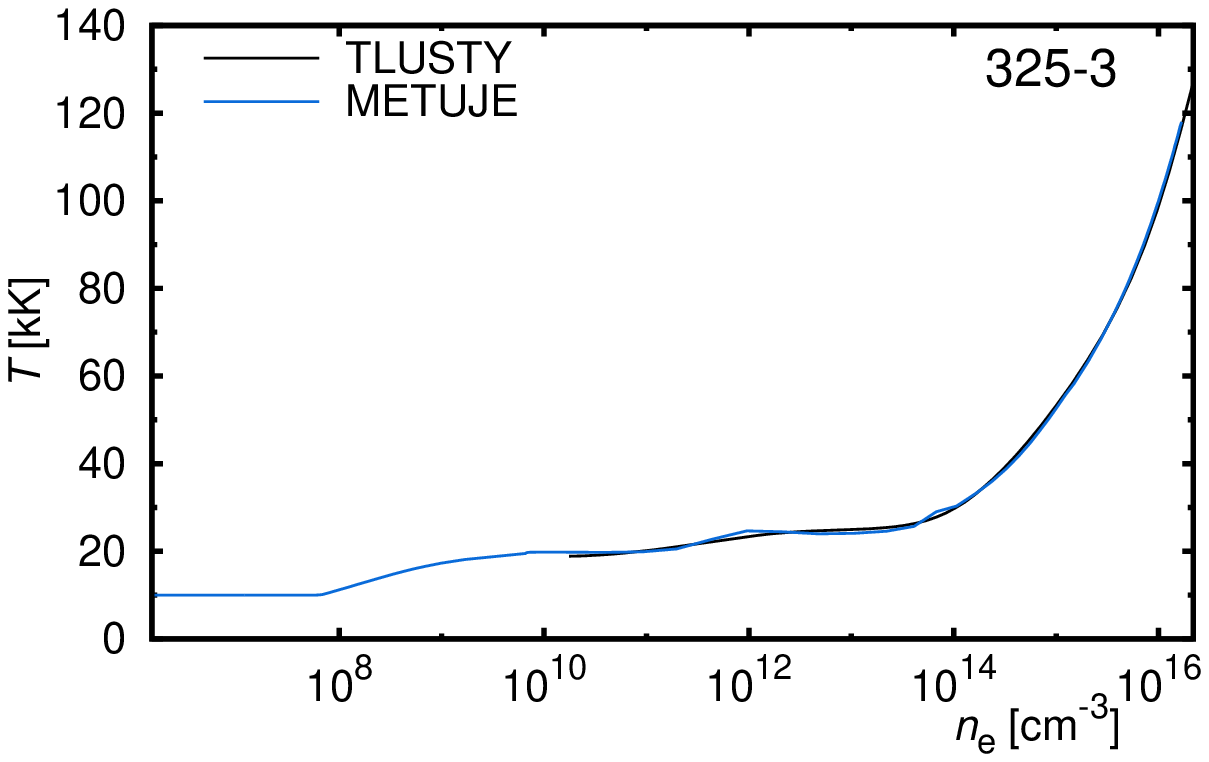}}
\resizebox{0.310\hsize}{!}{\includegraphics{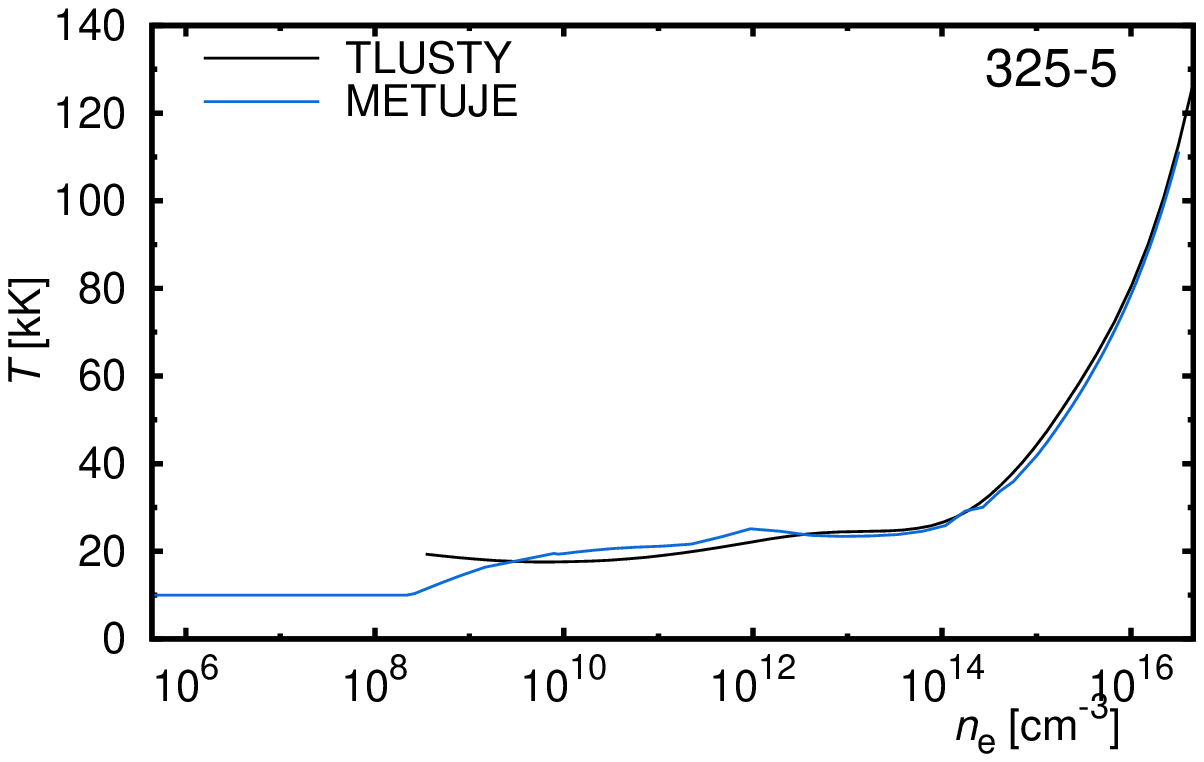}}\\
\resizebox{0.310\hsize}{!}{\includegraphics{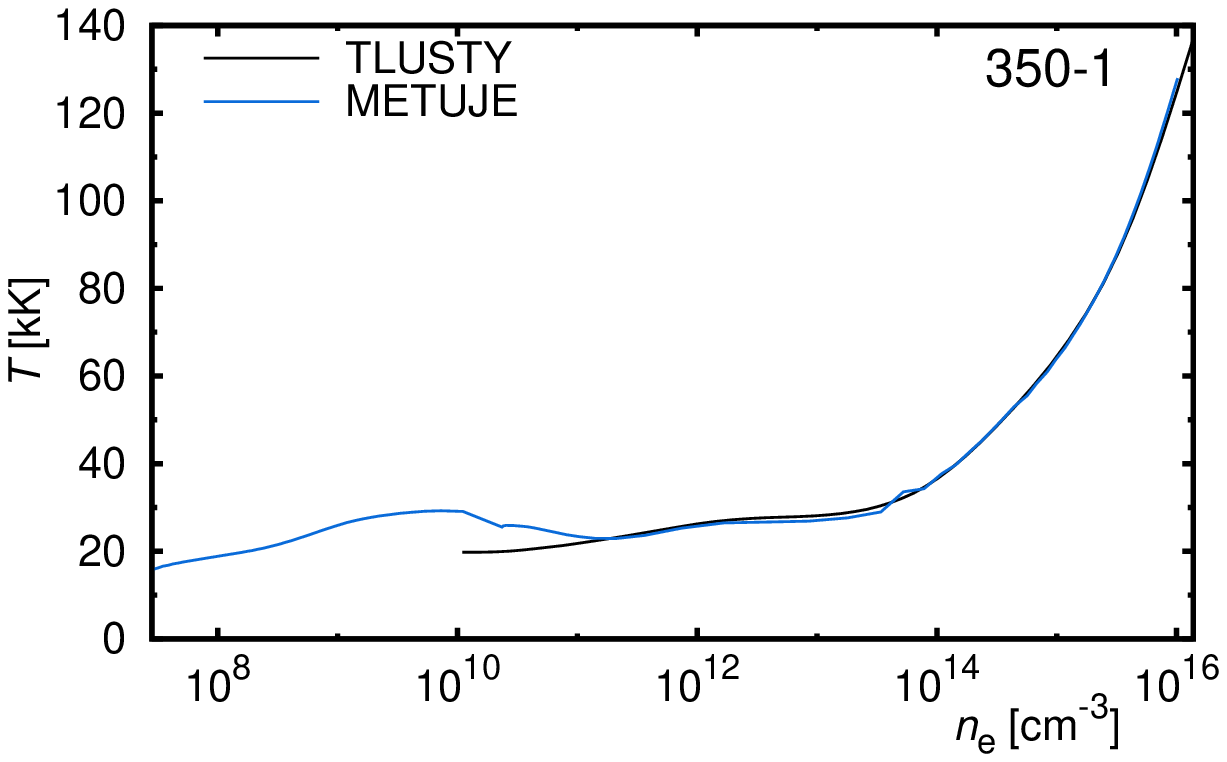}}
\resizebox{0.310\hsize}{!}{\includegraphics{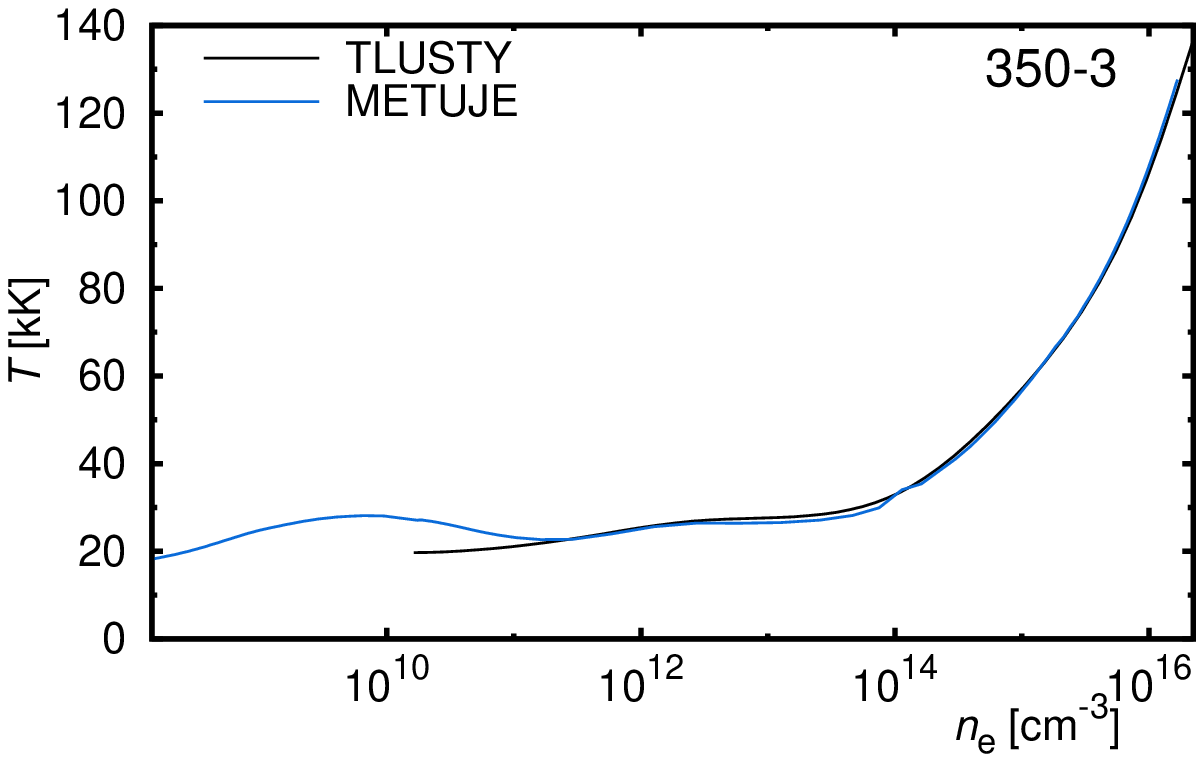}}
\resizebox{0.310\hsize}{!}{\includegraphics{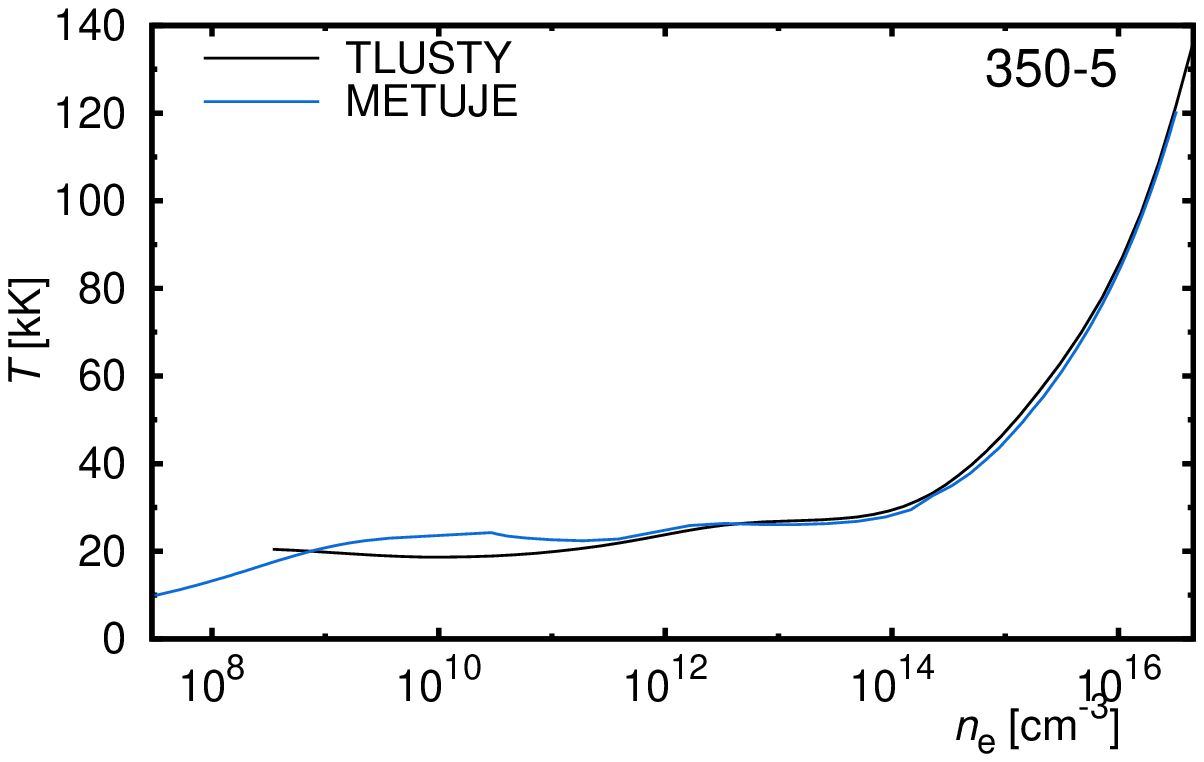}}\\
\resizebox{0.310\hsize}{!}{\includegraphics{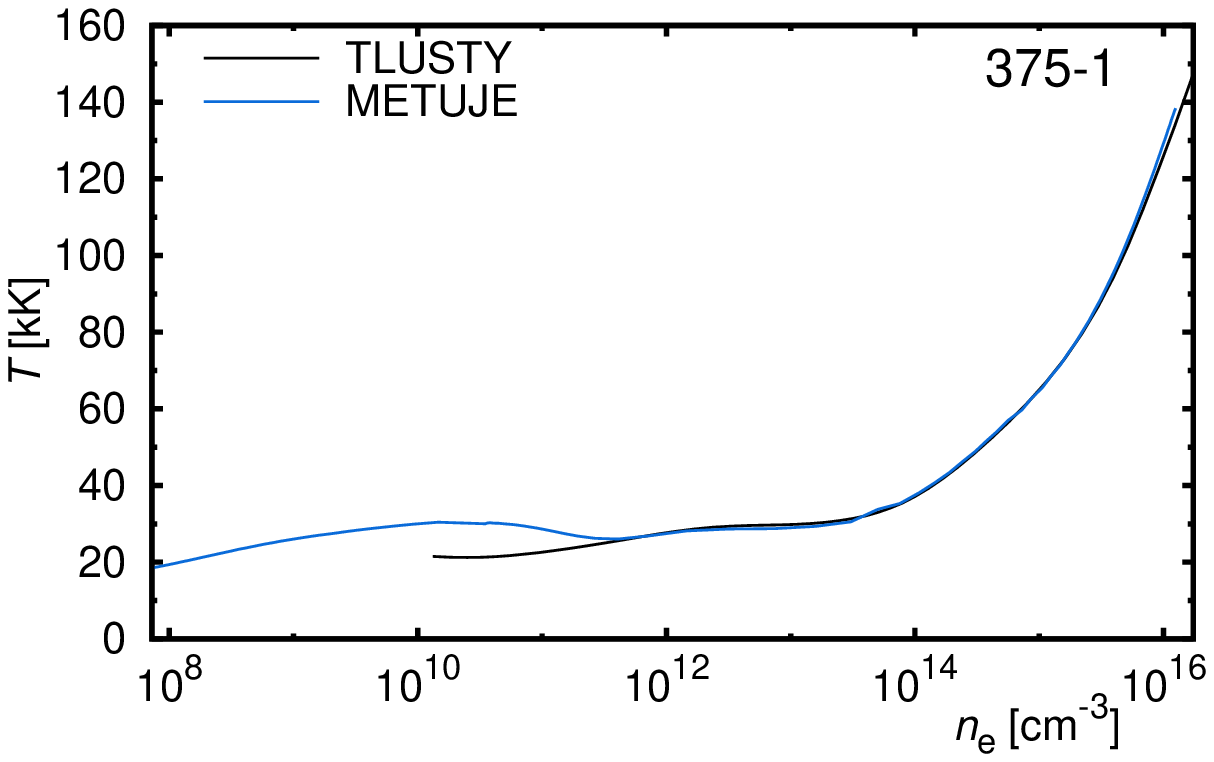}}
\resizebox{0.310\hsize}{!}{\includegraphics{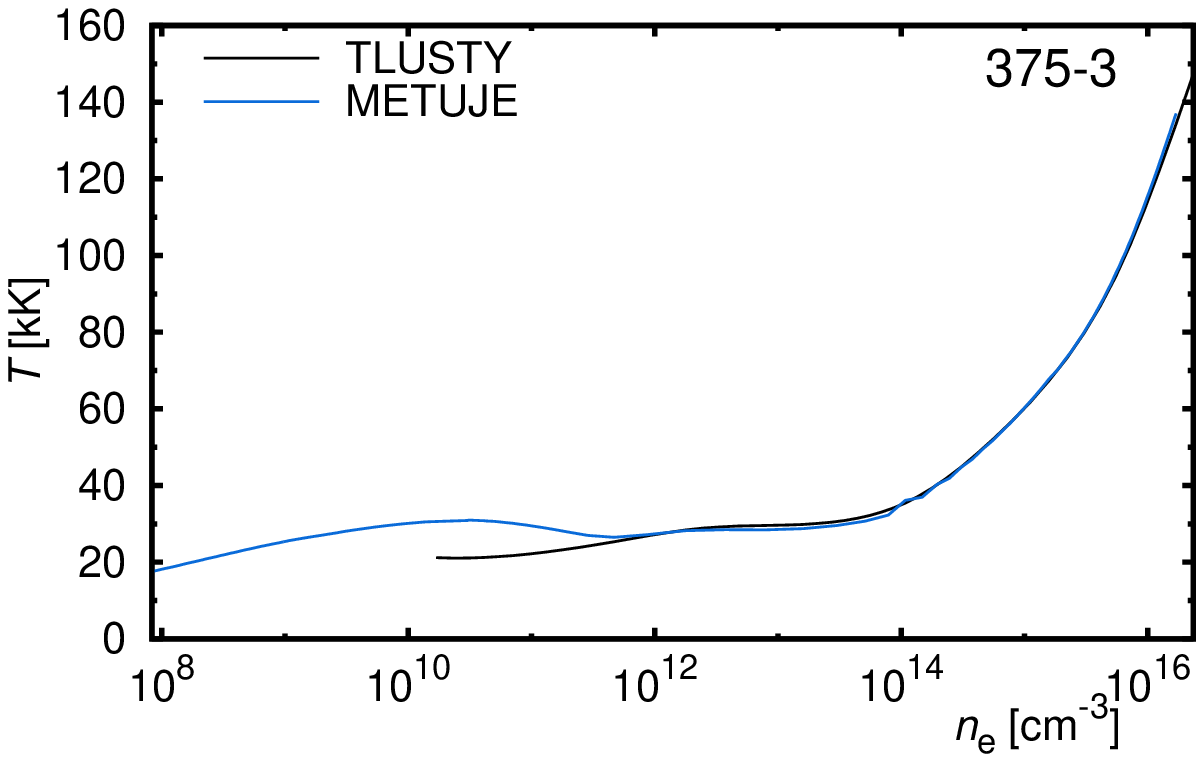}}
\resizebox{0.310\hsize}{!}{\includegraphics{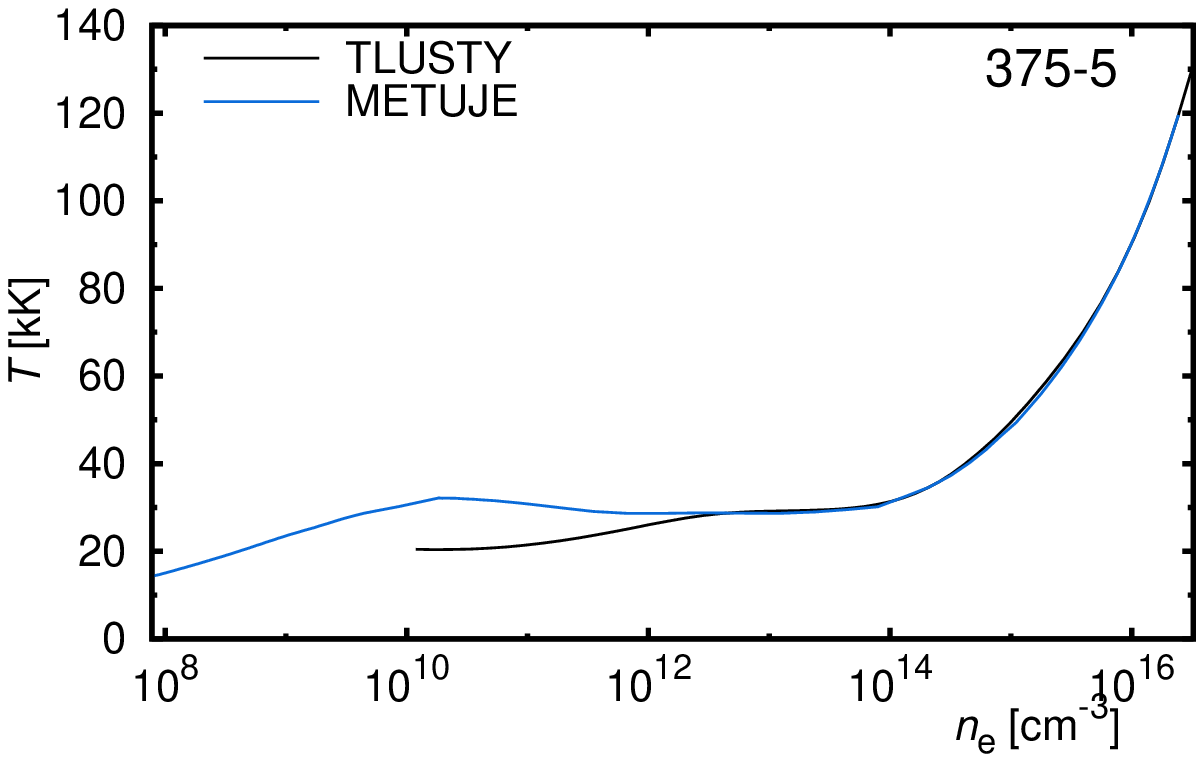}}\\
\resizebox{0.310\hsize}{!}{\includegraphics{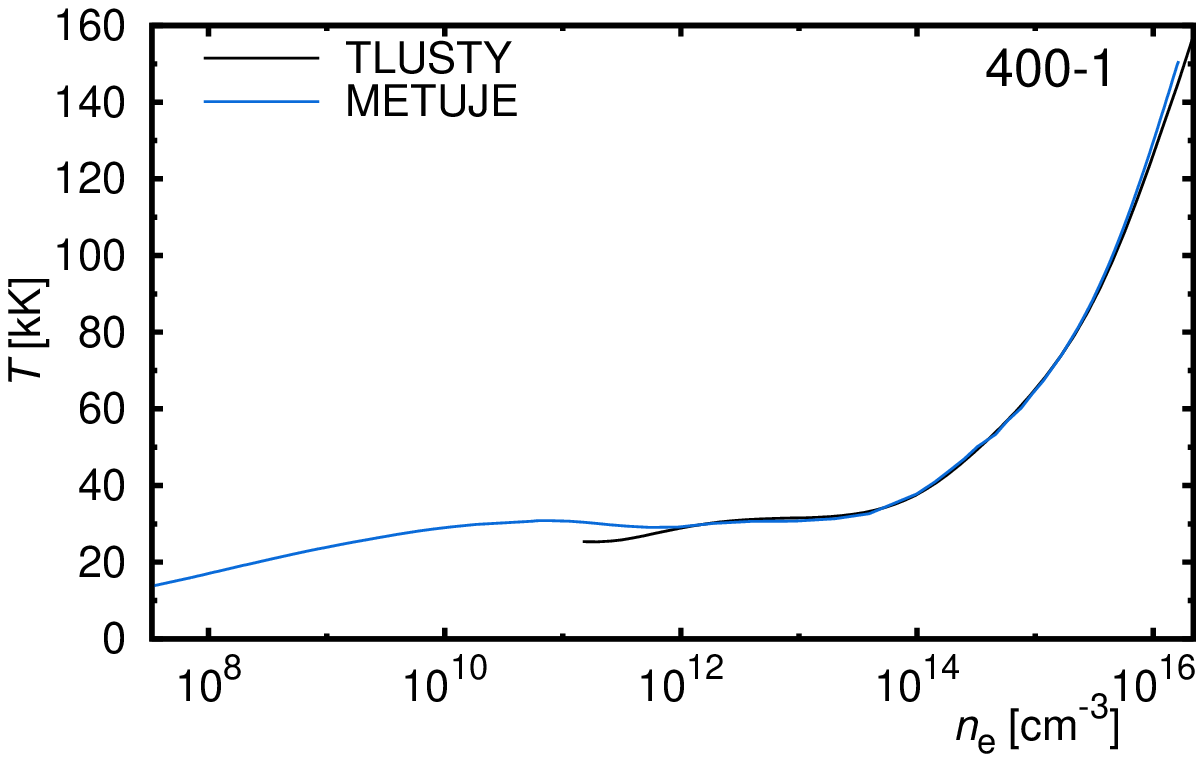}}
\resizebox{0.310\hsize}{!}{\includegraphics{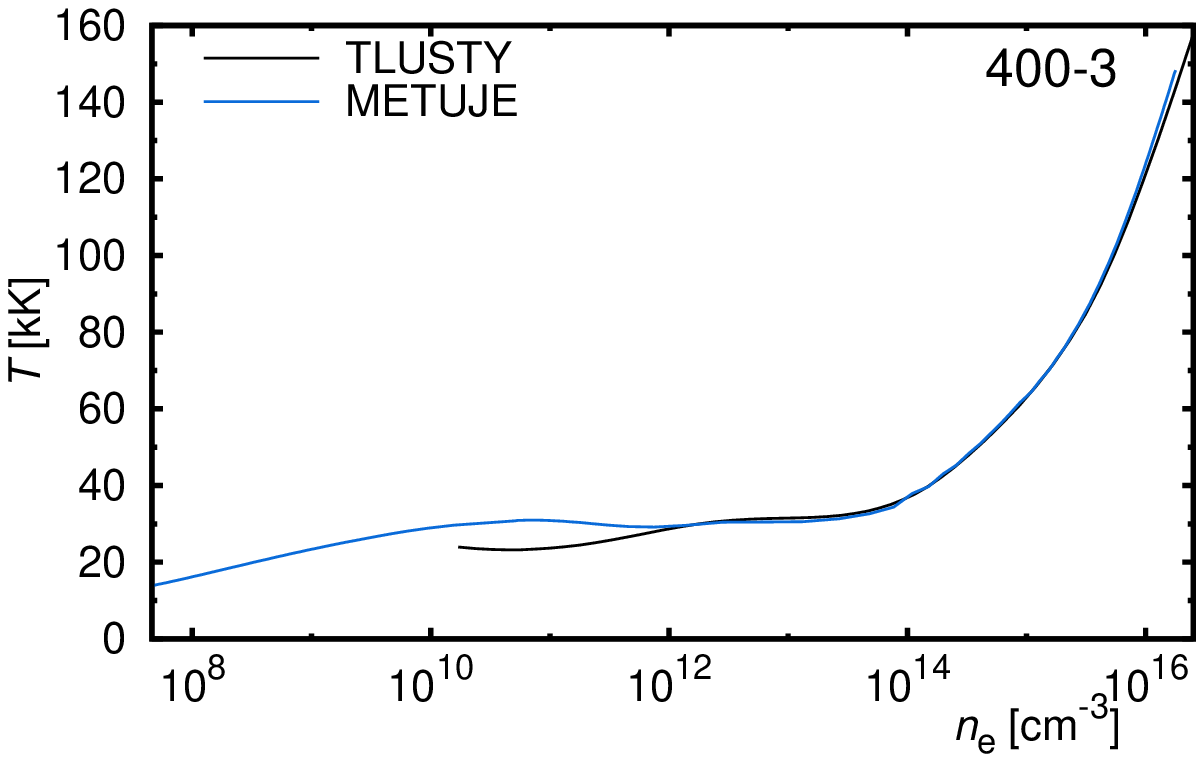}}
\resizebox{0.310\hsize}{!}{\includegraphics{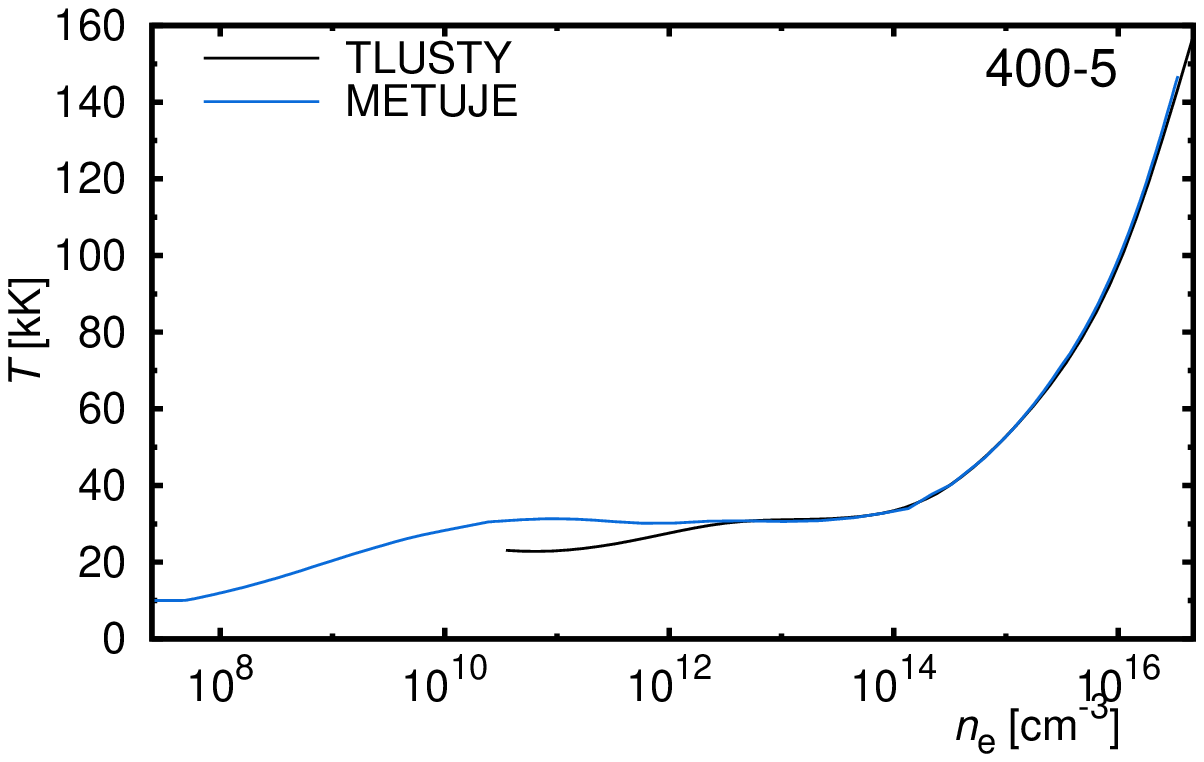}}\\
\resizebox{0.310\hsize}{!}{\includegraphics{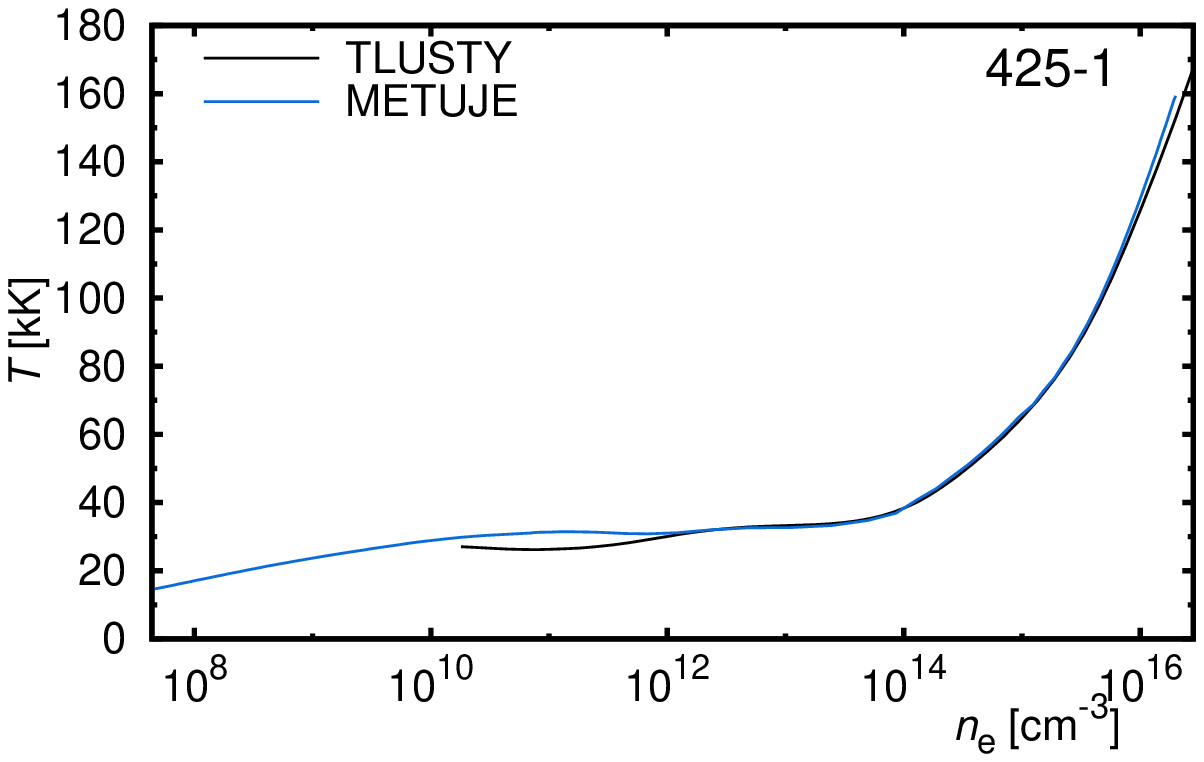}}
\resizebox{0.310\hsize}{!}{\includegraphics{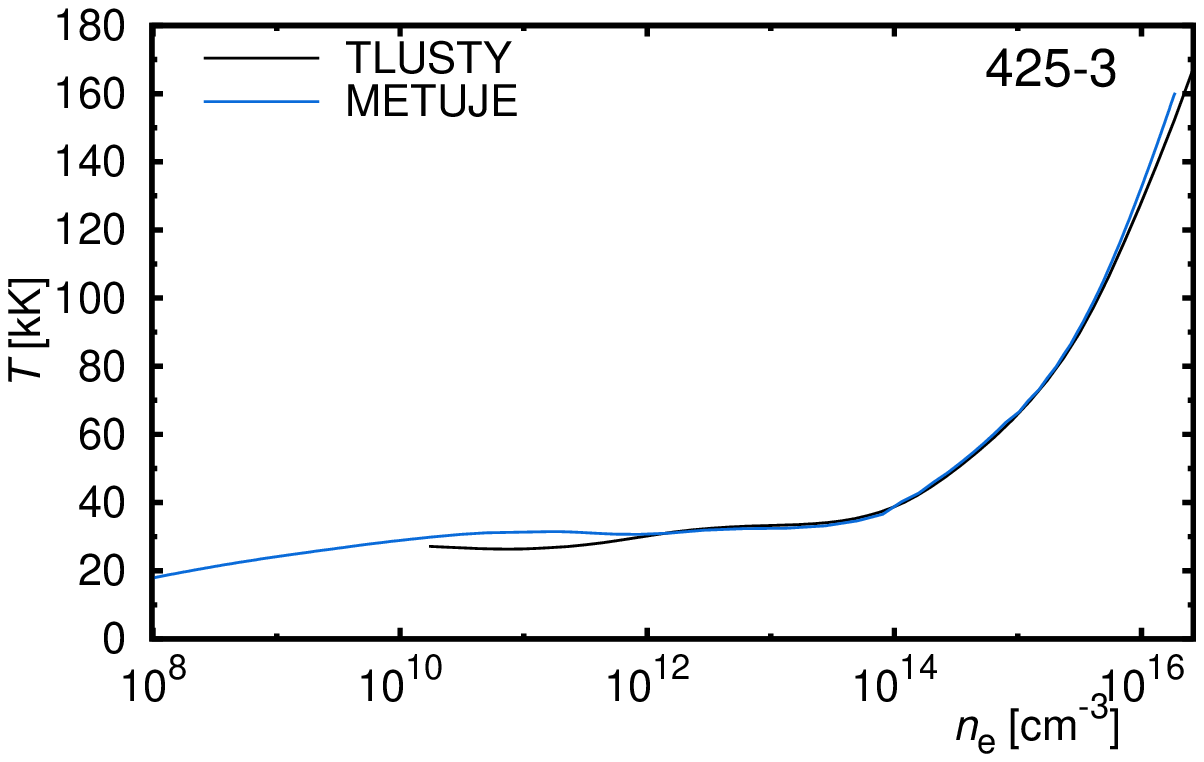}}
\resizebox{0.310\hsize}{!}{\includegraphics{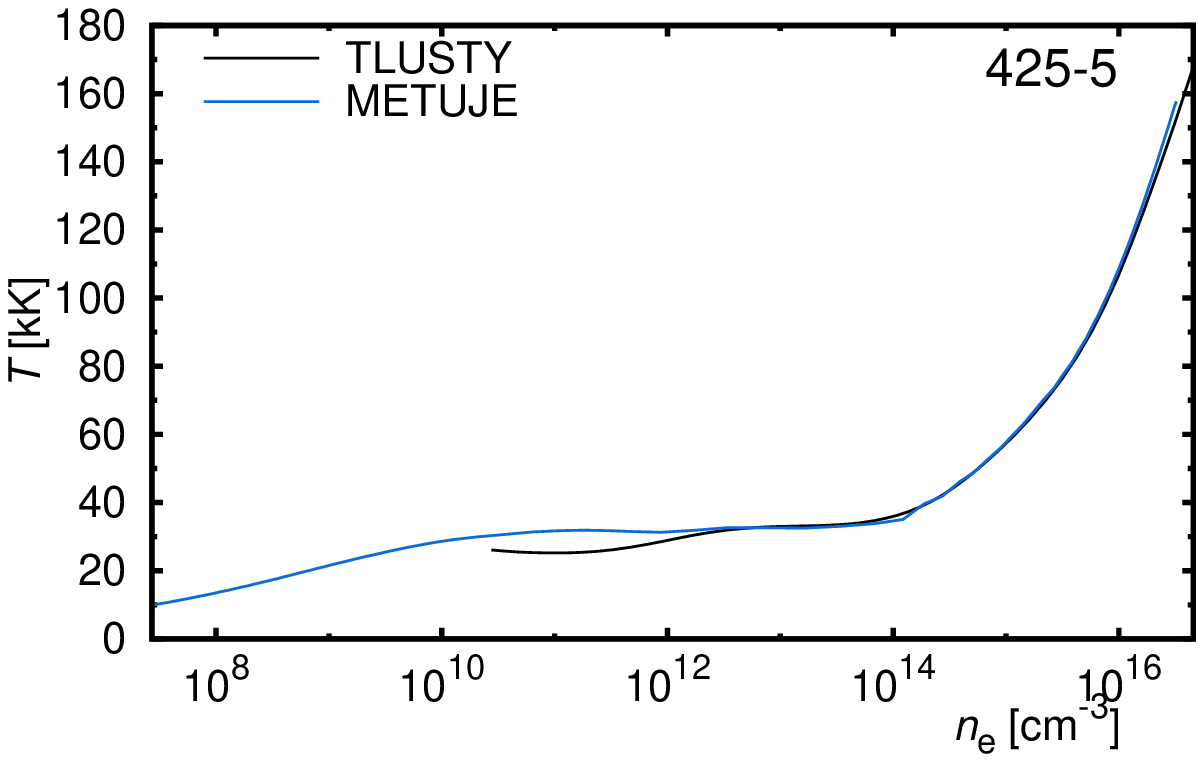}}\\
\resizebox{0.310\hsize}{!}{\includegraphics{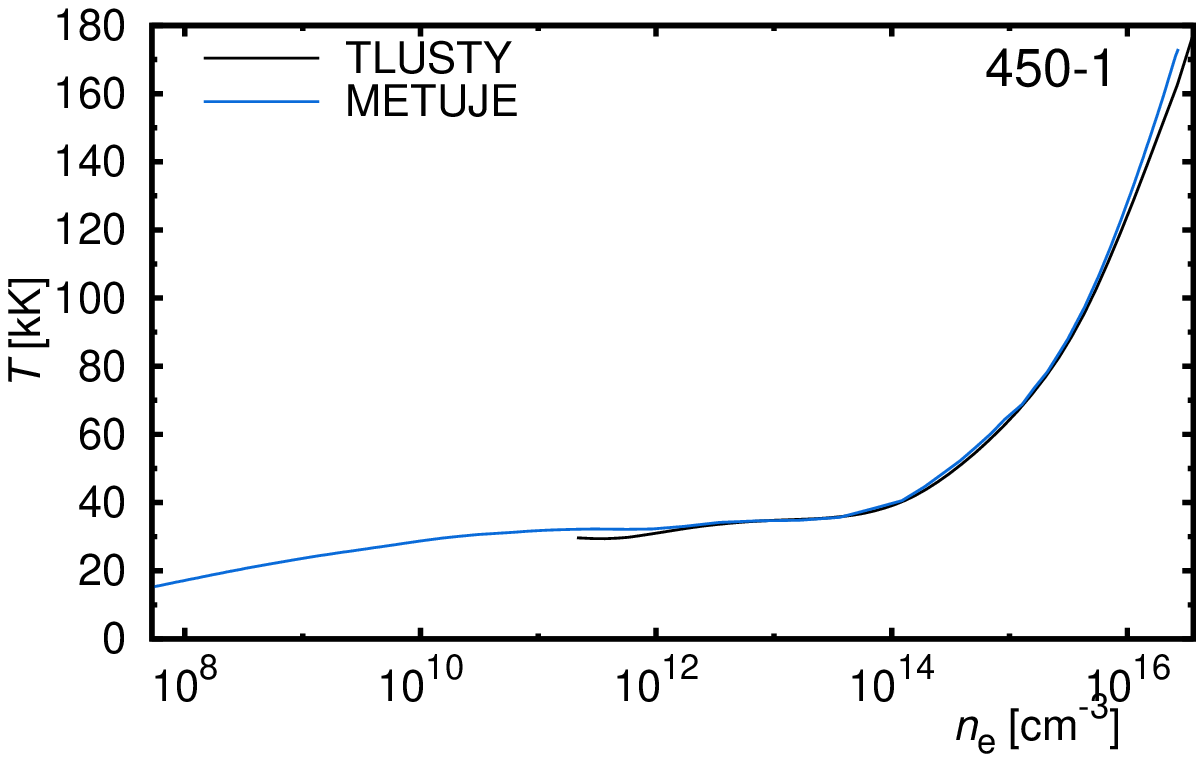}}
\resizebox{0.310\hsize}{!}{\includegraphics{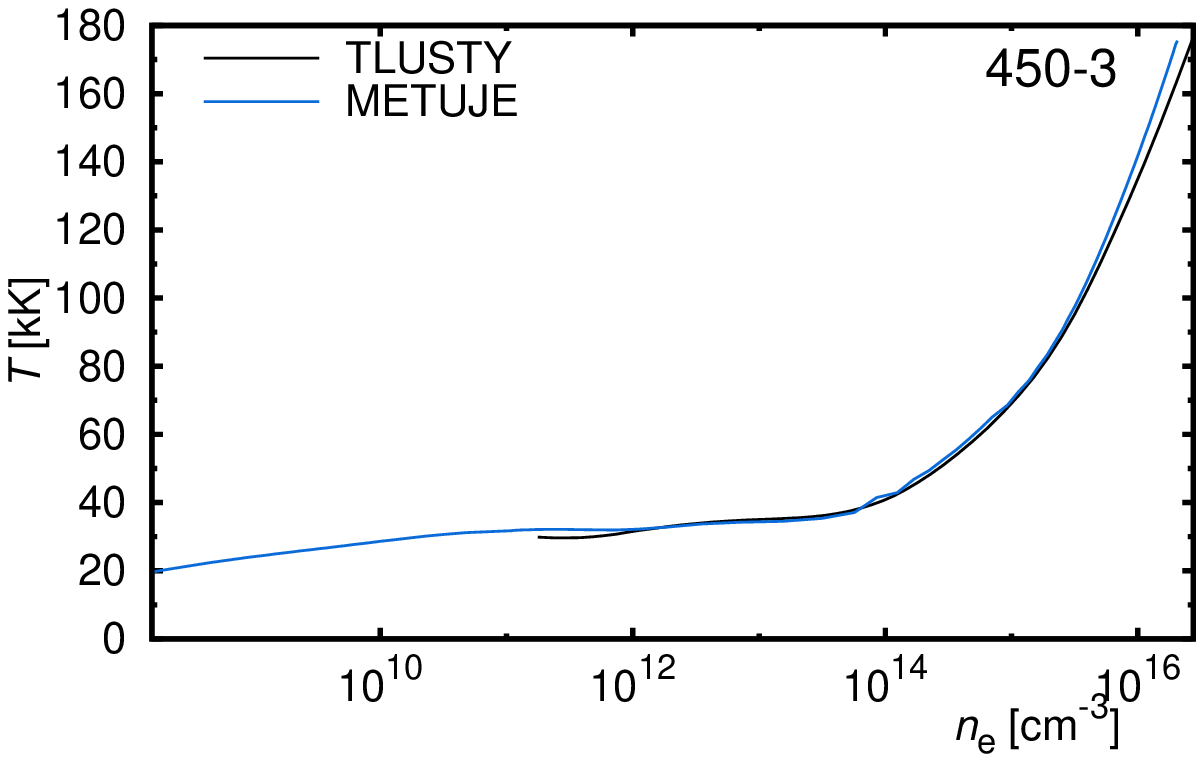}}
\resizebox{0.310\hsize}{!}{\includegraphics{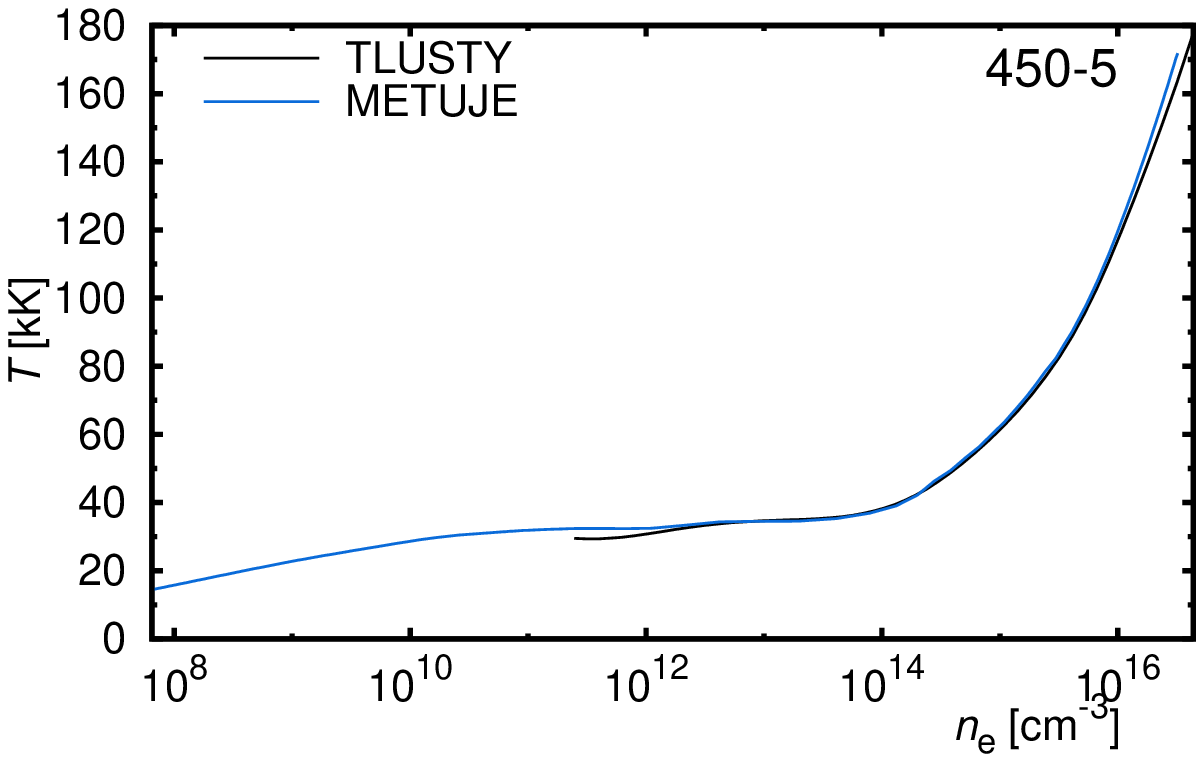}}
\caption{Comparison of the dependence of temperature on electron density in
TLUSTY and METUJE models for SMC stars. The graphs are plotted for individual
model stars from Table~\ref{ohvezpar} (denoted in the graphs).}
\label{tepmetlum}
\end{figure*}

\begin{figure*}[tp]
\centering
\resizebox{0.310\hsize}{!}{\includegraphics{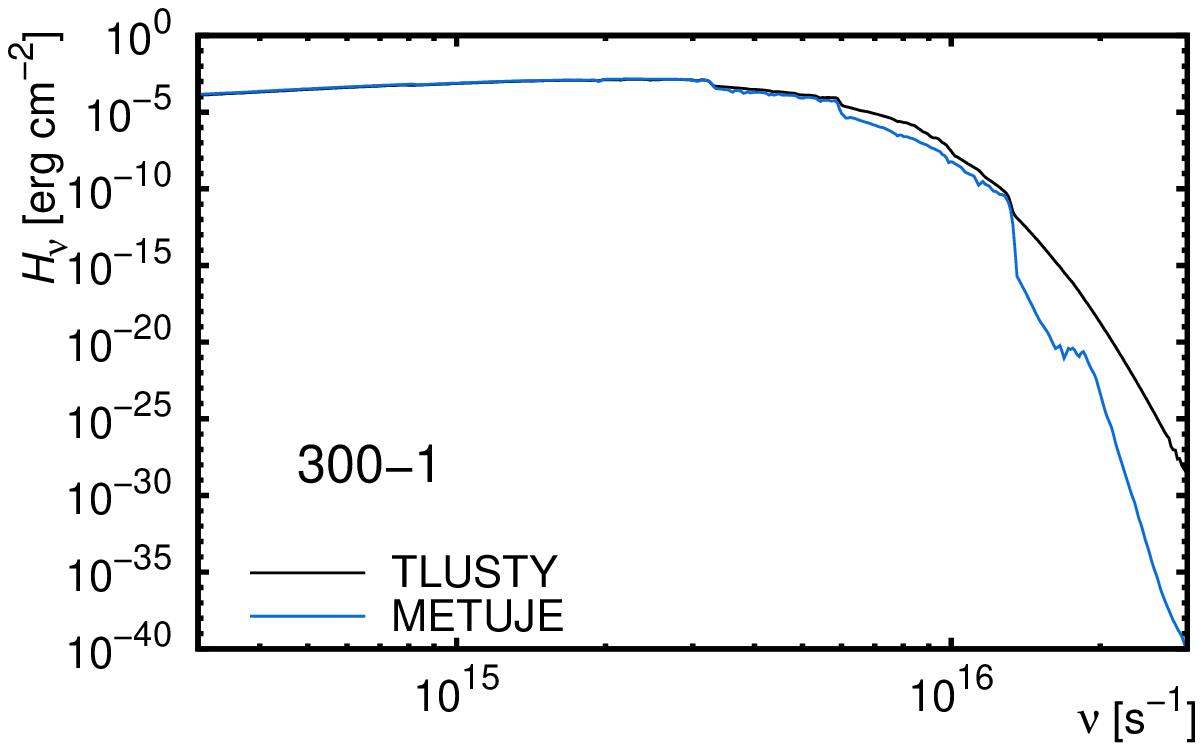}}
\resizebox{0.310\hsize}{!}{\includegraphics{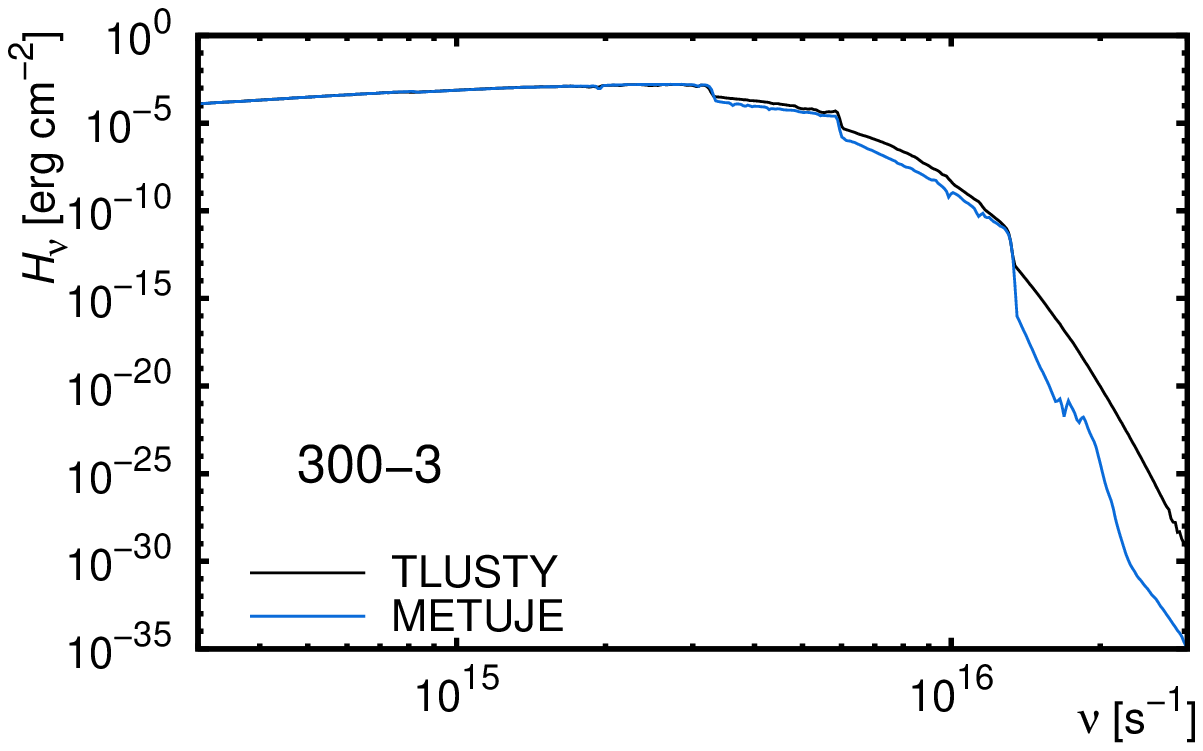}}
\resizebox{0.310\hsize}{!}{\includegraphics{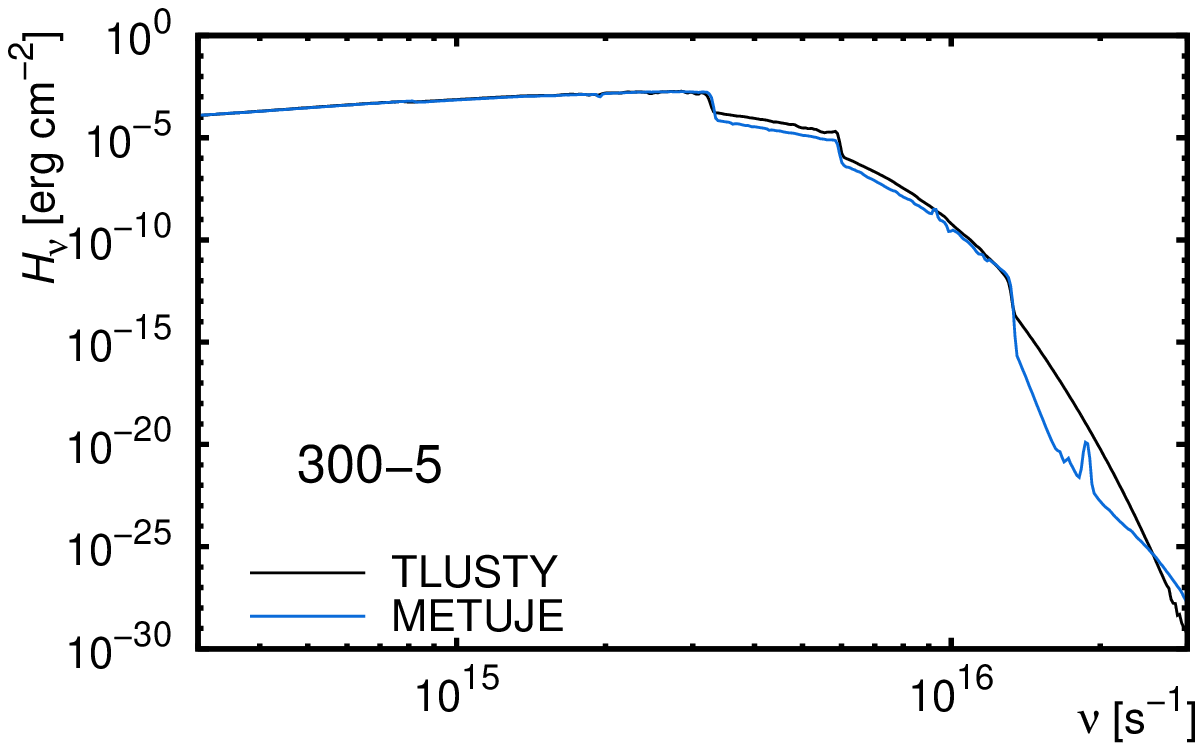}}\\
\resizebox{0.310\hsize}{!}{\includegraphics{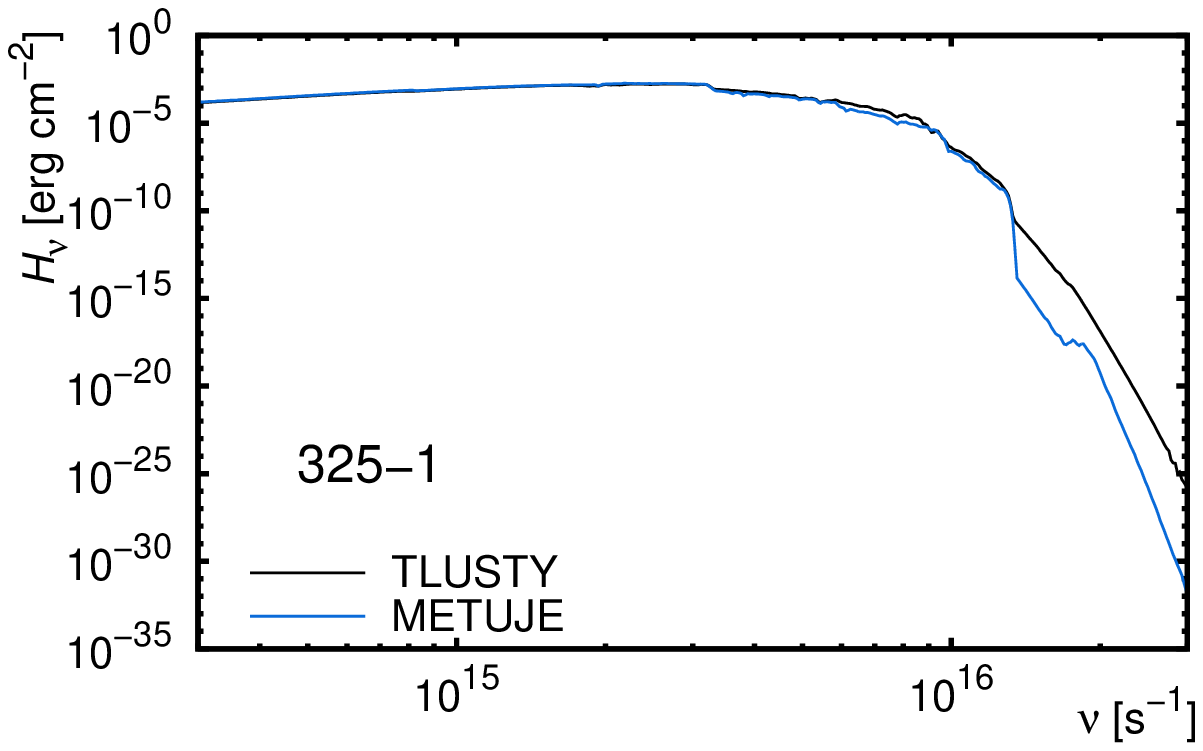}}
\resizebox{0.310\hsize}{!}{\includegraphics{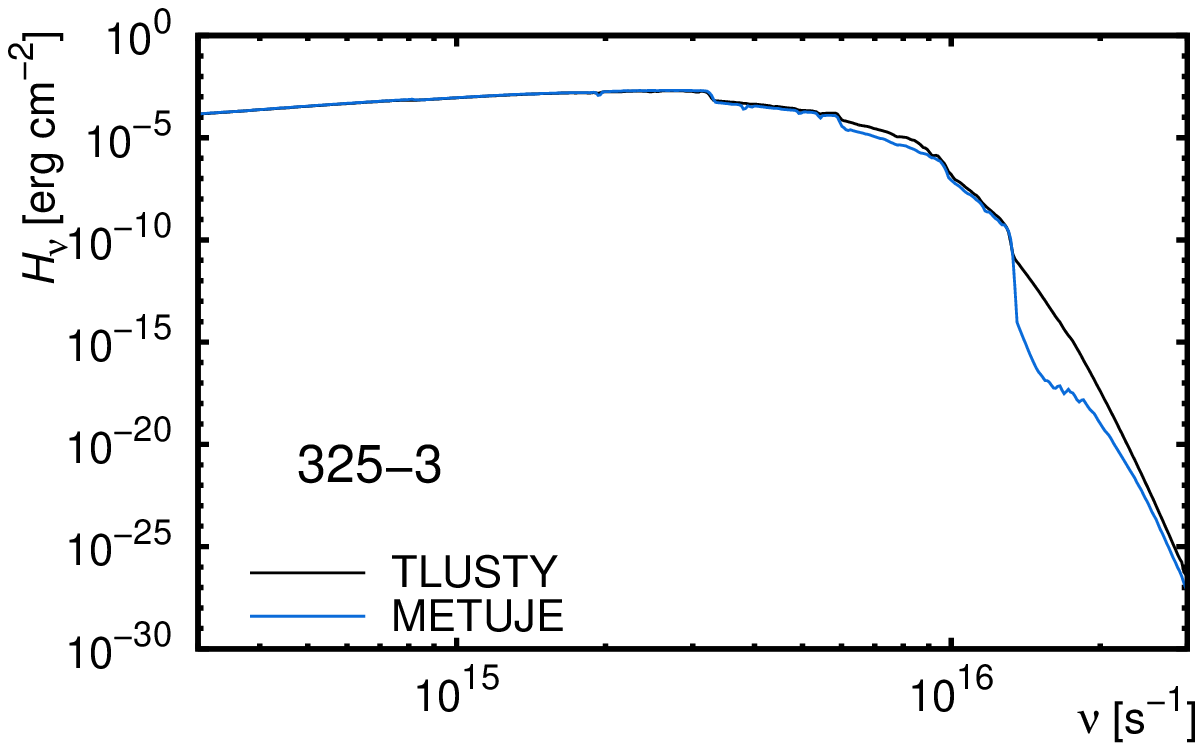}}
\resizebox{0.310\hsize}{!}{\includegraphics{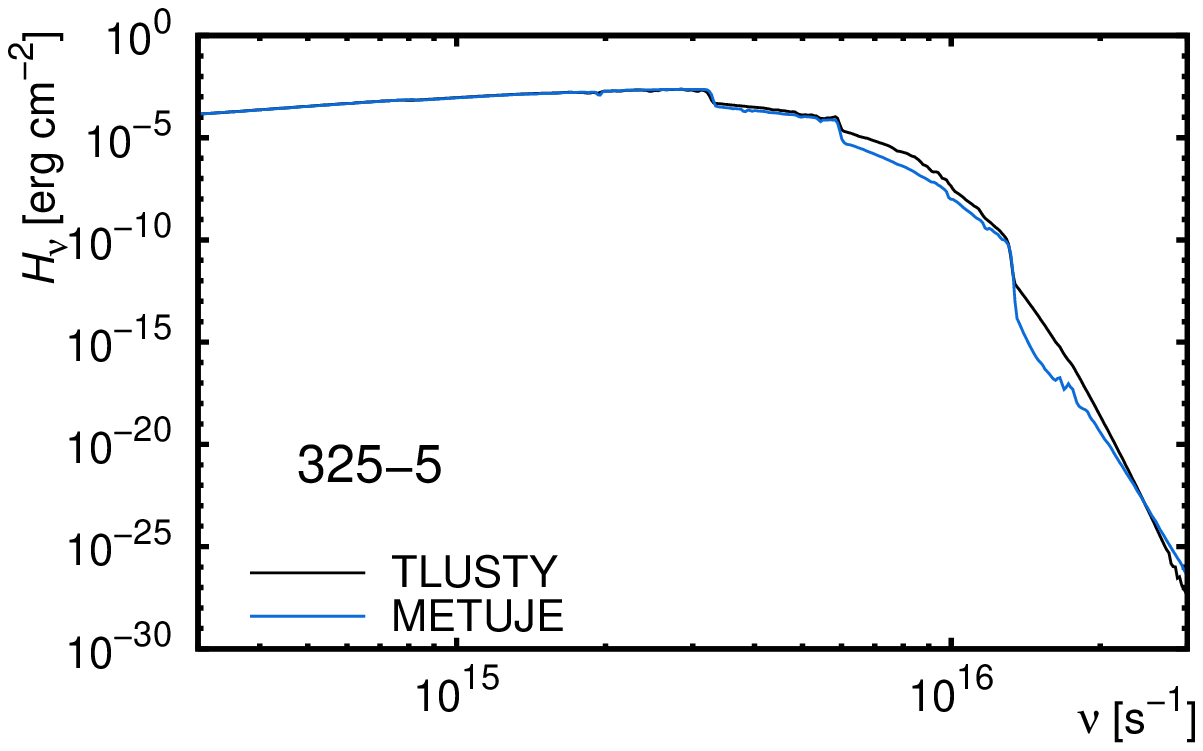}}\\
\resizebox{0.310\hsize}{!}{\includegraphics{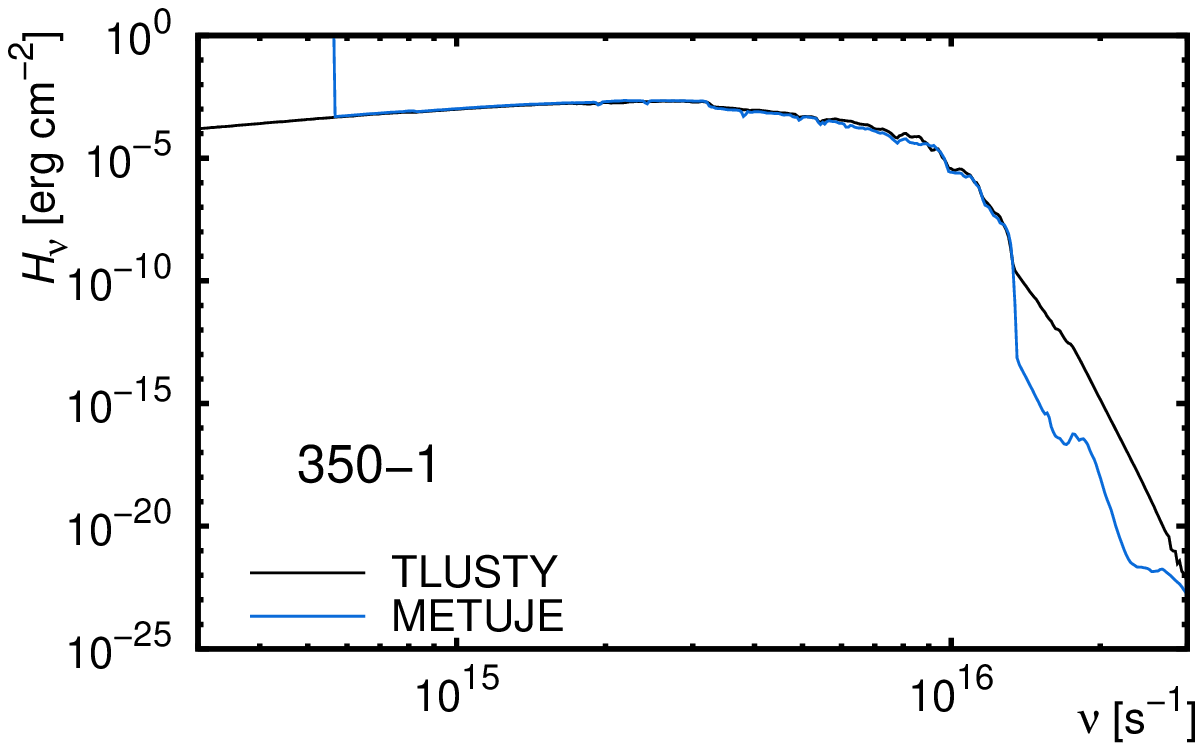}}
\resizebox{0.310\hsize}{!}{\includegraphics{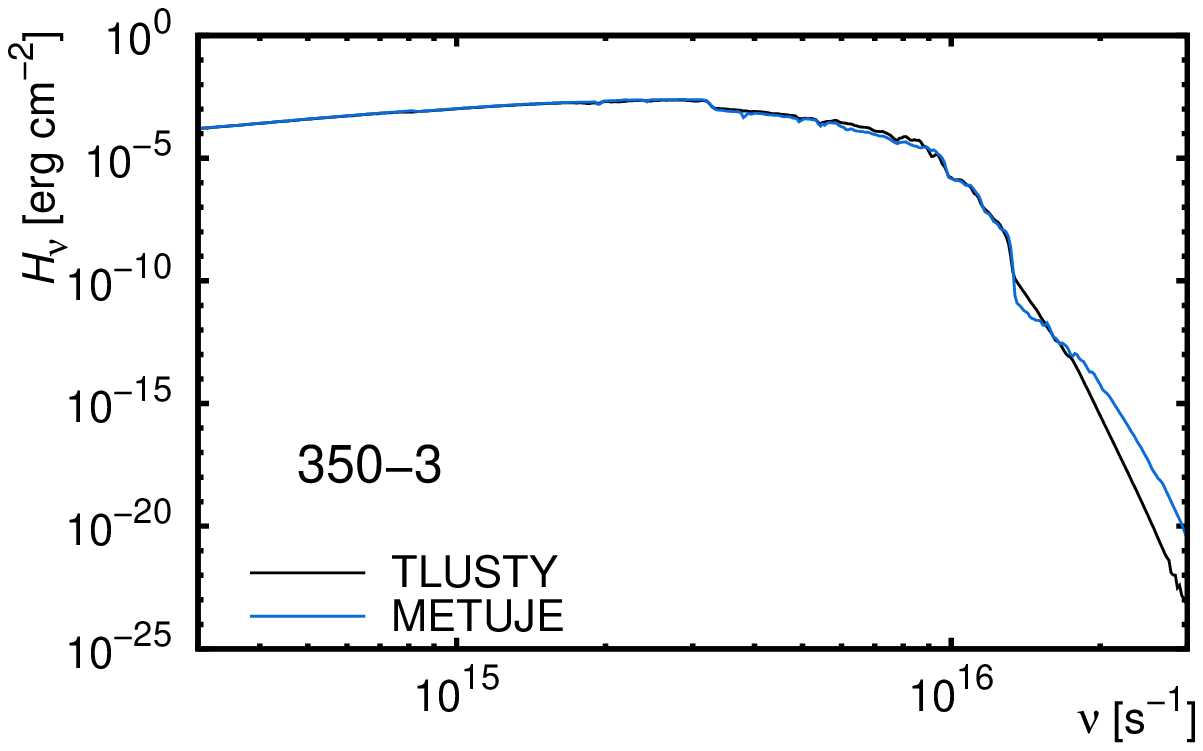}}
\resizebox{0.310\hsize}{!}{\includegraphics{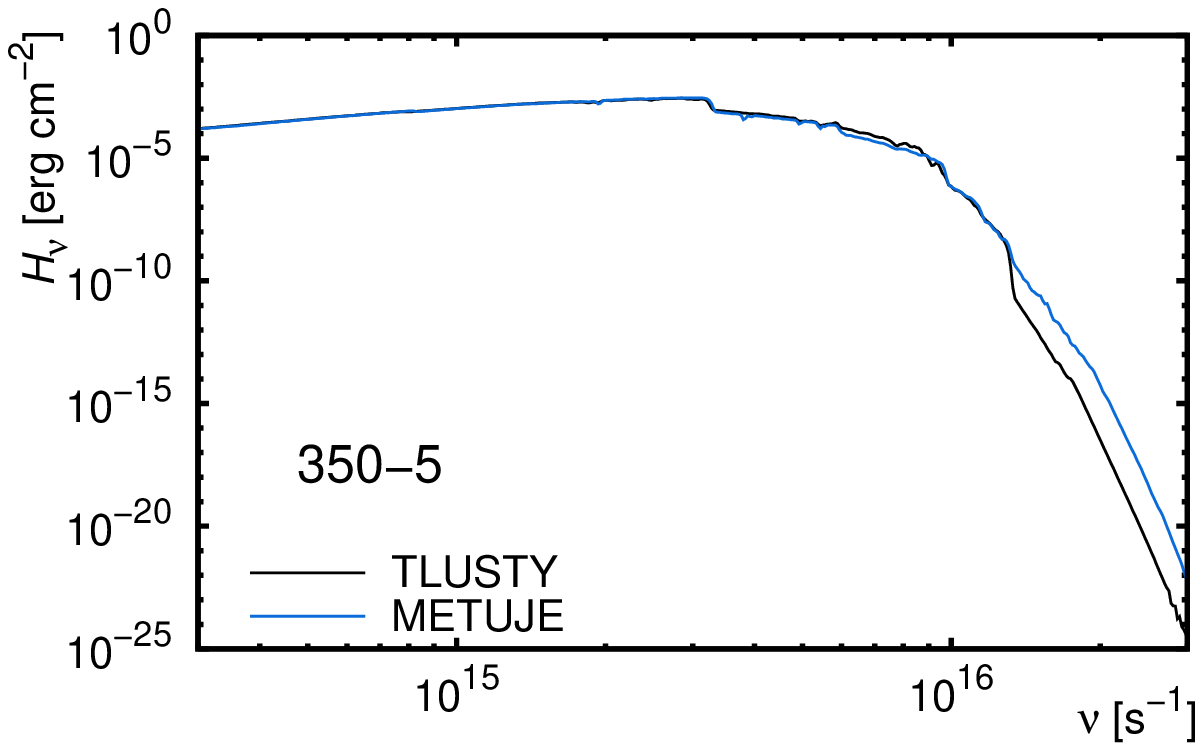}}\\
\resizebox{0.310\hsize}{!}{\includegraphics{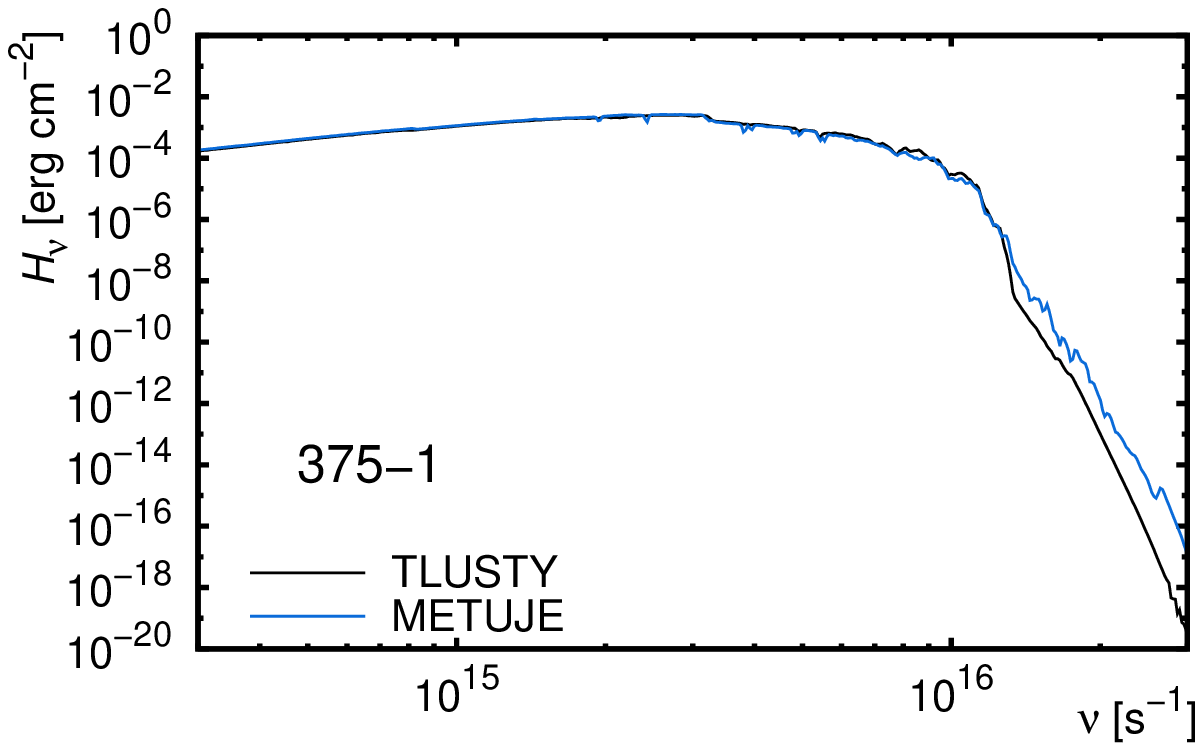}}
\resizebox{0.310\hsize}{!}{\includegraphics{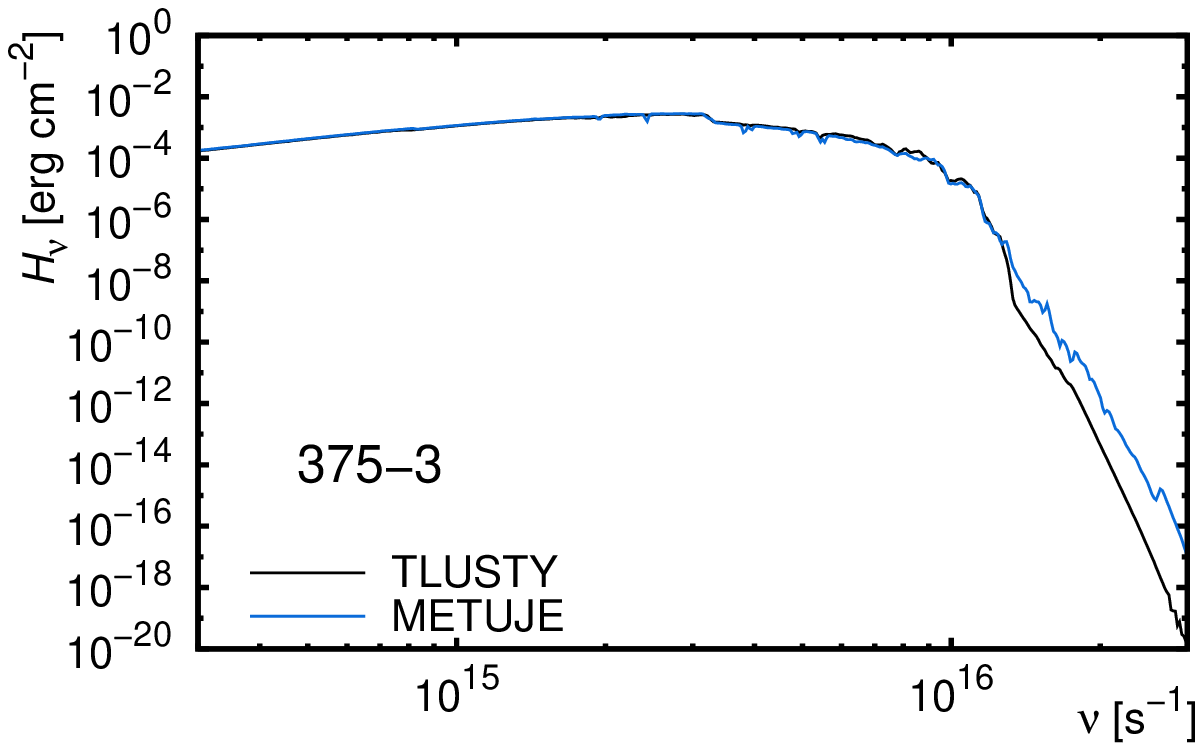}}
\resizebox{0.310\hsize}{!}{\includegraphics{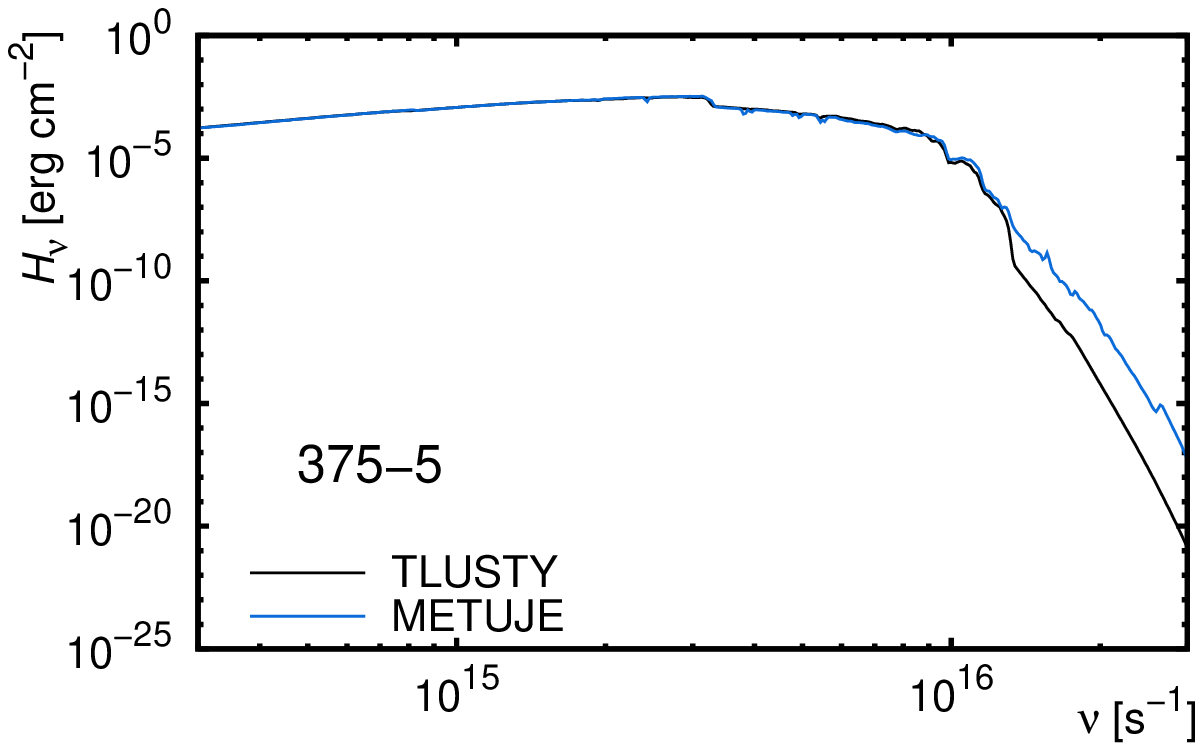}}\\
\resizebox{0.310\hsize}{!}{\includegraphics{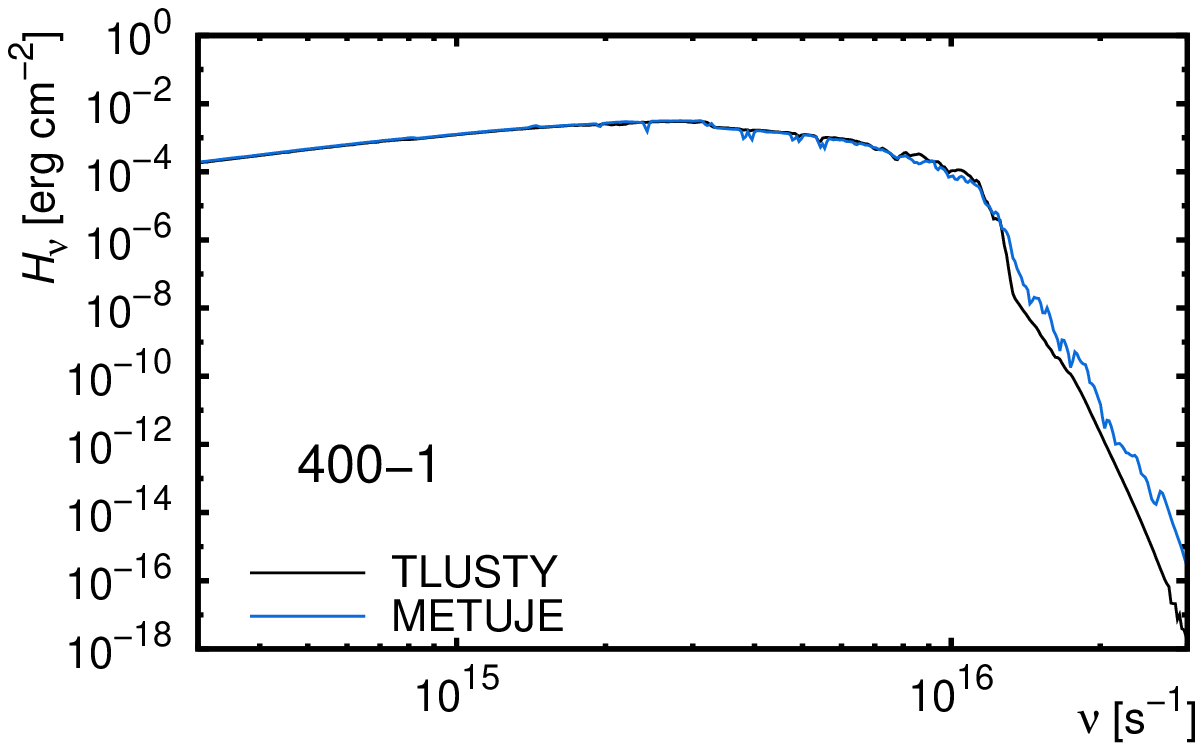}}
\resizebox{0.310\hsize}{!}{\includegraphics{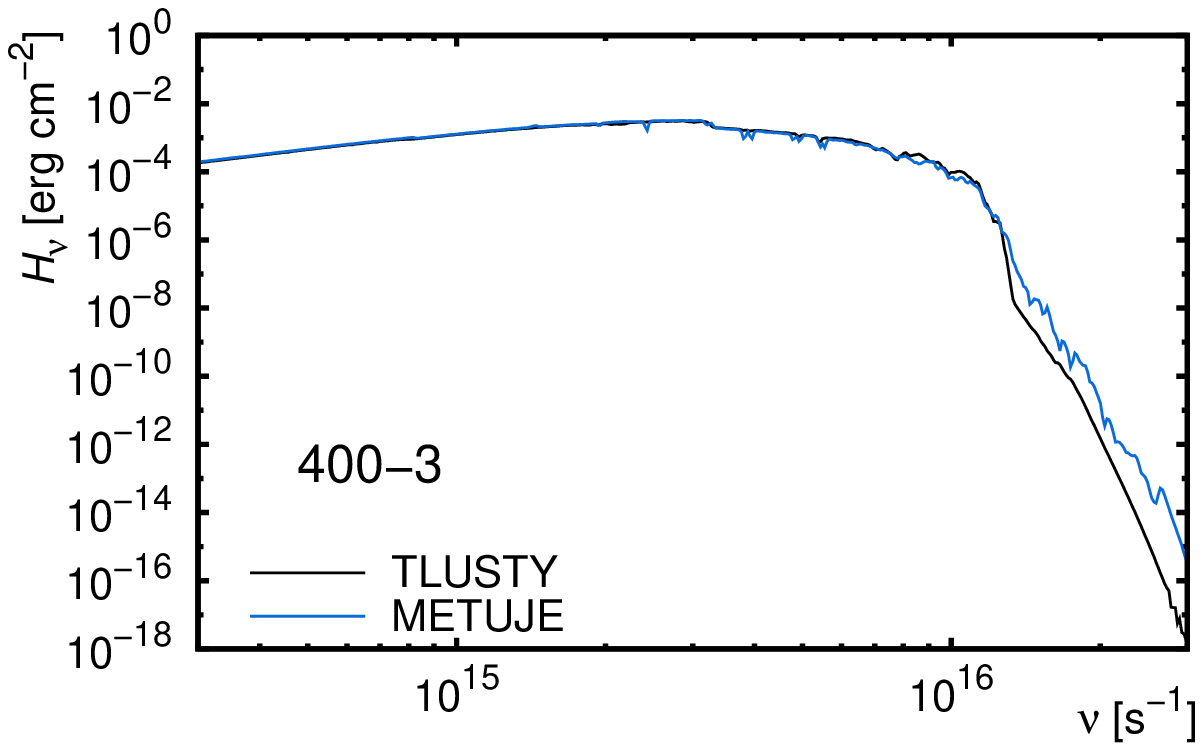}}
\resizebox{0.310\hsize}{!}{\includegraphics{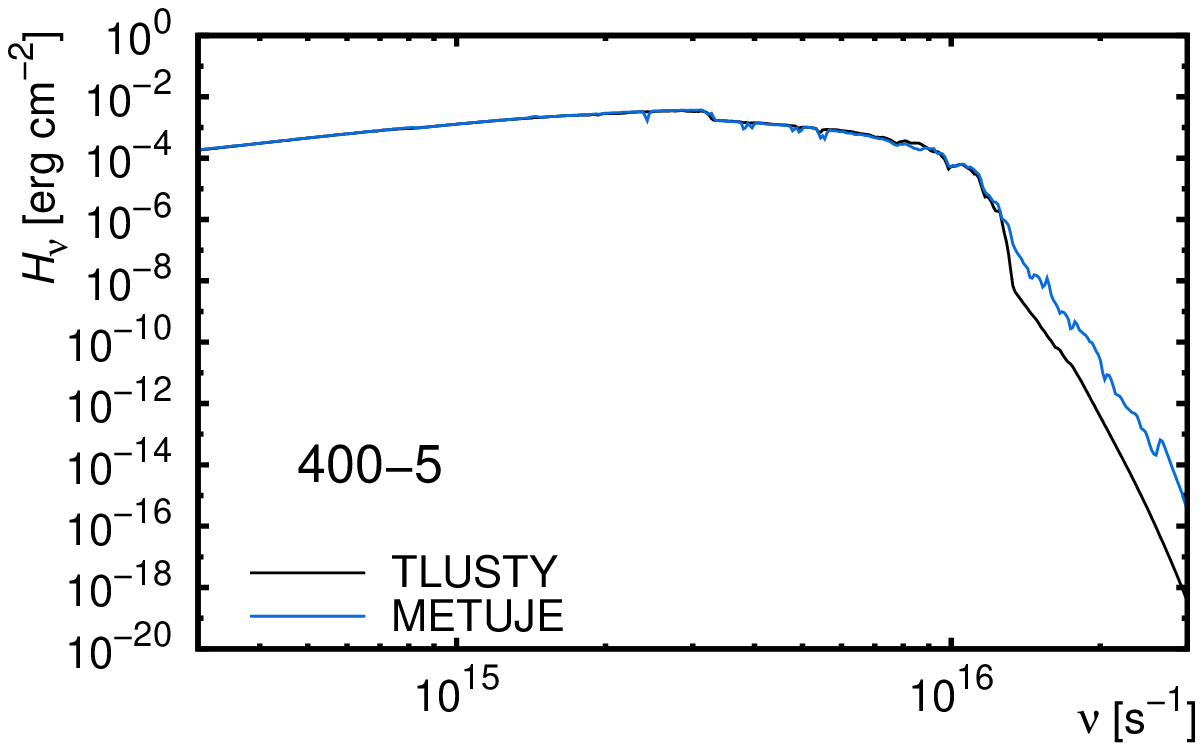}}\\
\resizebox{0.310\hsize}{!}{\includegraphics{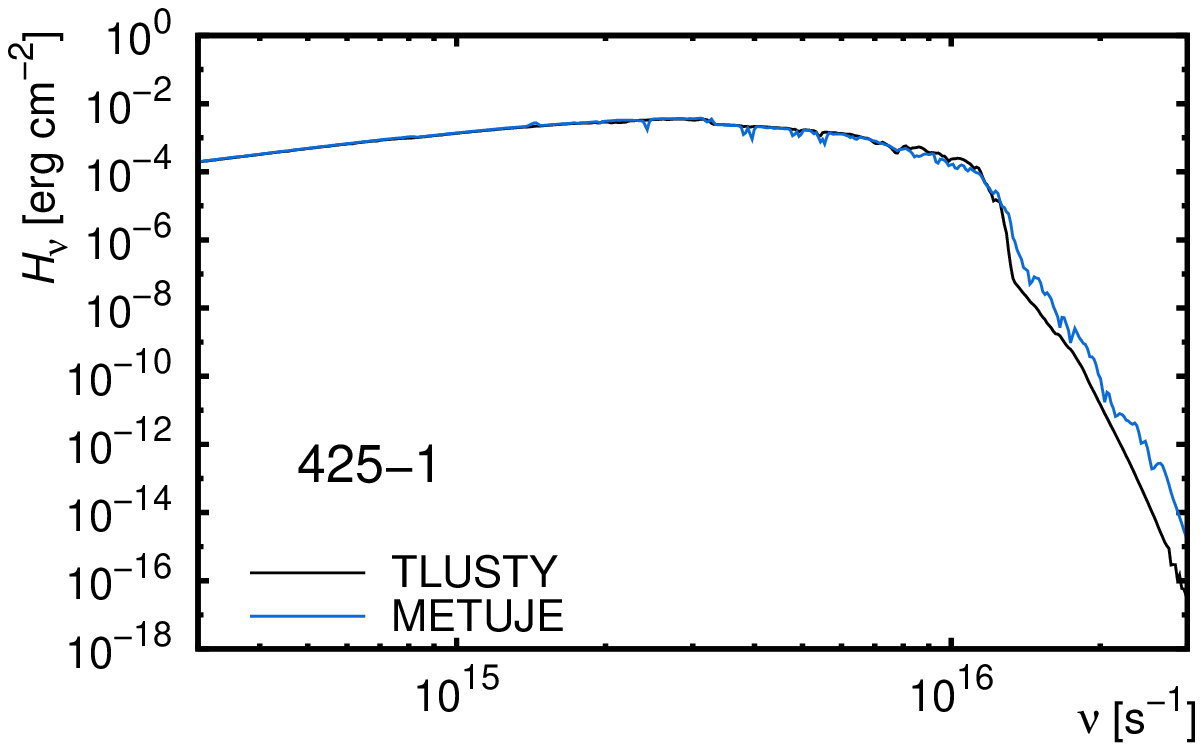}}
\resizebox{0.310\hsize}{!}{\includegraphics{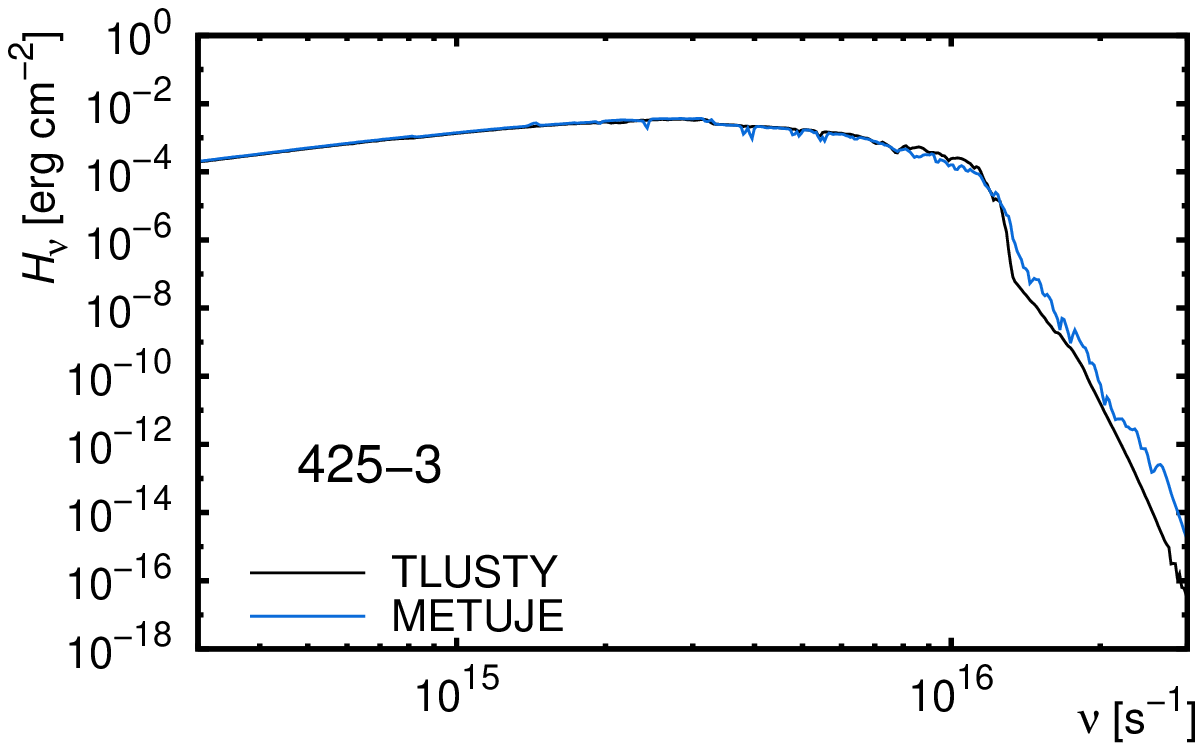}}
\resizebox{0.310\hsize}{!}{\includegraphics{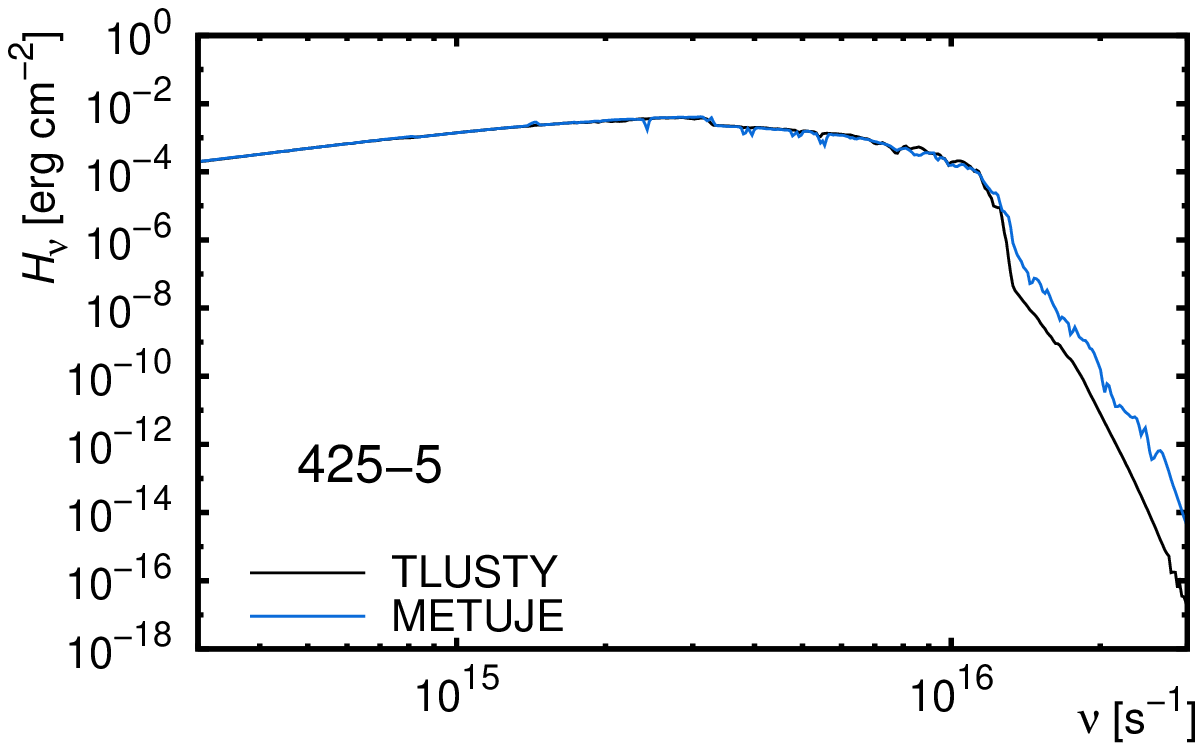}}\\
\resizebox{0.310\hsize}{!}{\includegraphics{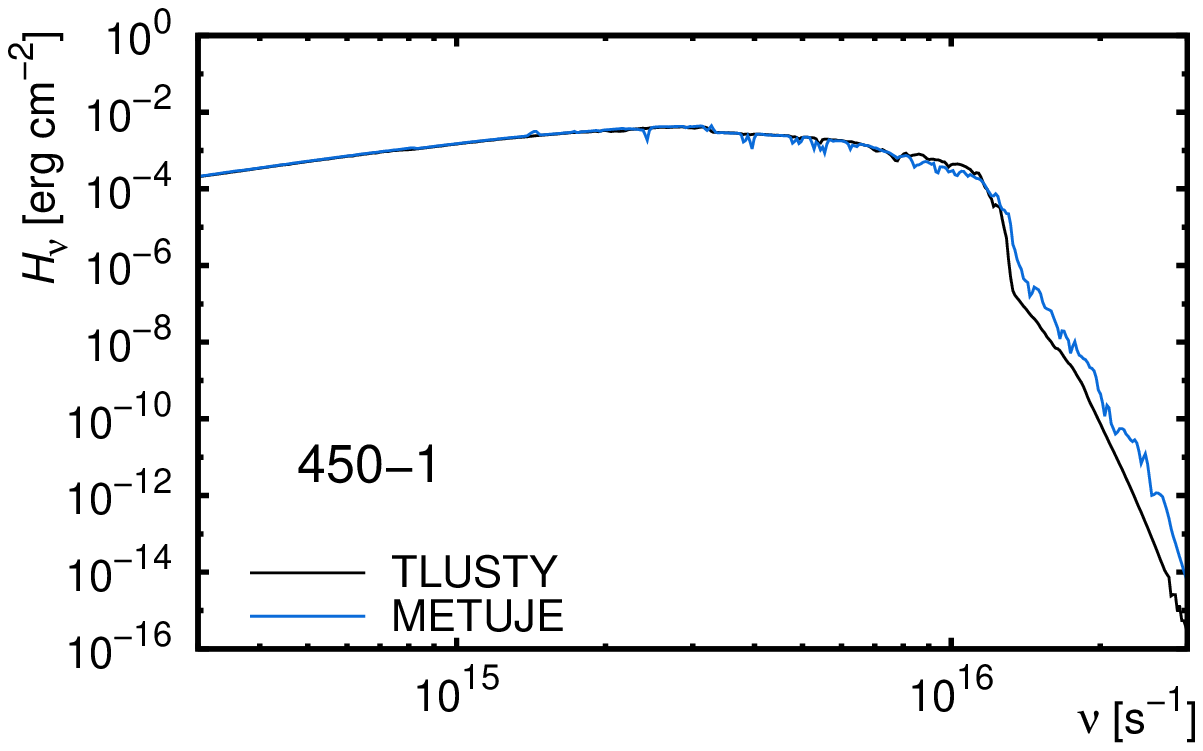}}
\resizebox{0.310\hsize}{!}{\includegraphics{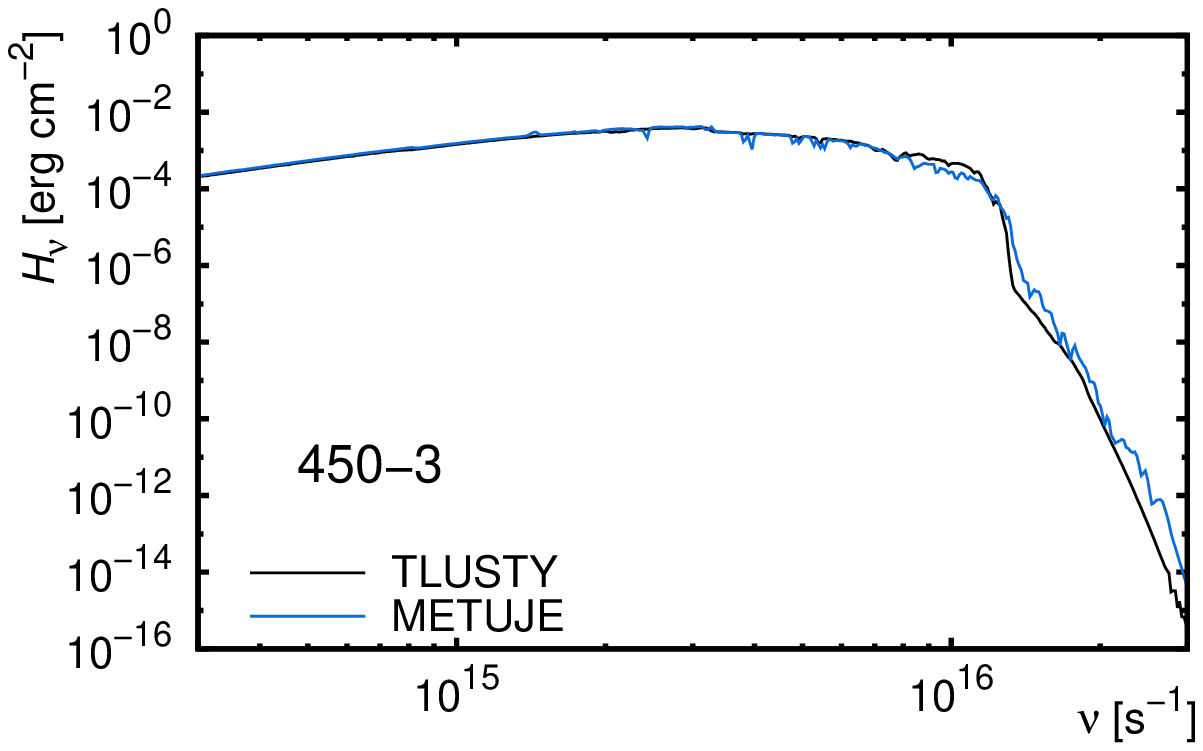}}
\resizebox{0.310\hsize}{!}{\includegraphics{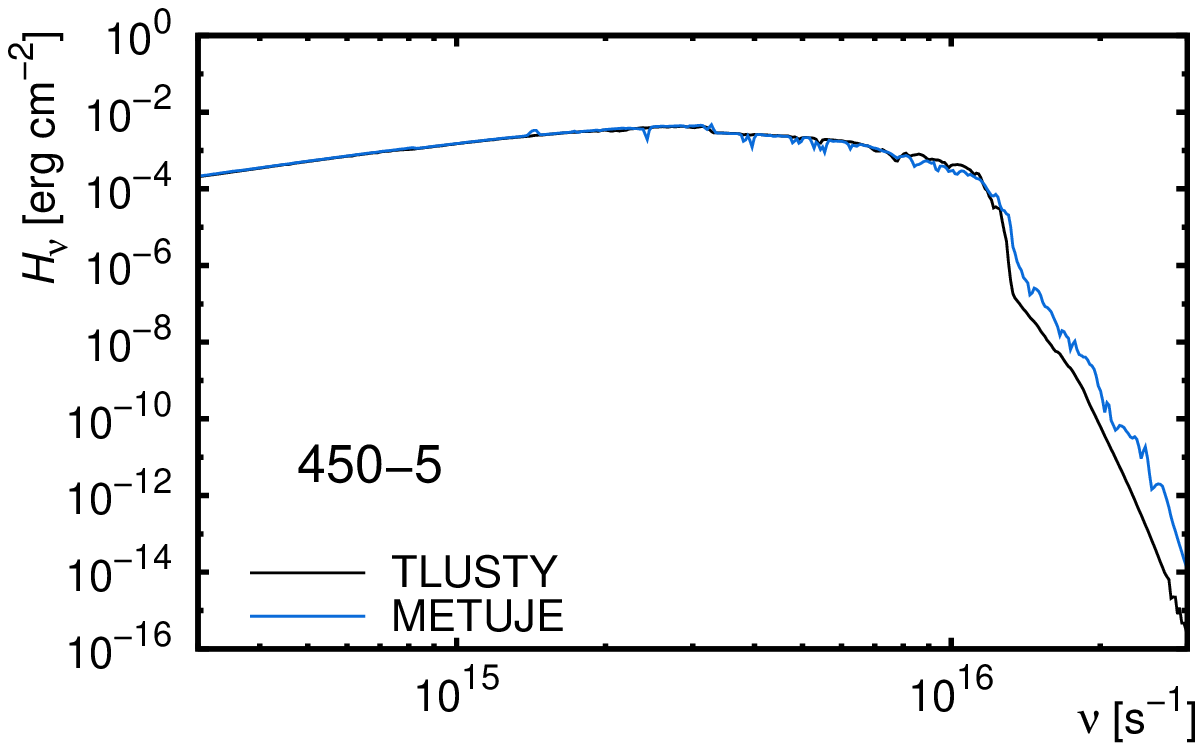}}
\caption{Comparison of the emergent flux from TLUSTY and METUJE models for SMC
stars. The graphs are plotted for individual model stars from
Table~\ref{ohvezpar} (denoted in the graphs).}
\label{tokmetlum}
\end{figure*}

\begin{figure*}[tp]
\centering
\resizebox{0.310\hsize}{!}{\includegraphics{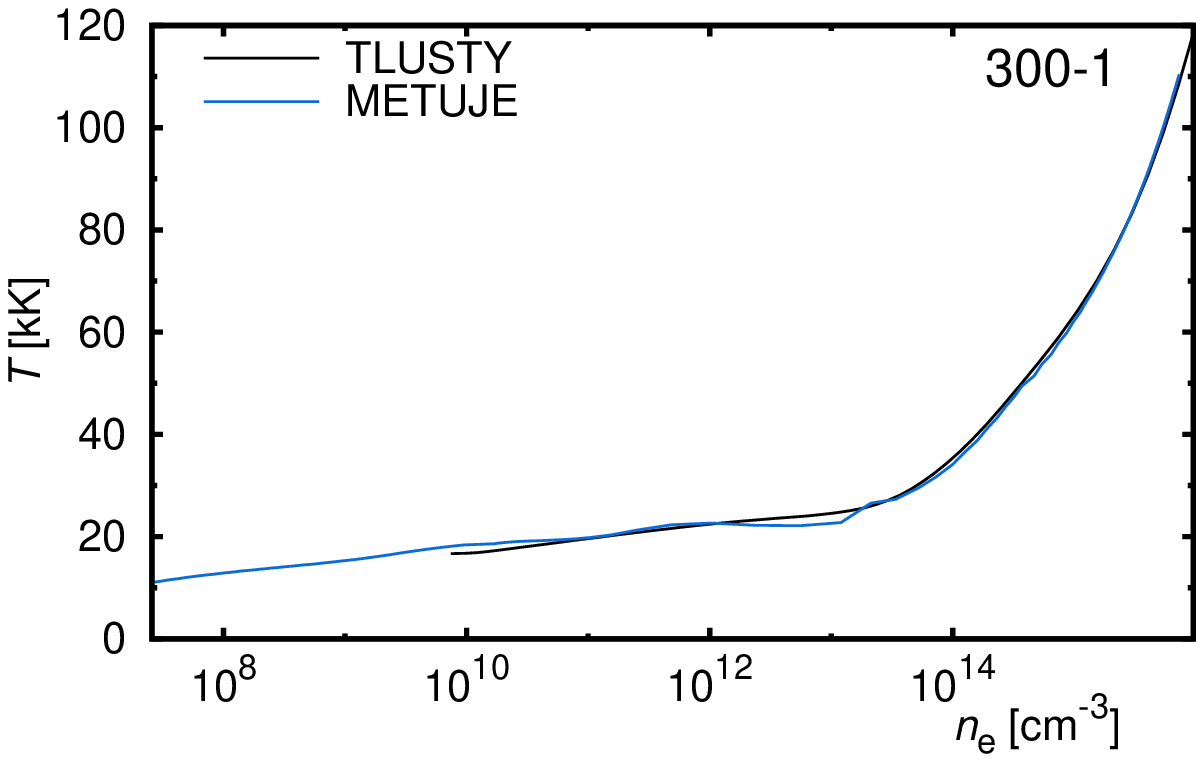}}
\resizebox{0.310\hsize}{!}{\includegraphics{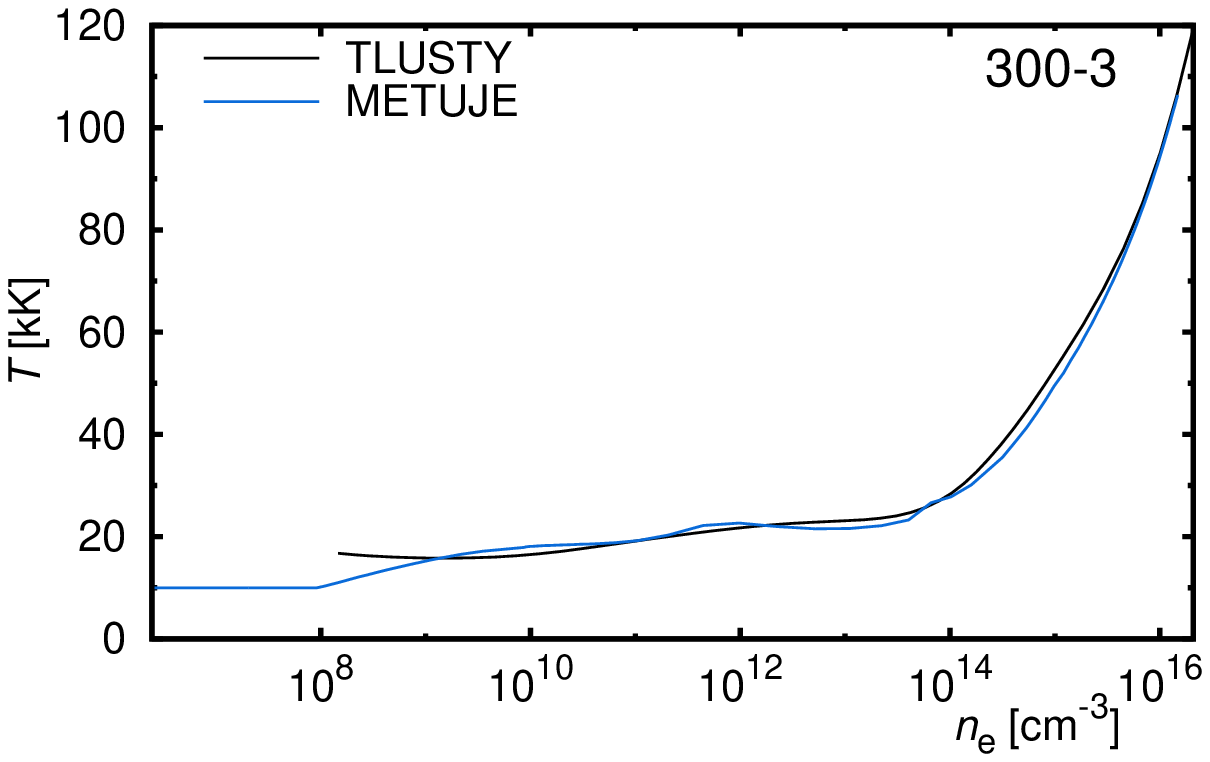}}
\resizebox{0.310\hsize}{!}{\includegraphics{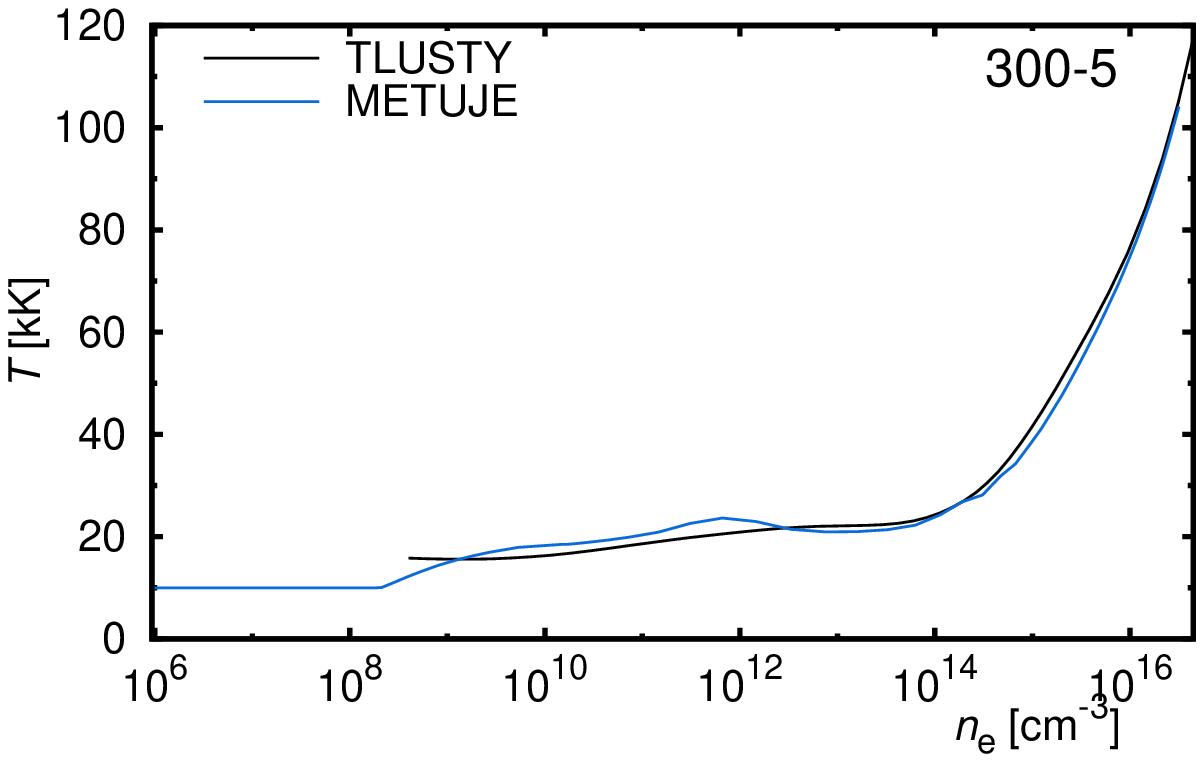}}\\
\resizebox{0.310\hsize}{!}{\includegraphics{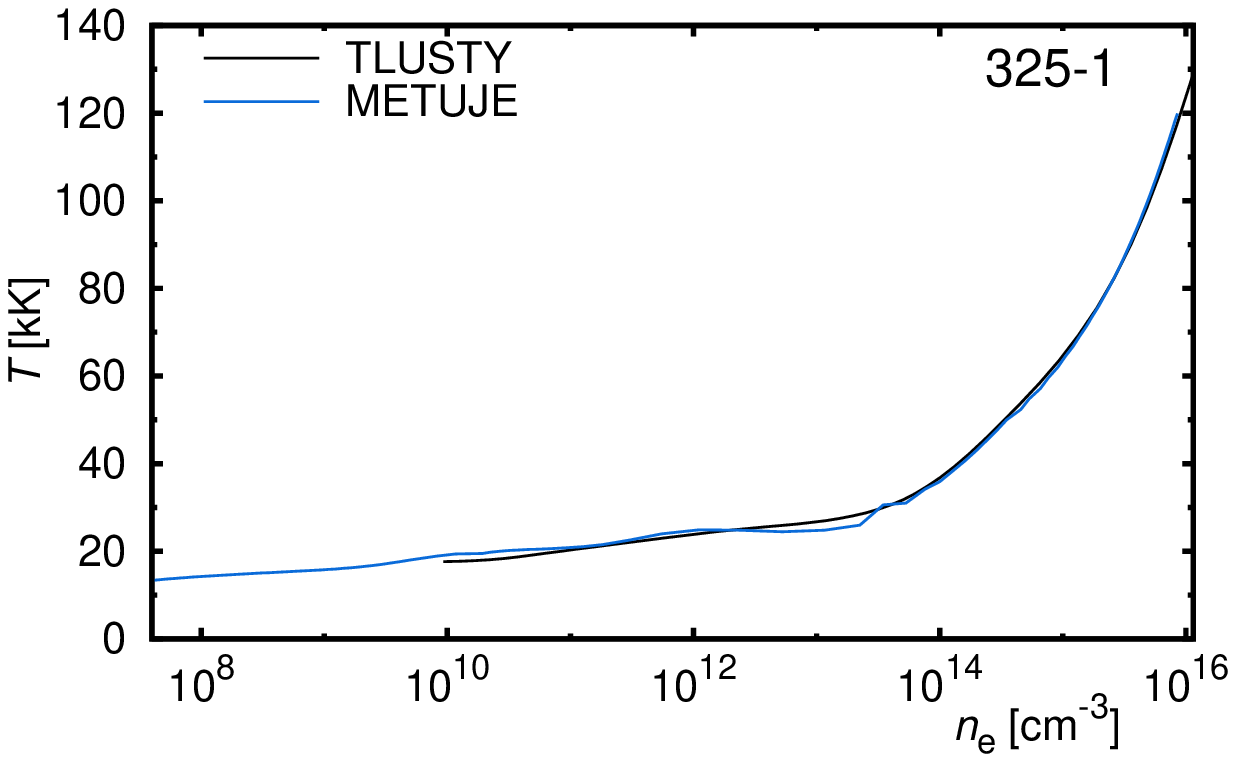}}
\resizebox{0.310\hsize}{!}{\includegraphics{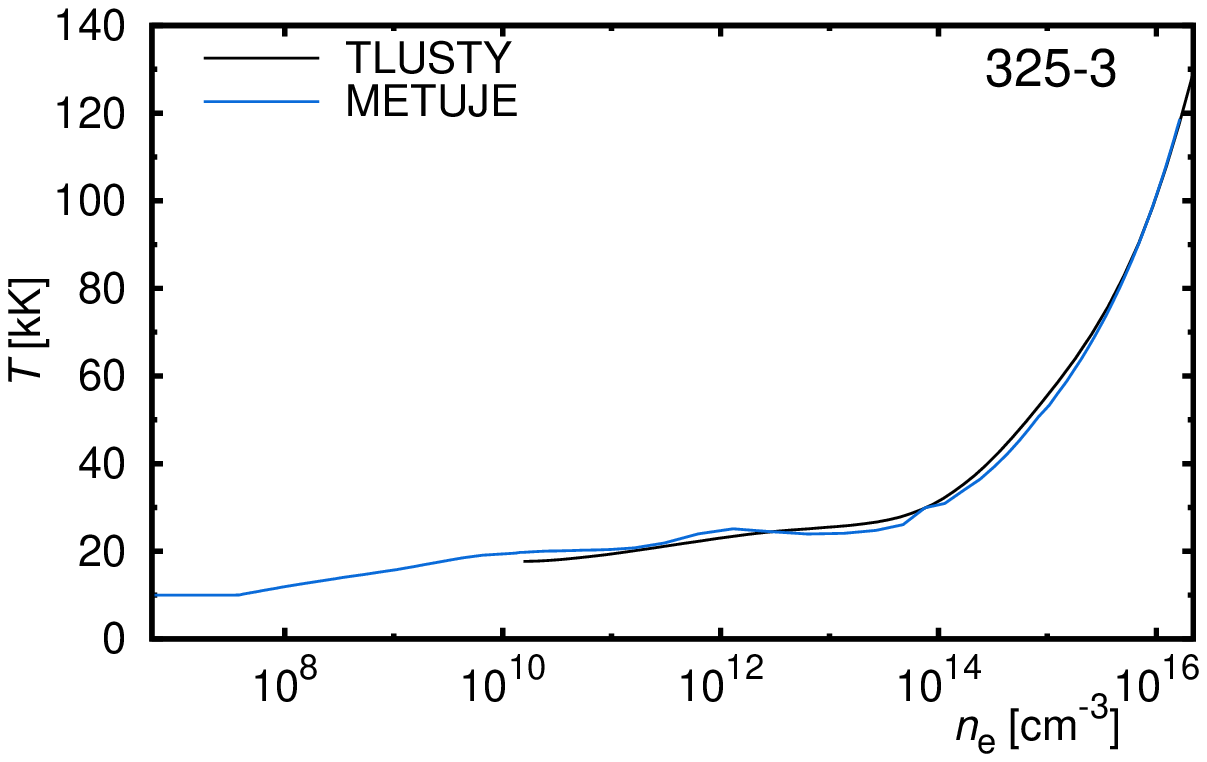}}
\resizebox{0.310\hsize}{!}{\includegraphics{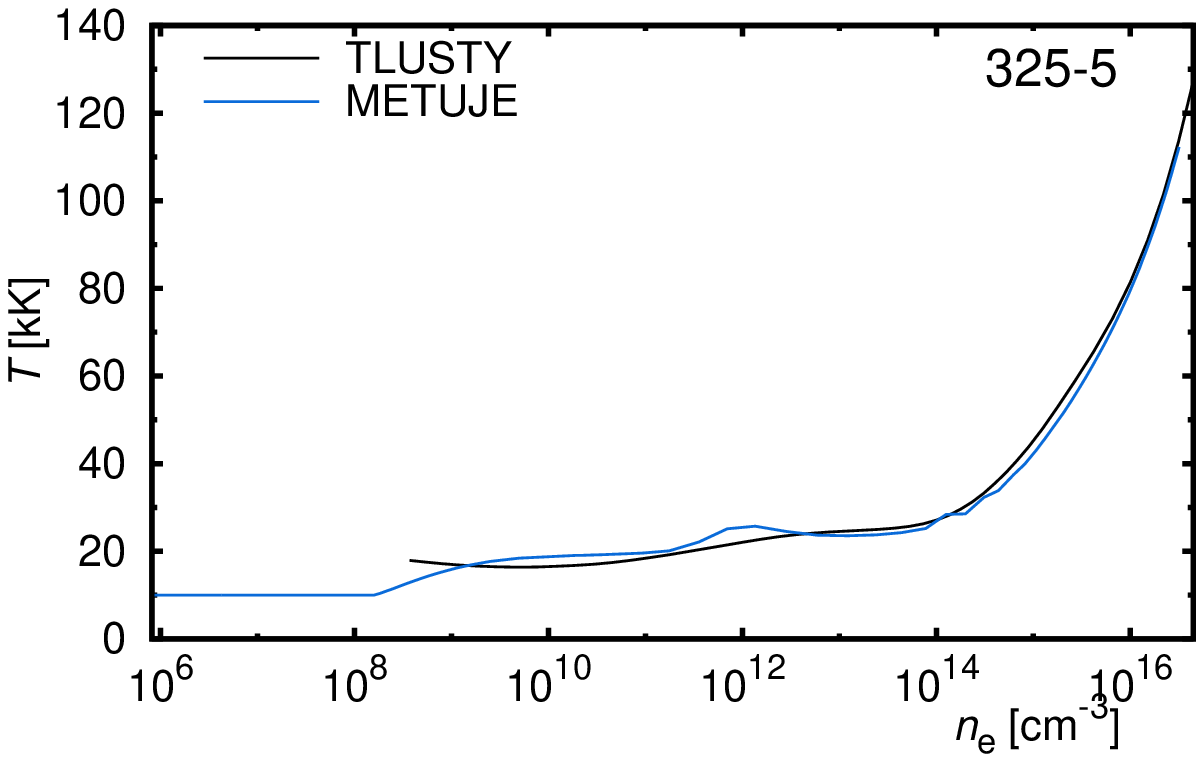}}\\
\resizebox{0.310\hsize}{!}{\includegraphics{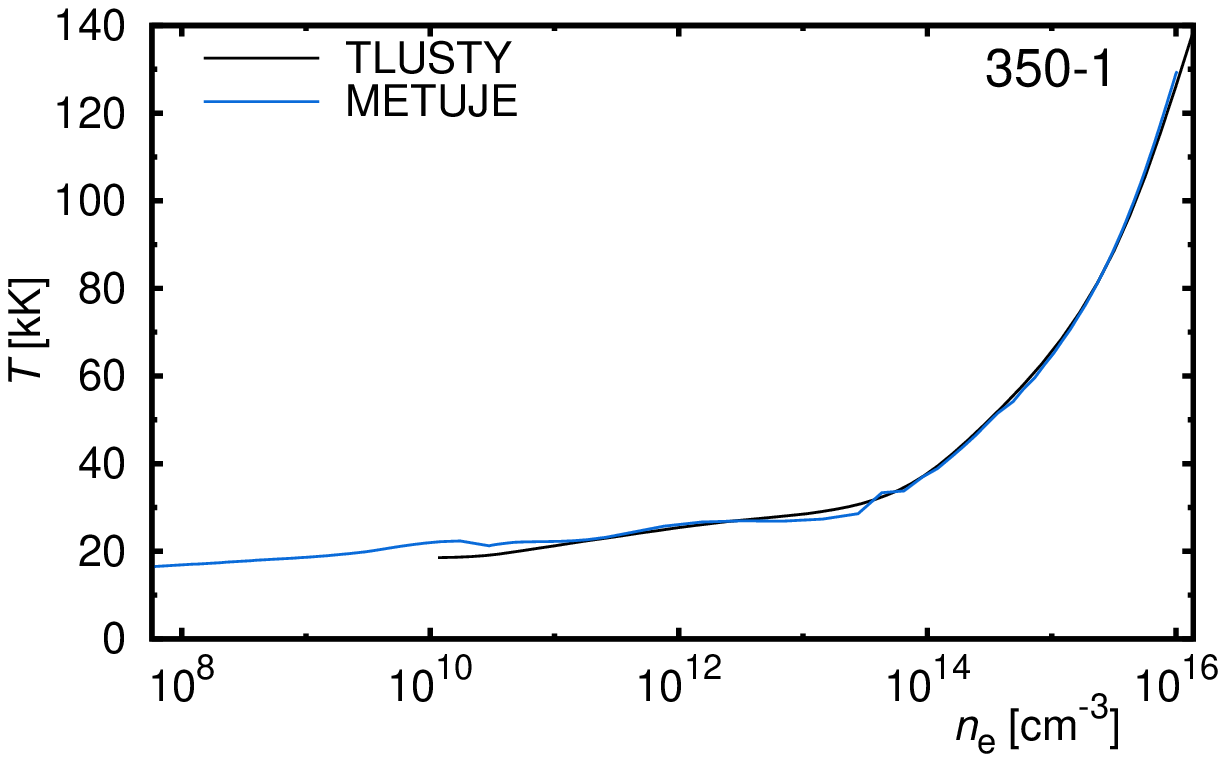}}
\resizebox{0.310\hsize}{!}{\includegraphics{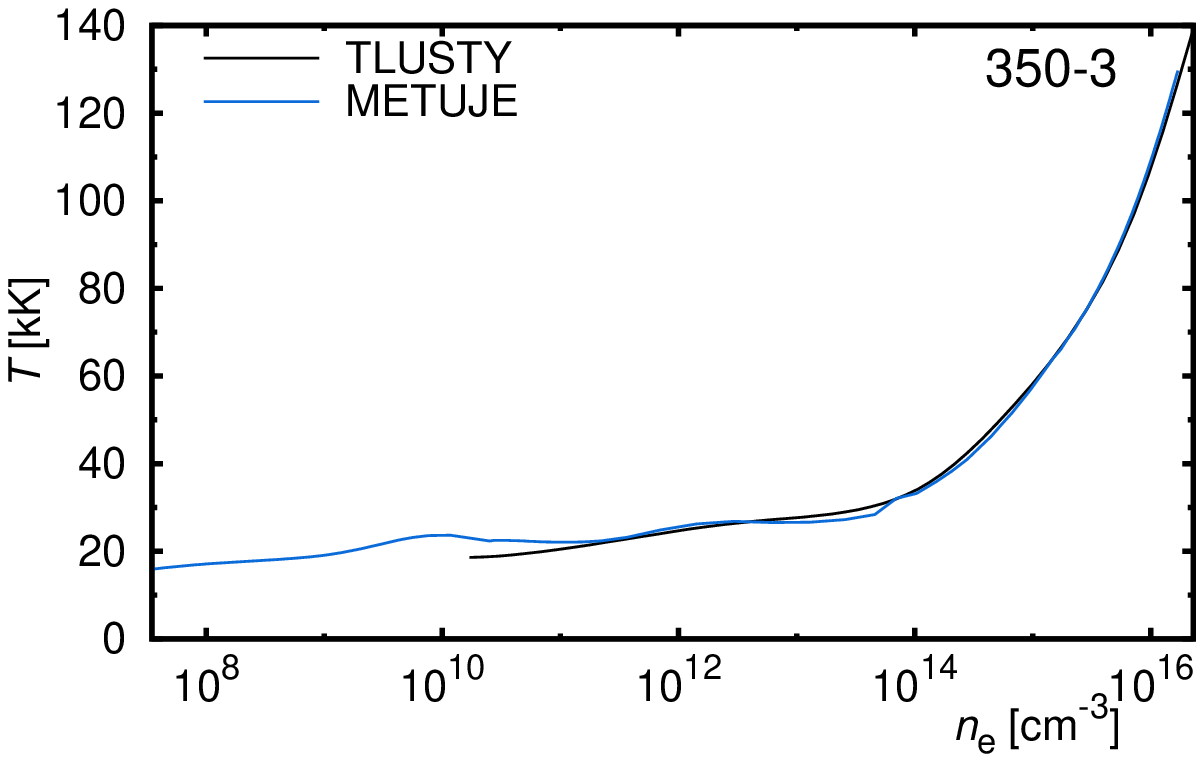}}
\resizebox{0.310\hsize}{!}{\includegraphics{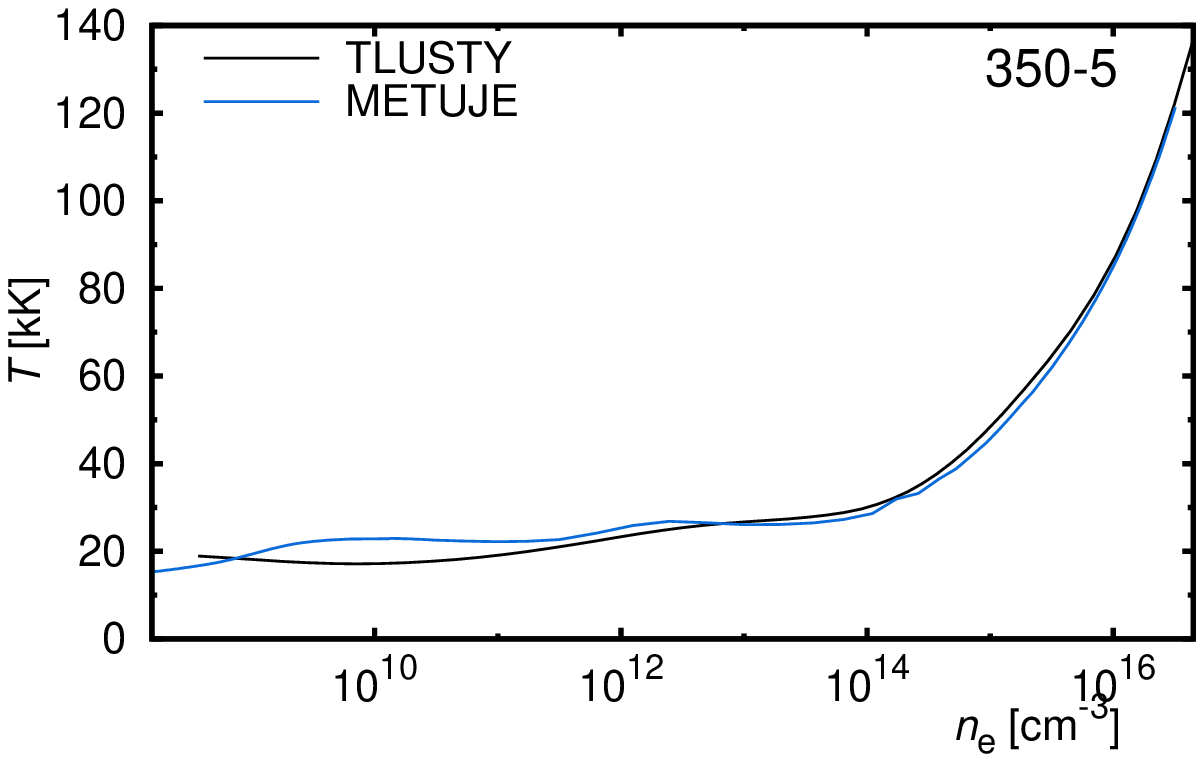}}\\
\resizebox{0.310\hsize}{!}{\includegraphics{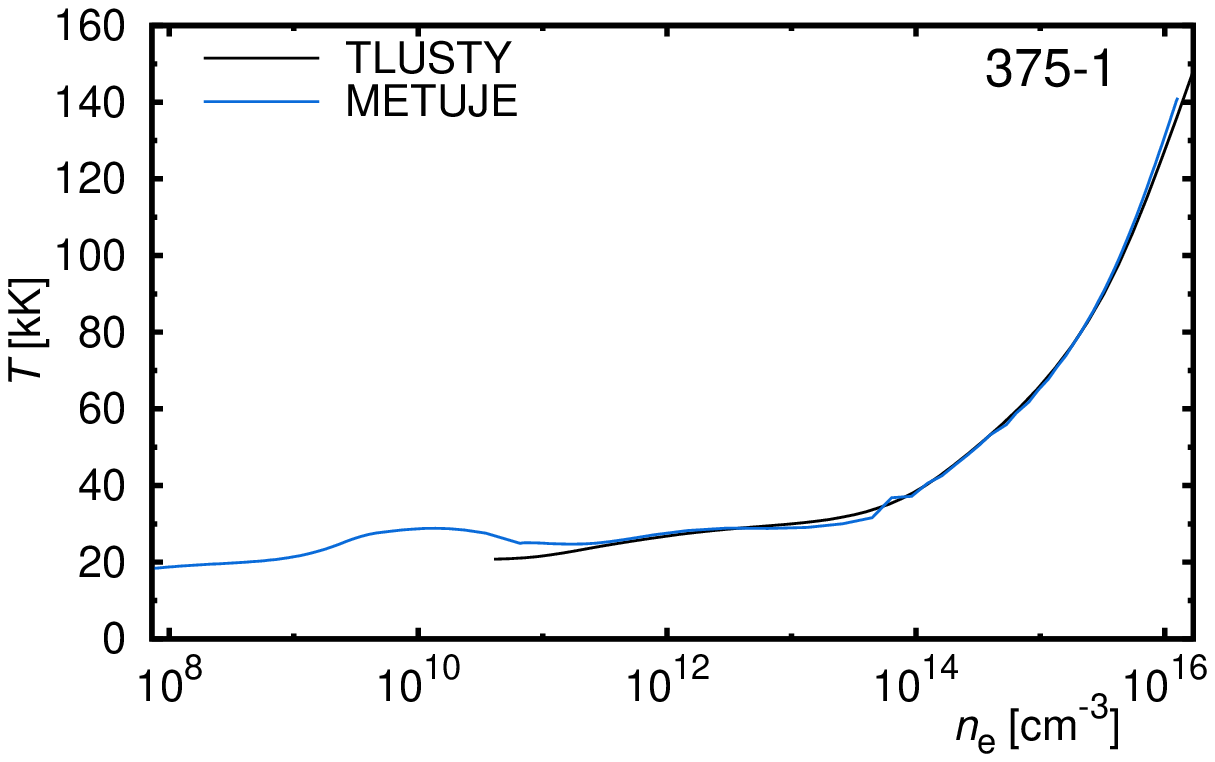}}
\resizebox{0.310\hsize}{!}{\includegraphics{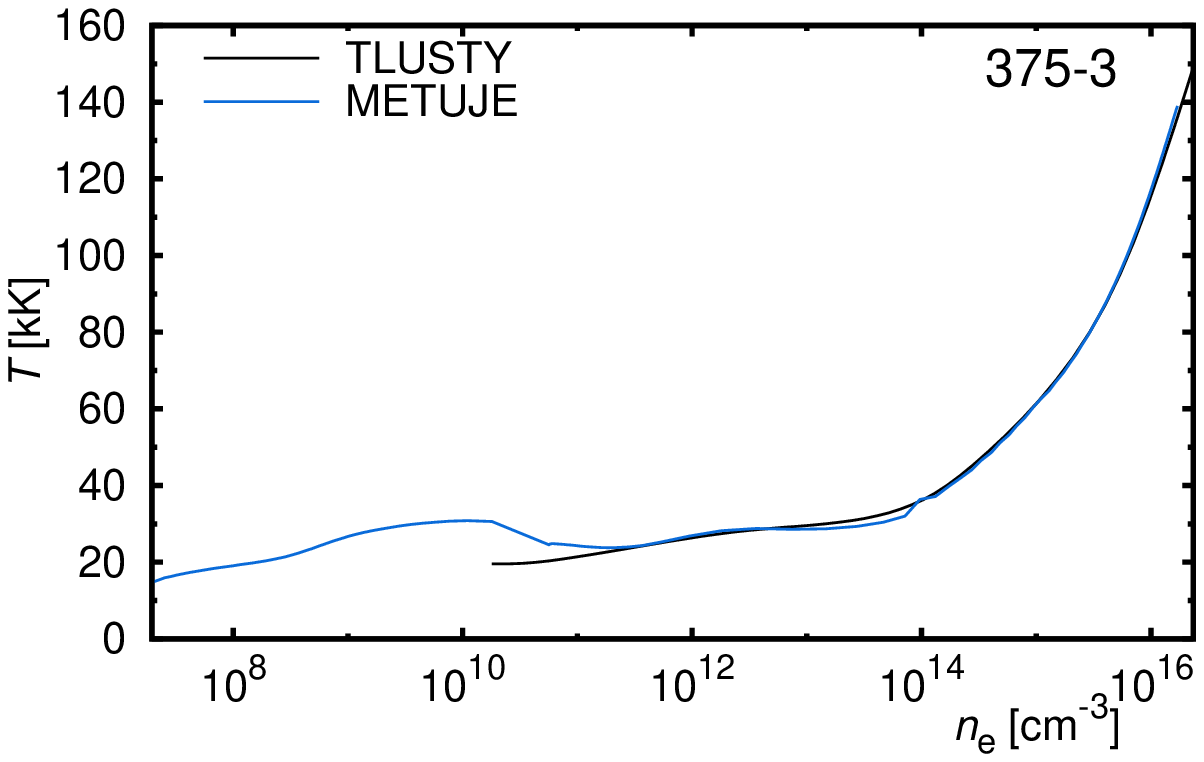}}
\resizebox{0.310\hsize}{!}{\includegraphics{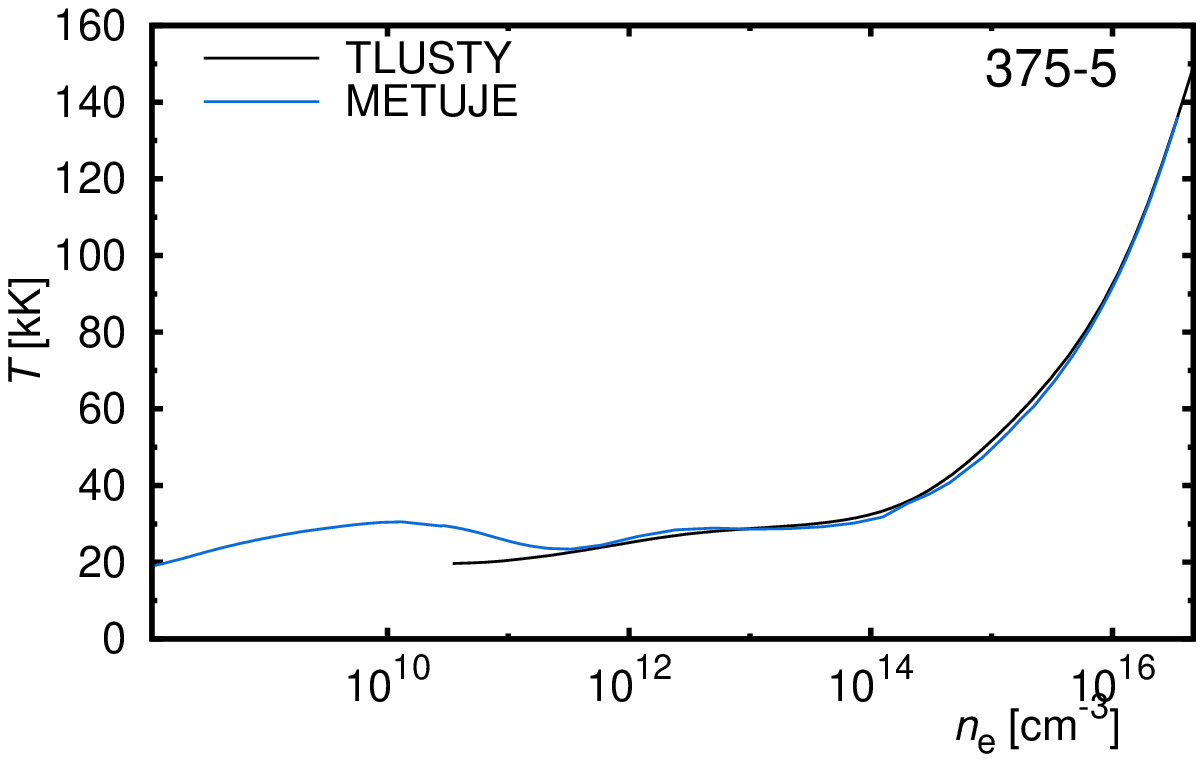}}\\
\resizebox{0.310\hsize}{!}{\includegraphics{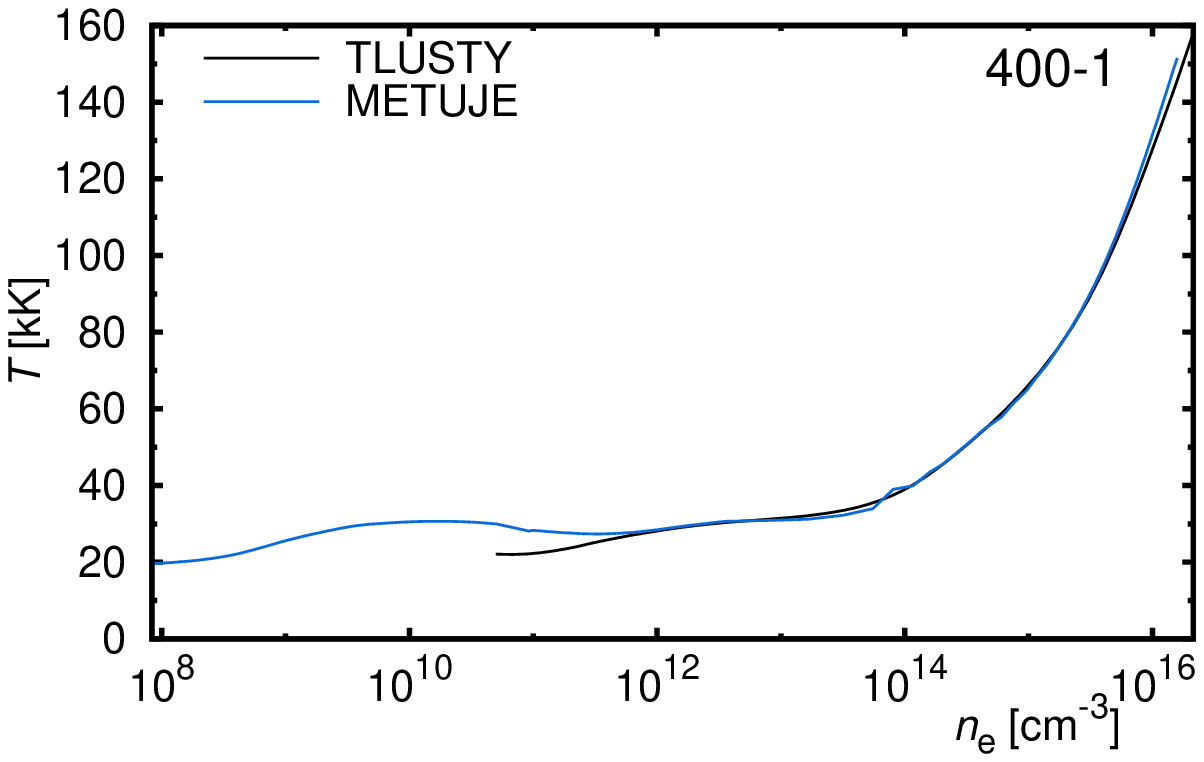}}
\resizebox{0.310\hsize}{!}{\includegraphics{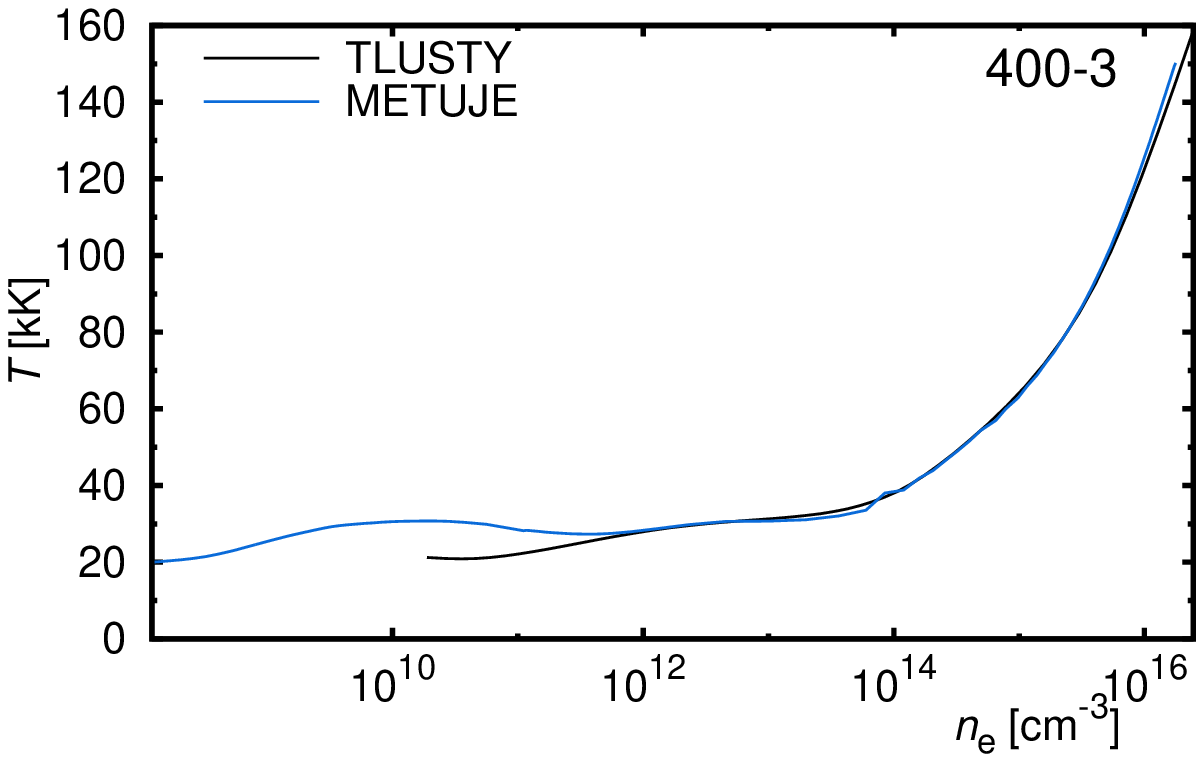}}
\resizebox{0.310\hsize}{!}{\includegraphics{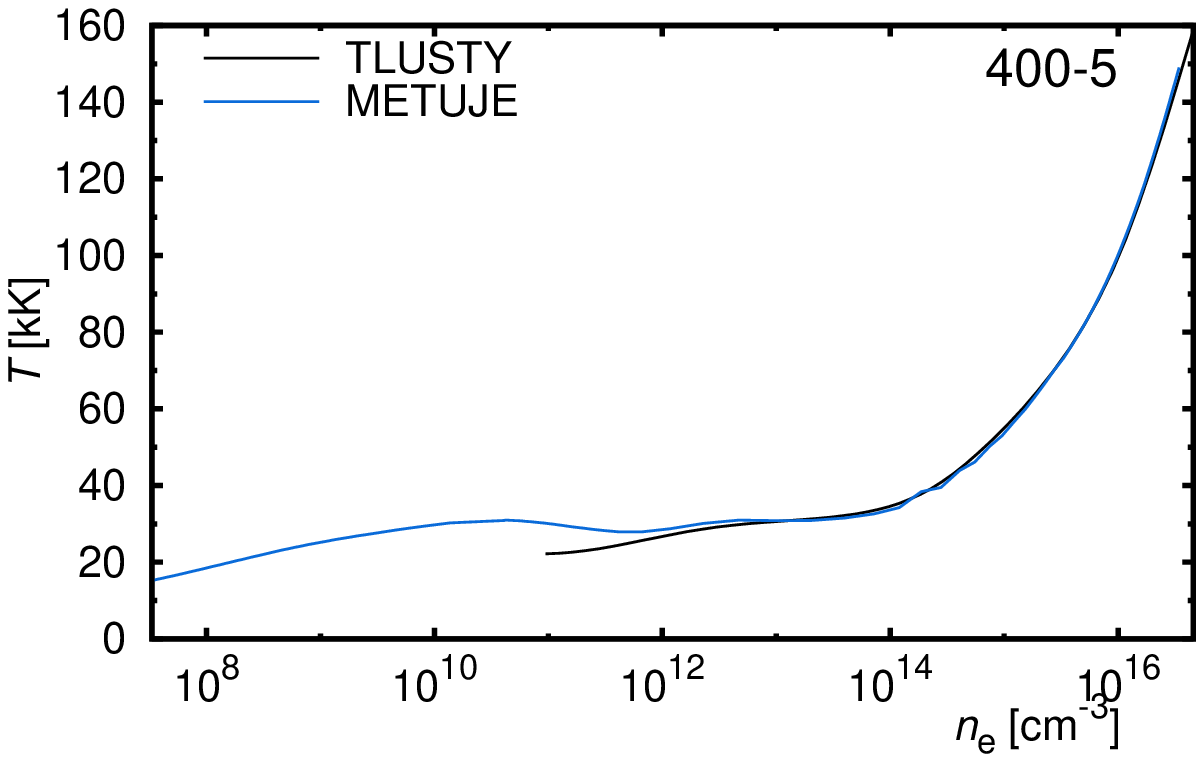}}\\
\resizebox{0.310\hsize}{!}{\includegraphics{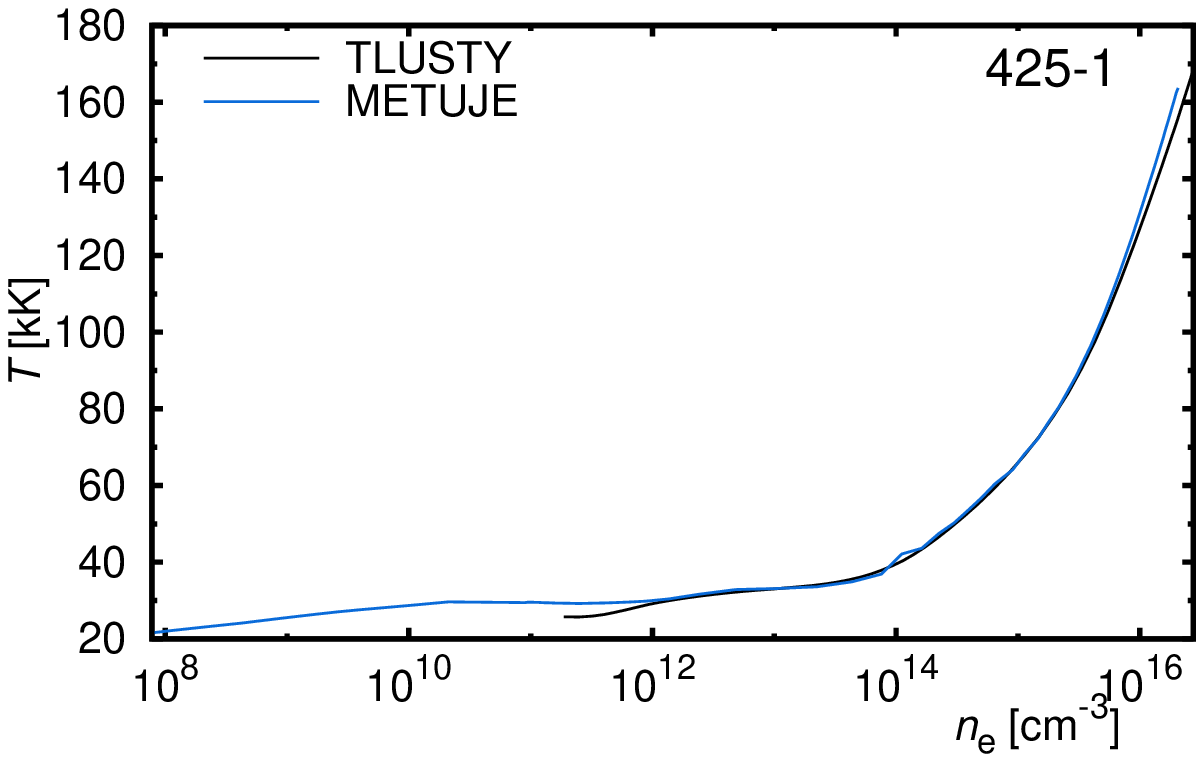}}
\resizebox{0.310\hsize}{!}{\includegraphics{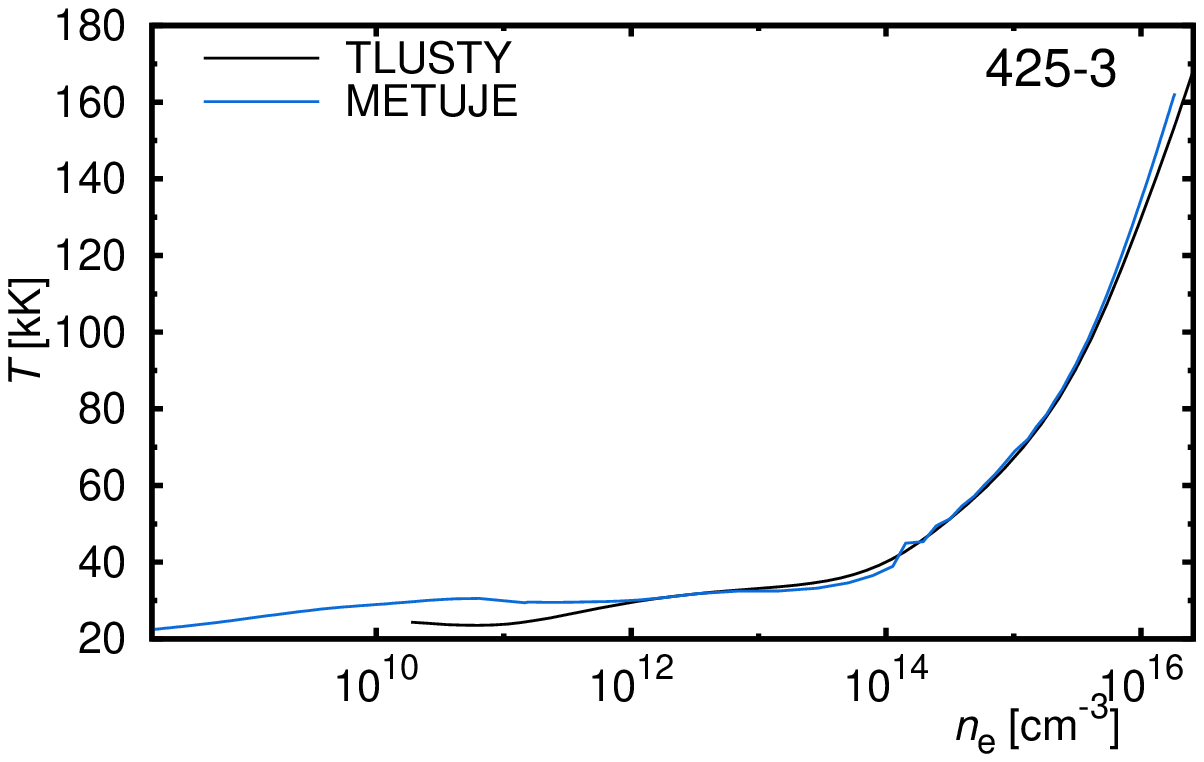}}
\resizebox{0.310\hsize}{!}{\includegraphics{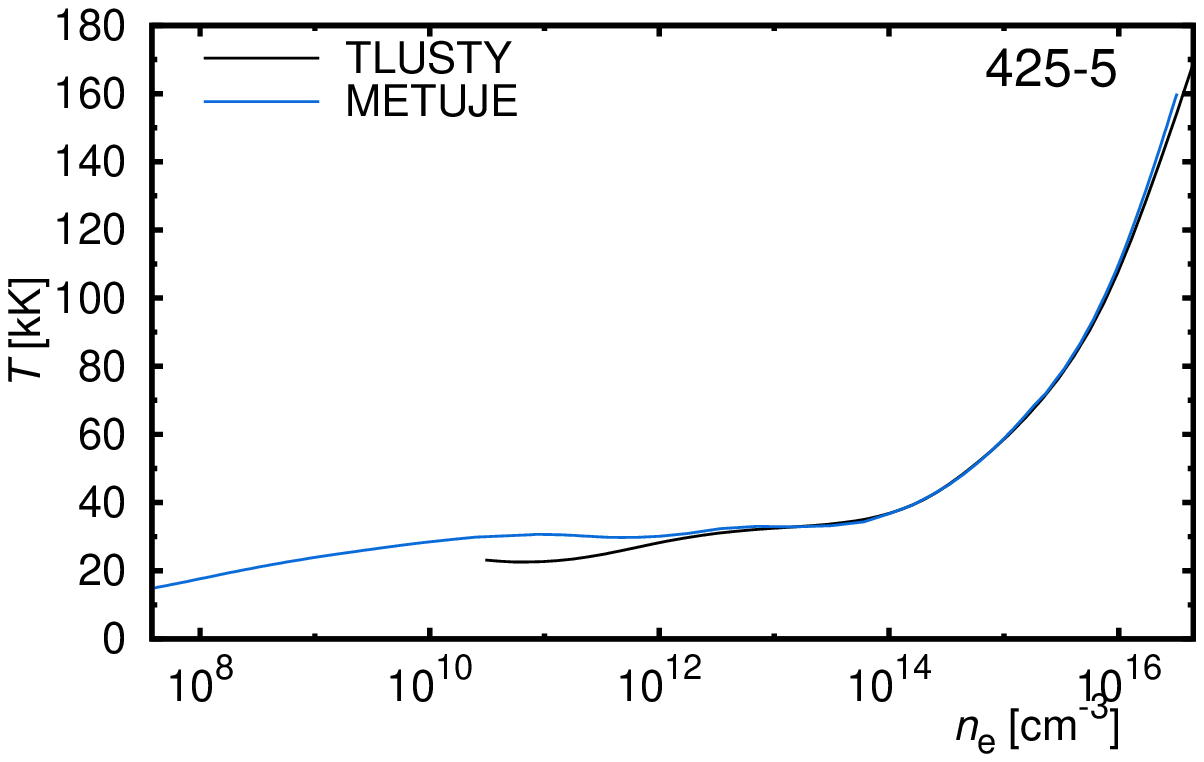}}\\
\resizebox{0.310\hsize}{!}{\includegraphics{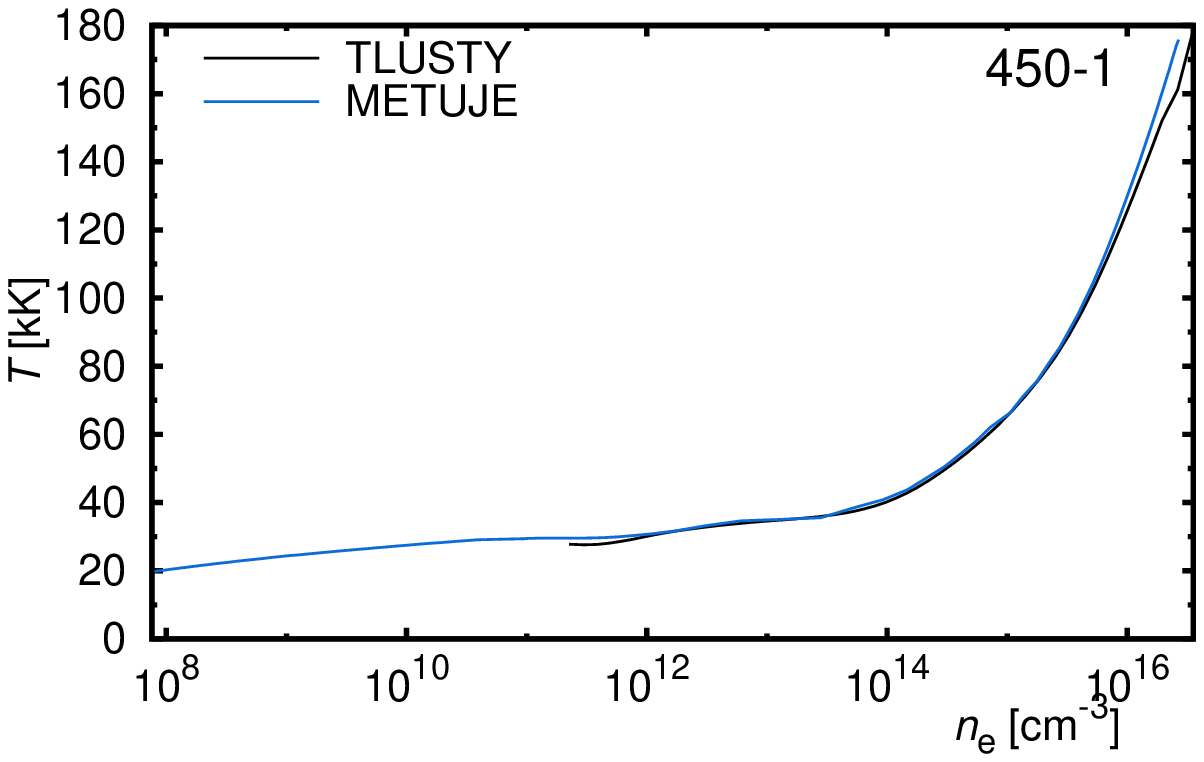}}
\resizebox{0.310\hsize}{!}{\includegraphics{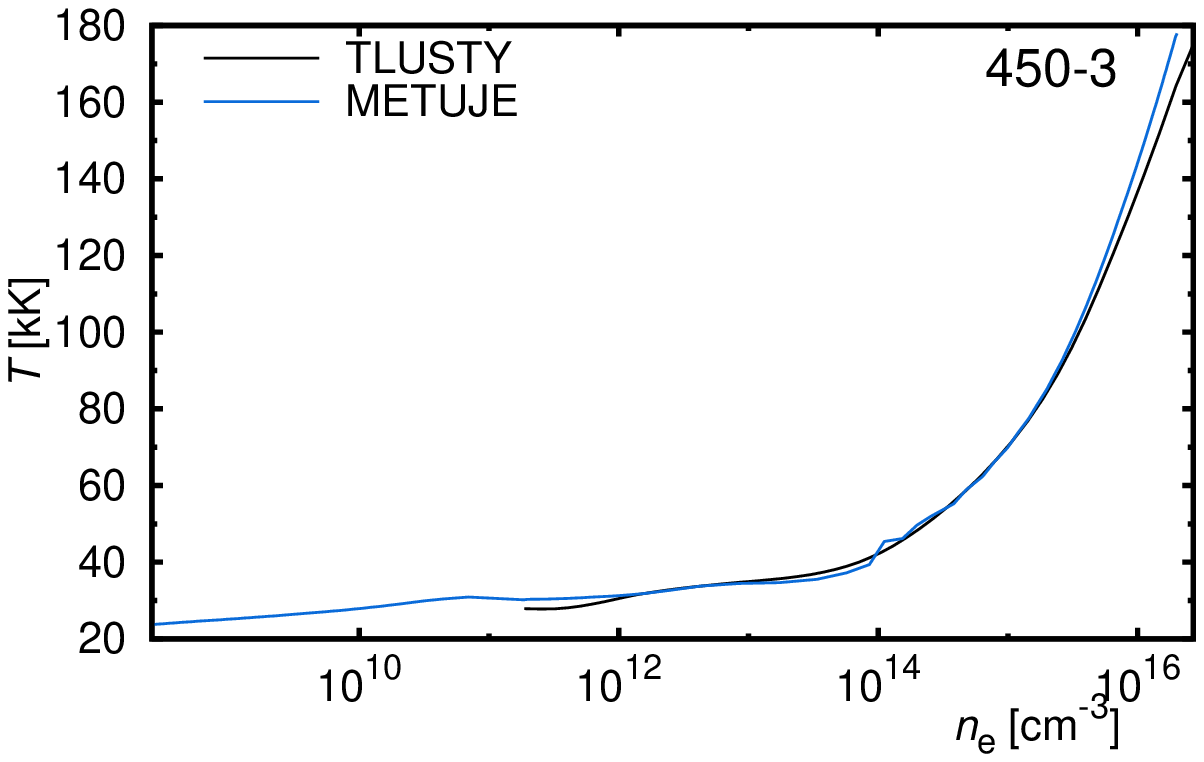}}
\resizebox{0.310\hsize}{!}{\includegraphics{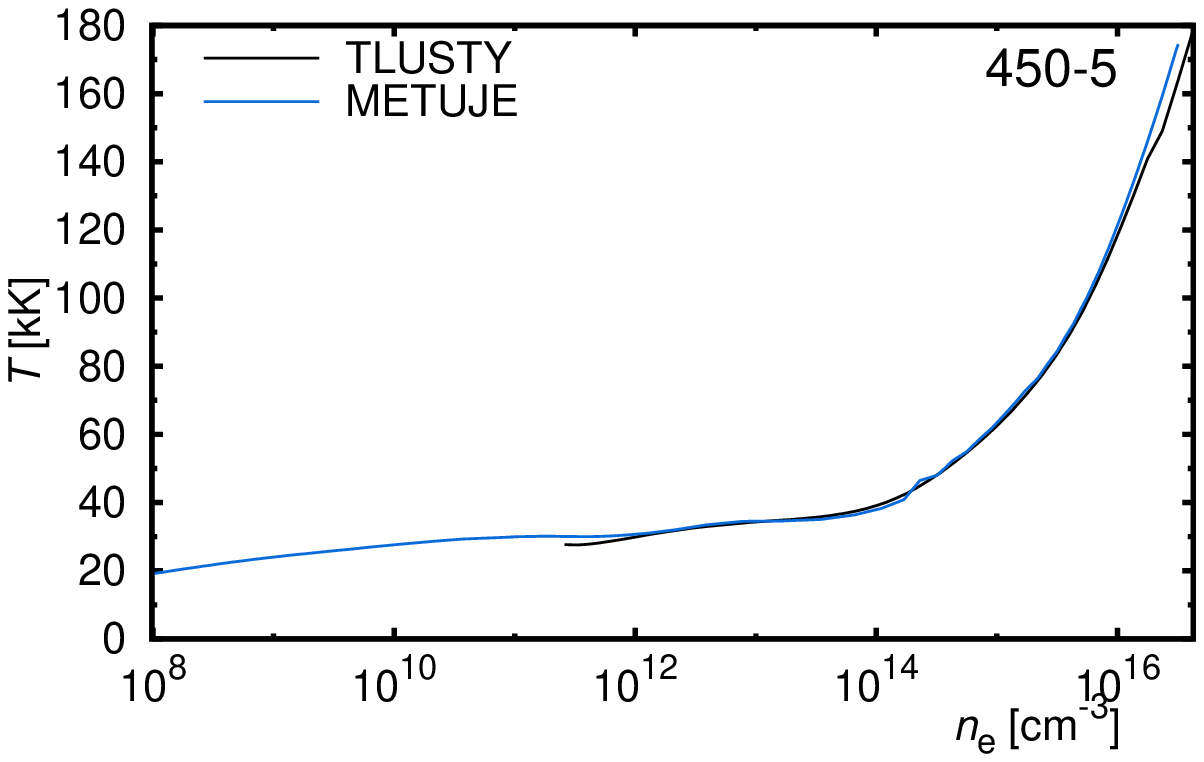}}
\caption{Comparison of the dependence of temperature on electron density in
TLUSTY and METUJE models for LMC stars. The graphs are plotted for individual
model stars from Table~\ref{ohvezpar} (denoted in the graphs).}
\label{tepmetluv}
\end{figure*}

\begin{figure*}[tp]
\centering
\resizebox{0.310\hsize}{!}{\includegraphics{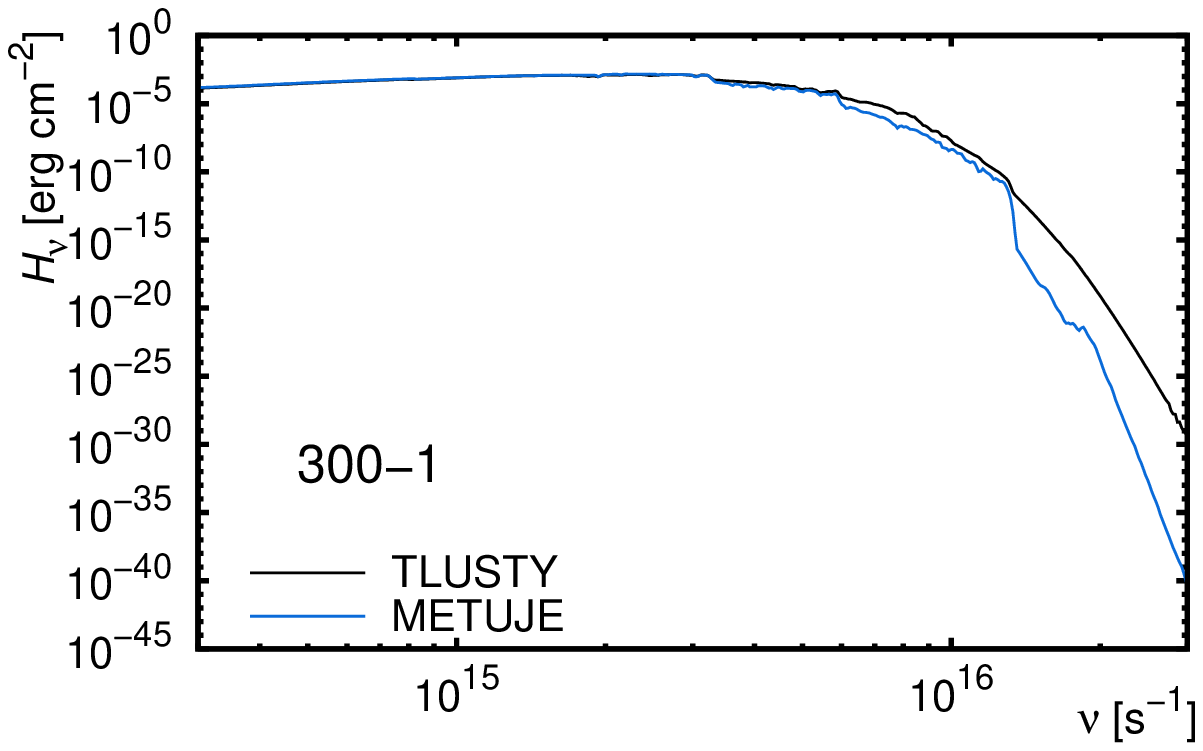}}
\resizebox{0.310\hsize}{!}{\includegraphics{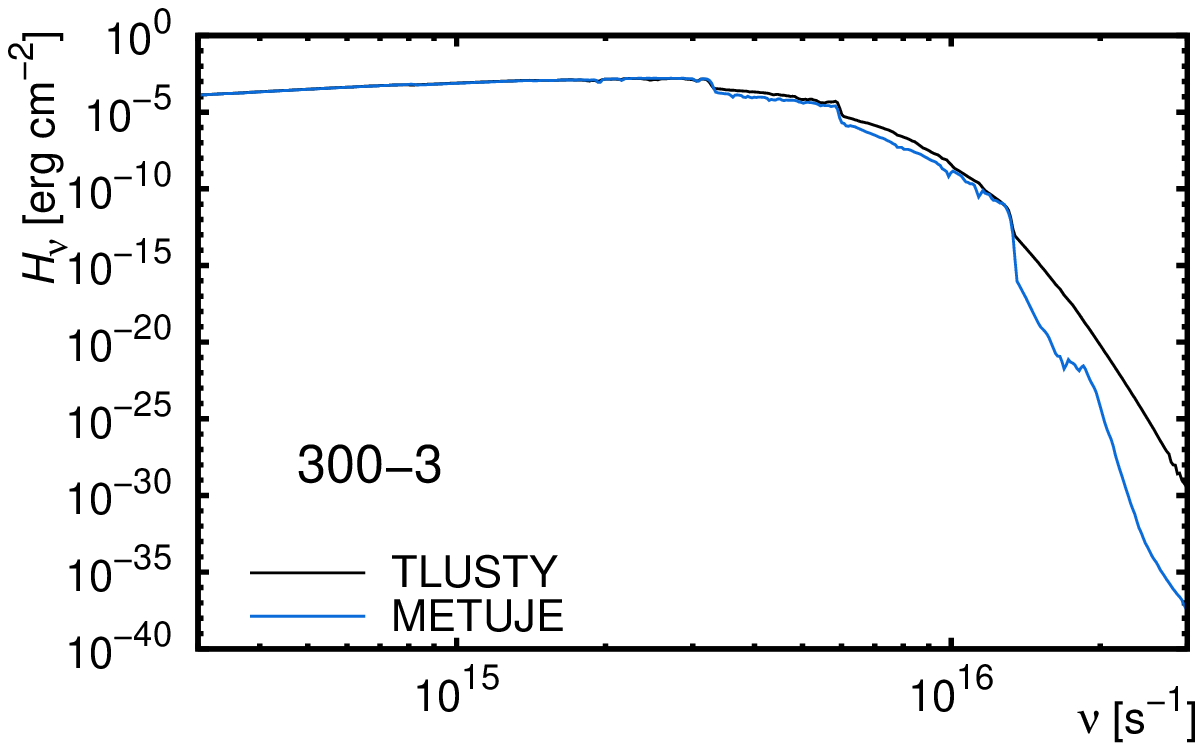}}
\resizebox{0.310\hsize}{!}{\includegraphics{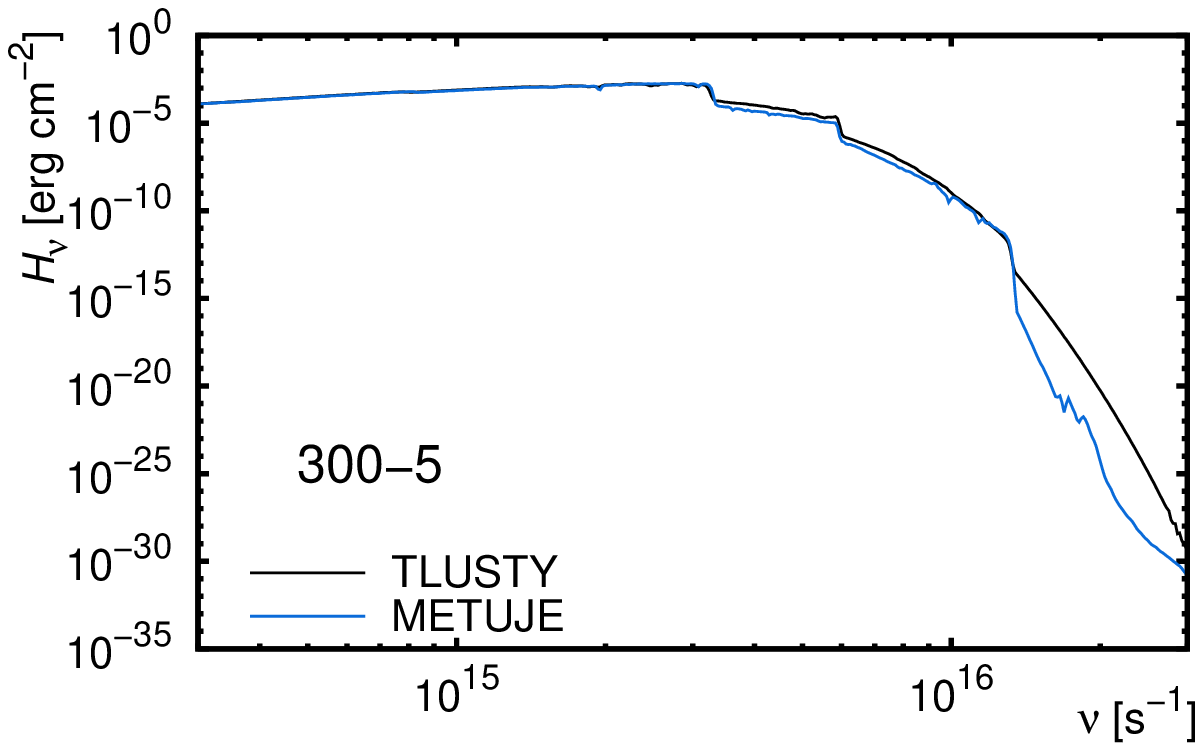}}\\
\resizebox{0.310\hsize}{!}{\includegraphics{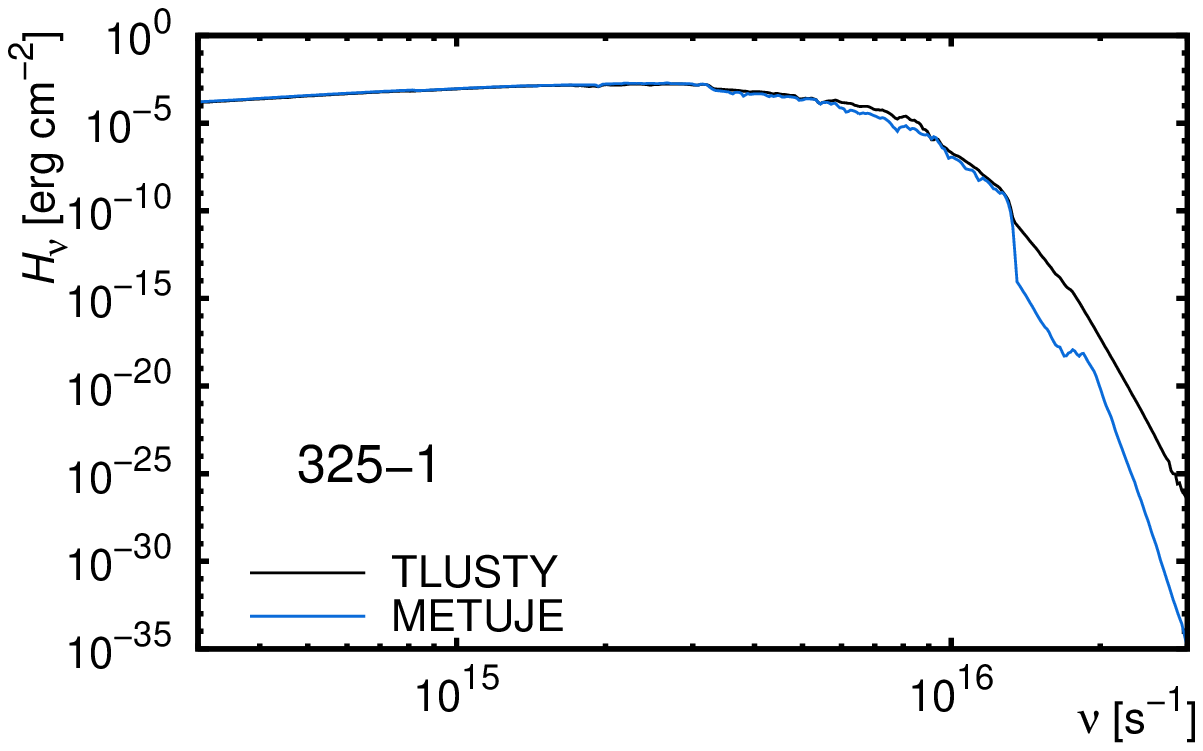}}
\resizebox{0.310\hsize}{!}{\includegraphics{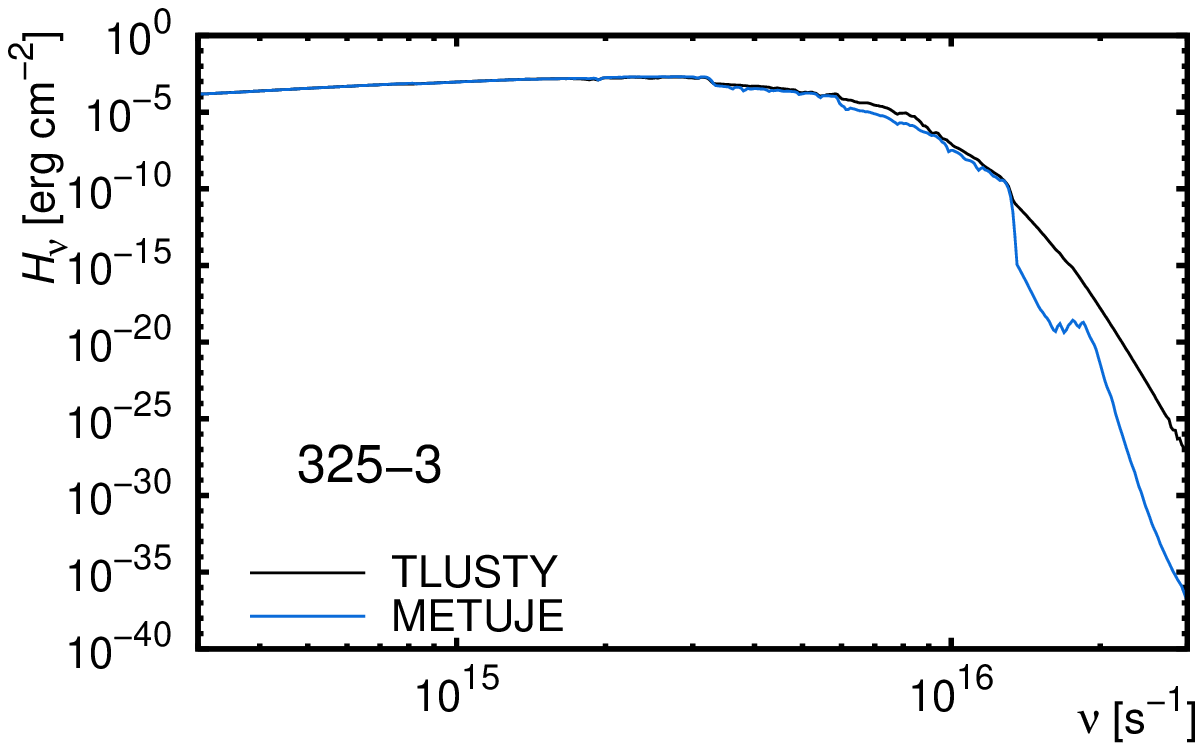}}
\resizebox{0.310\hsize}{!}{\includegraphics{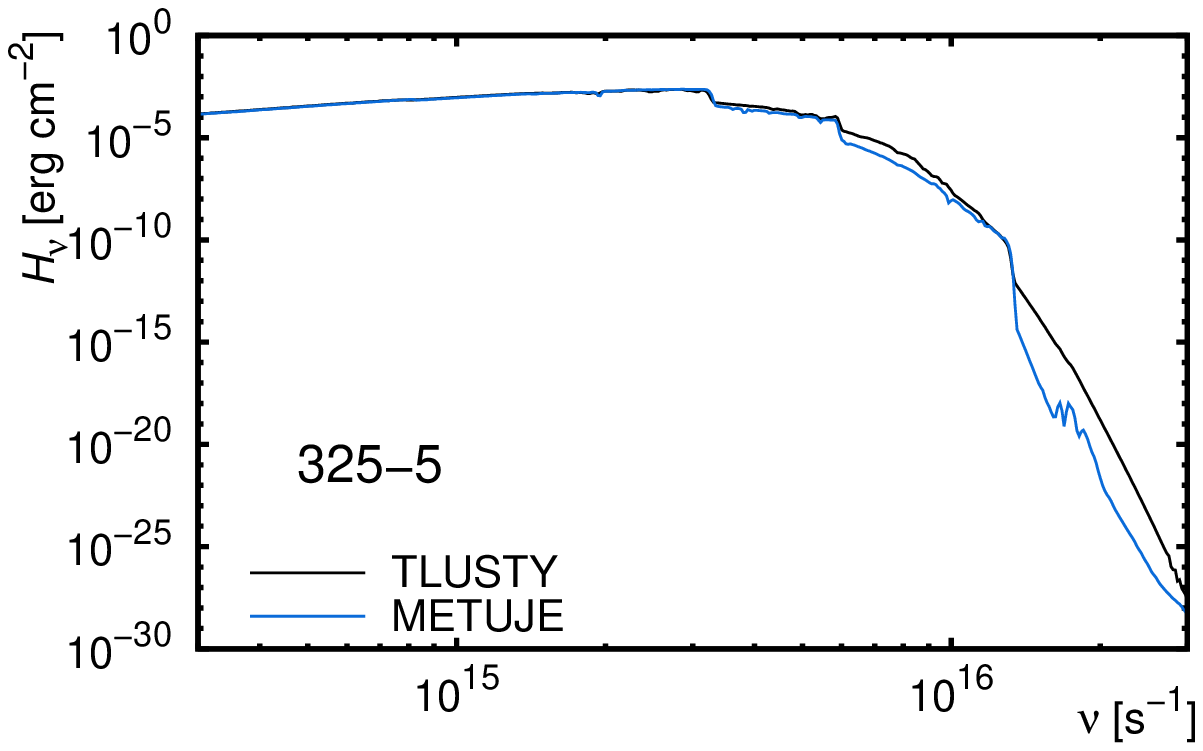}}\\
\resizebox{0.310\hsize}{!}{\includegraphics{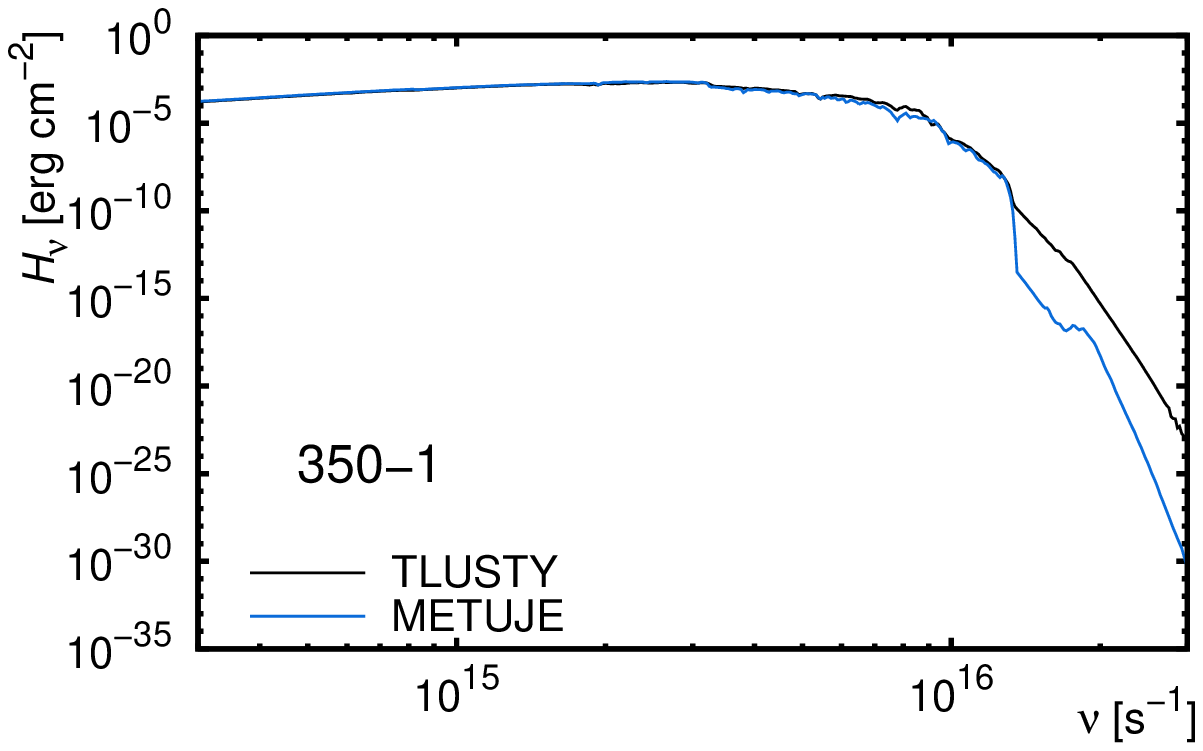}}
\resizebox{0.310\hsize}{!}{\includegraphics{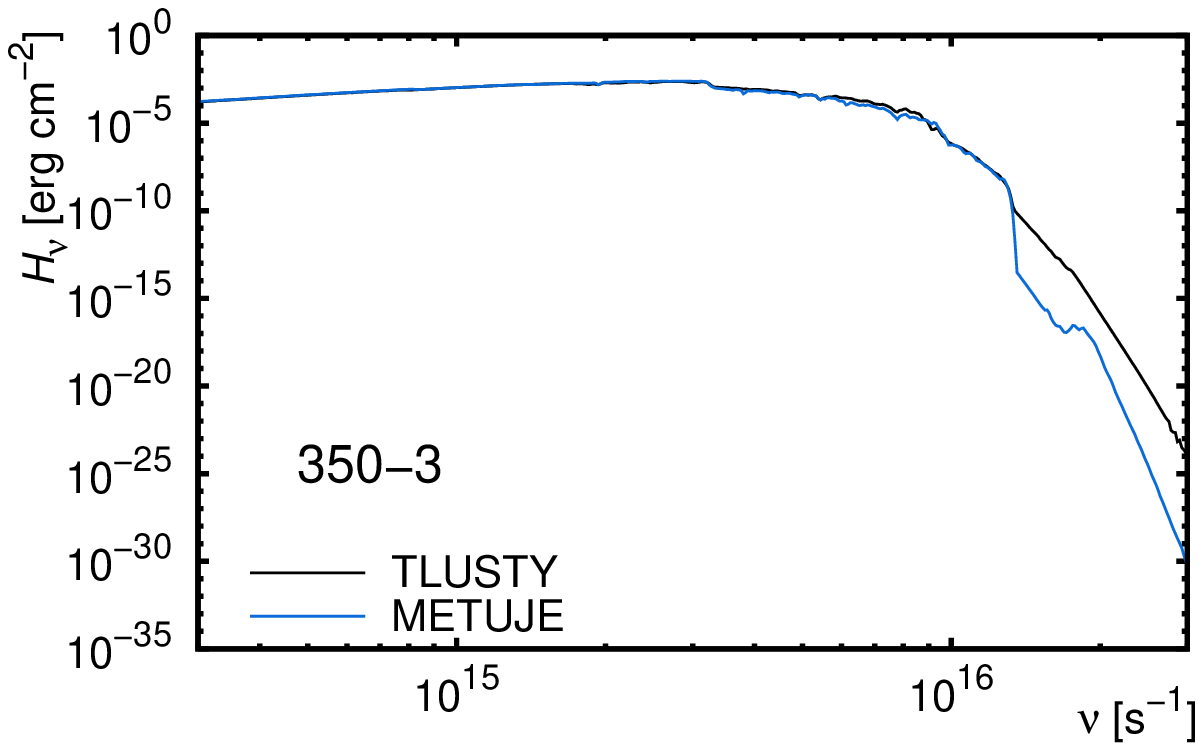}}
\resizebox{0.310\hsize}{!}{\includegraphics{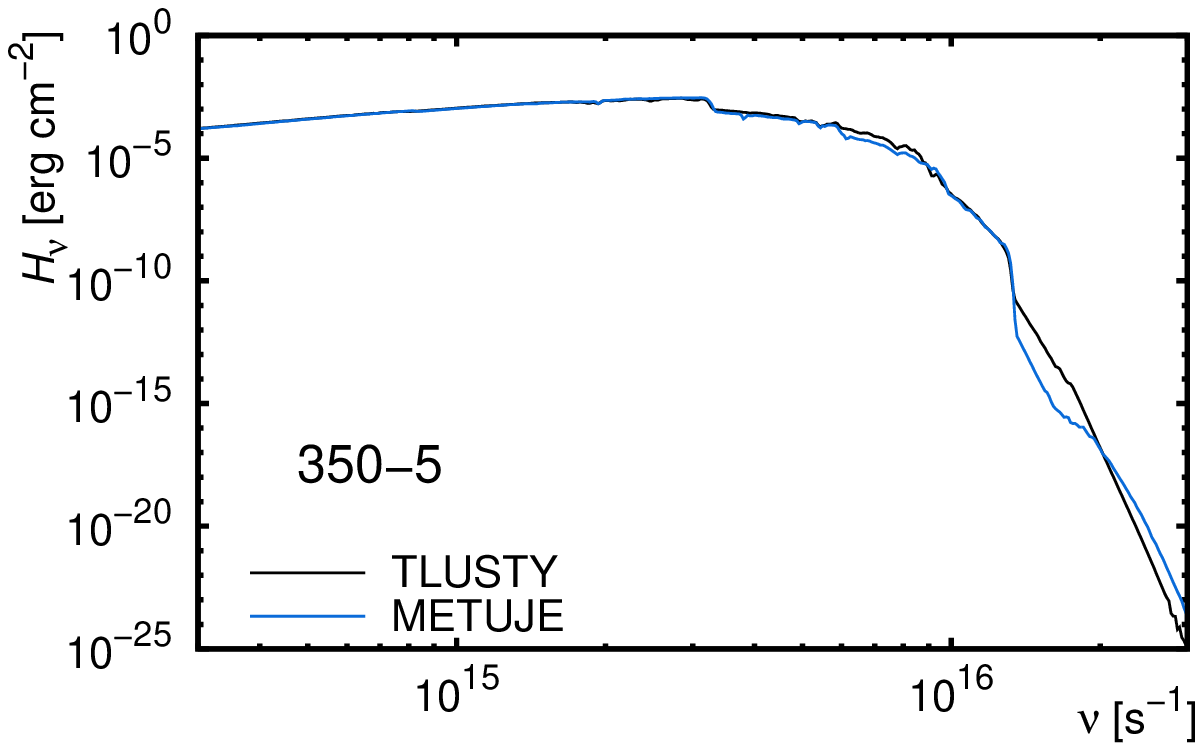}}\\
\resizebox{0.310\hsize}{!}{\includegraphics{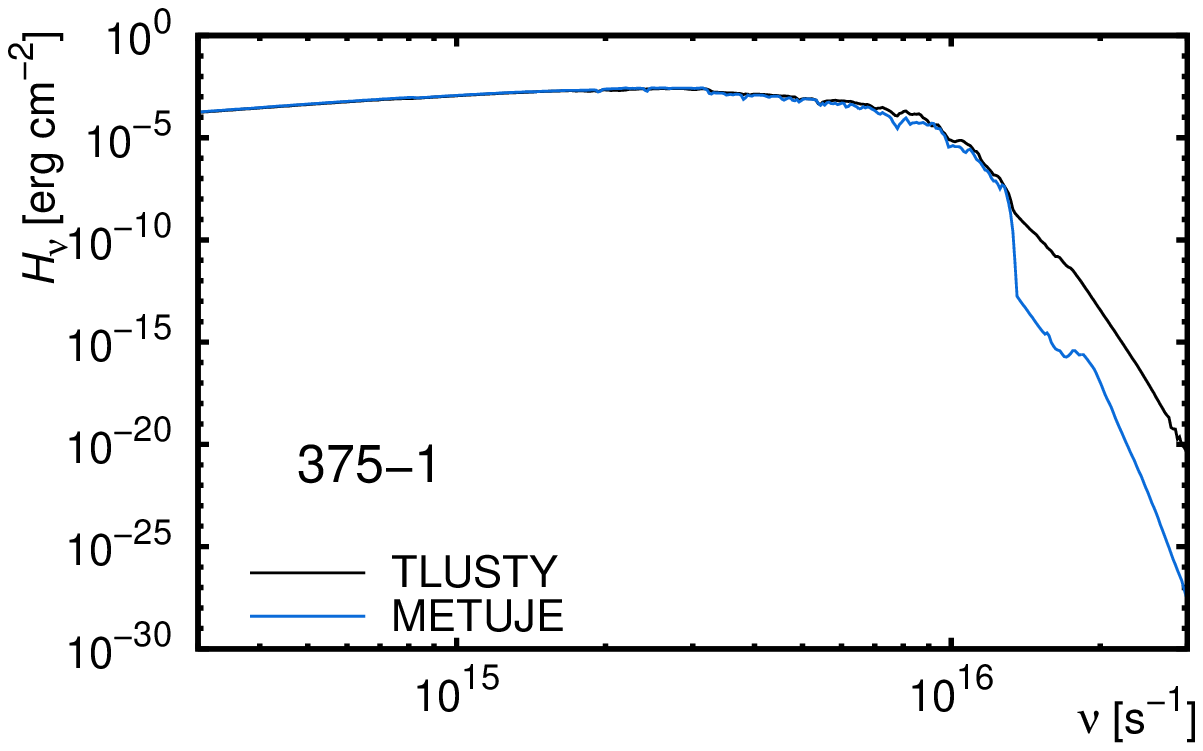}}
\resizebox{0.310\hsize}{!}{\includegraphics{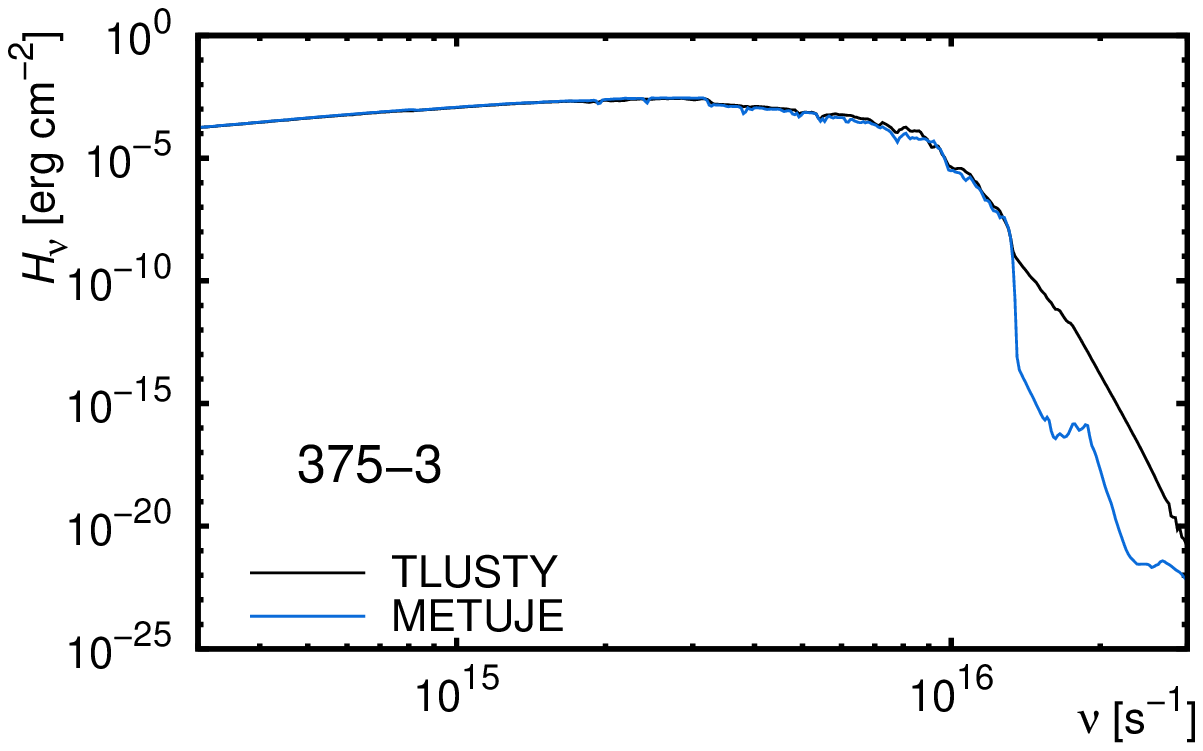}}
\resizebox{0.310\hsize}{!}{\includegraphics{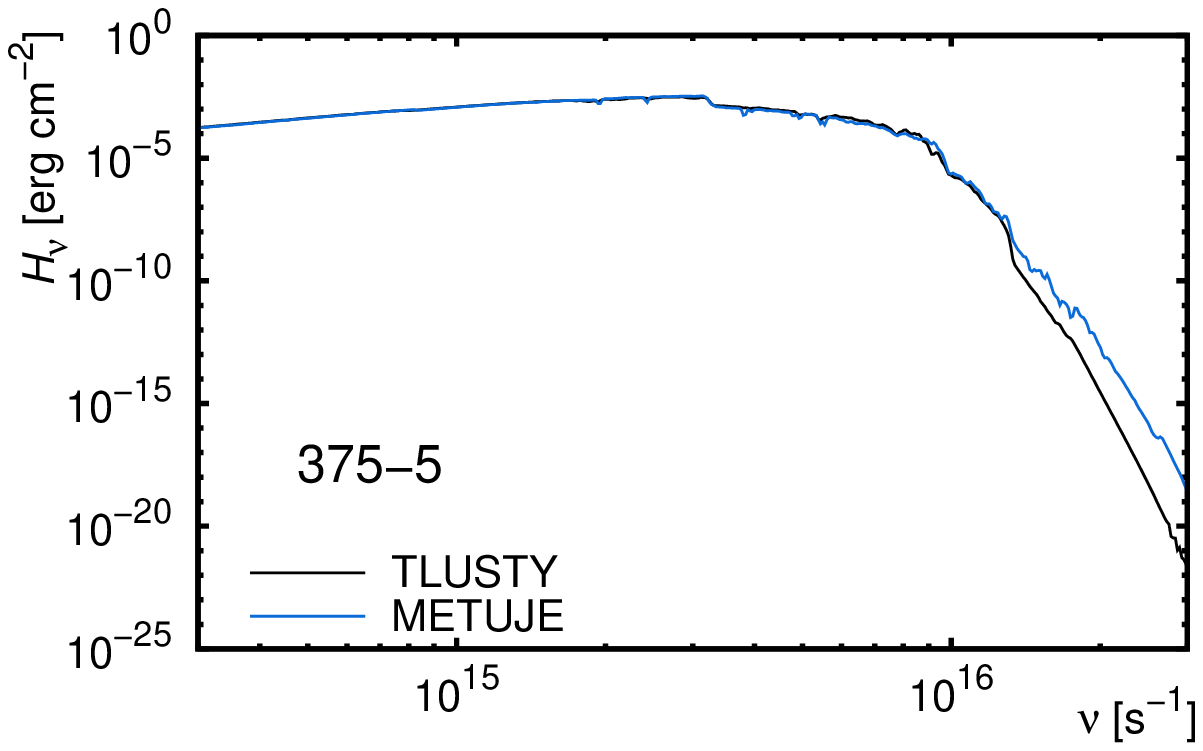}}\\
\resizebox{0.310\hsize}{!}{\includegraphics{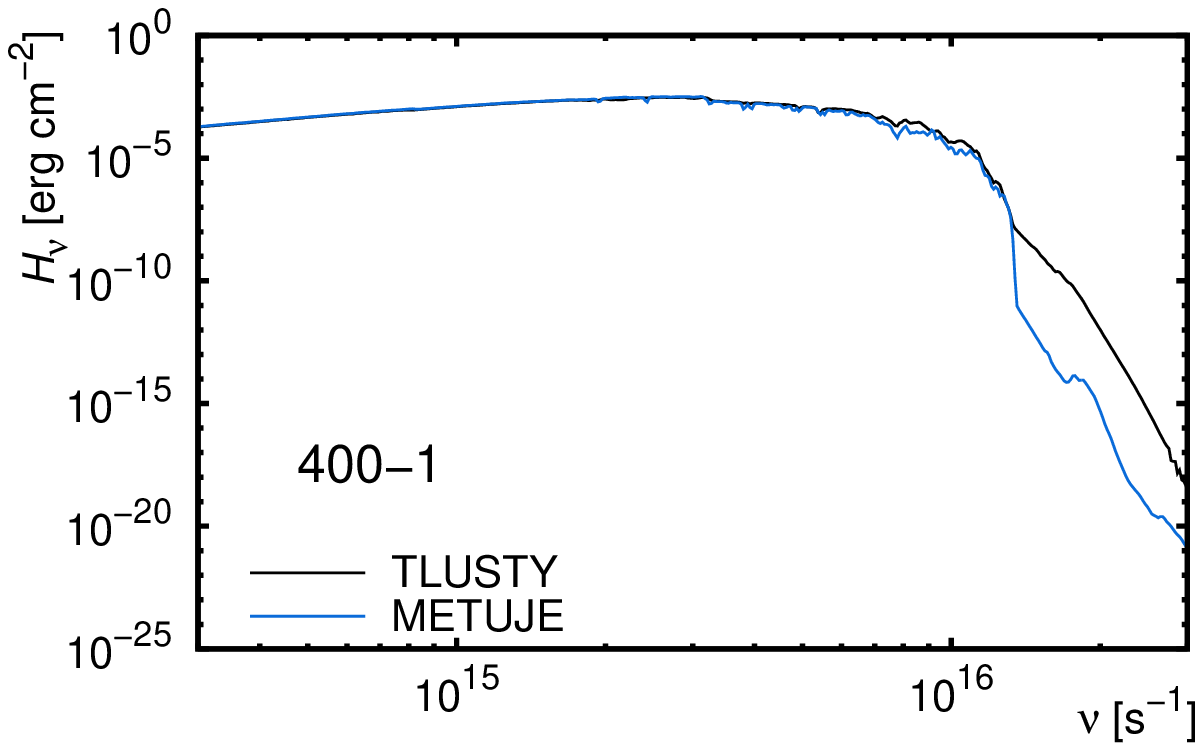}}
\resizebox{0.310\hsize}{!}{\includegraphics{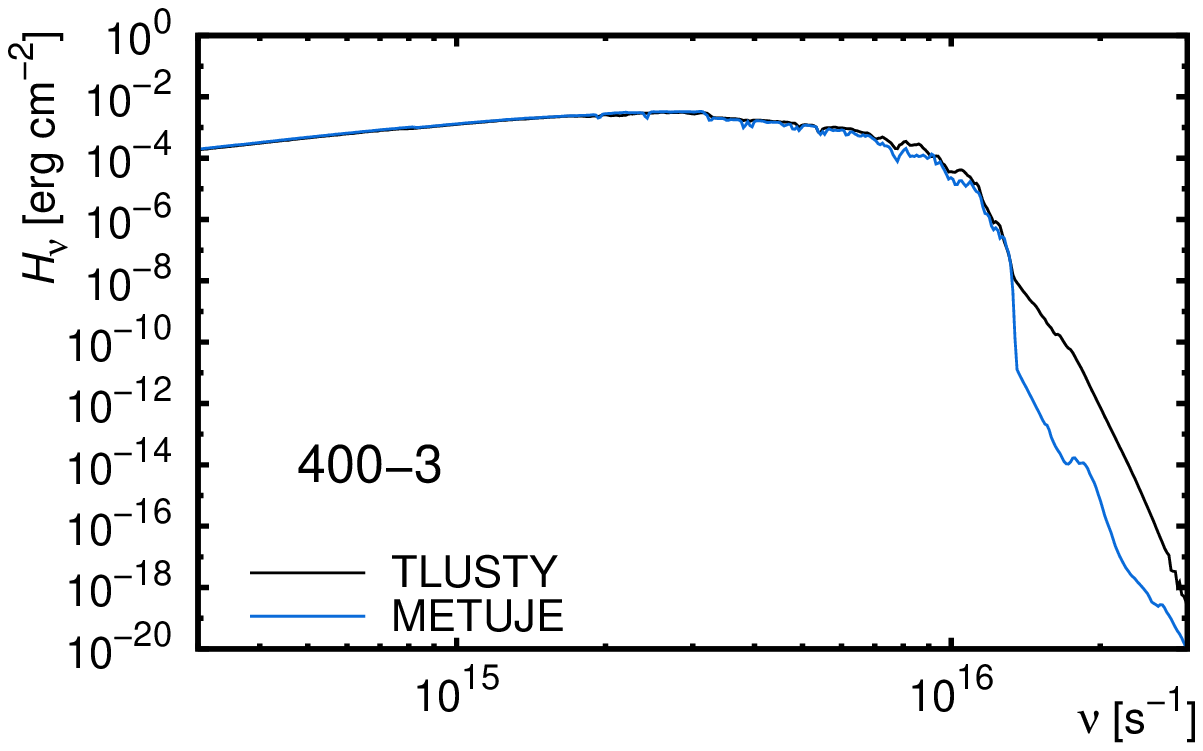}}
\resizebox{0.310\hsize}{!}{\includegraphics{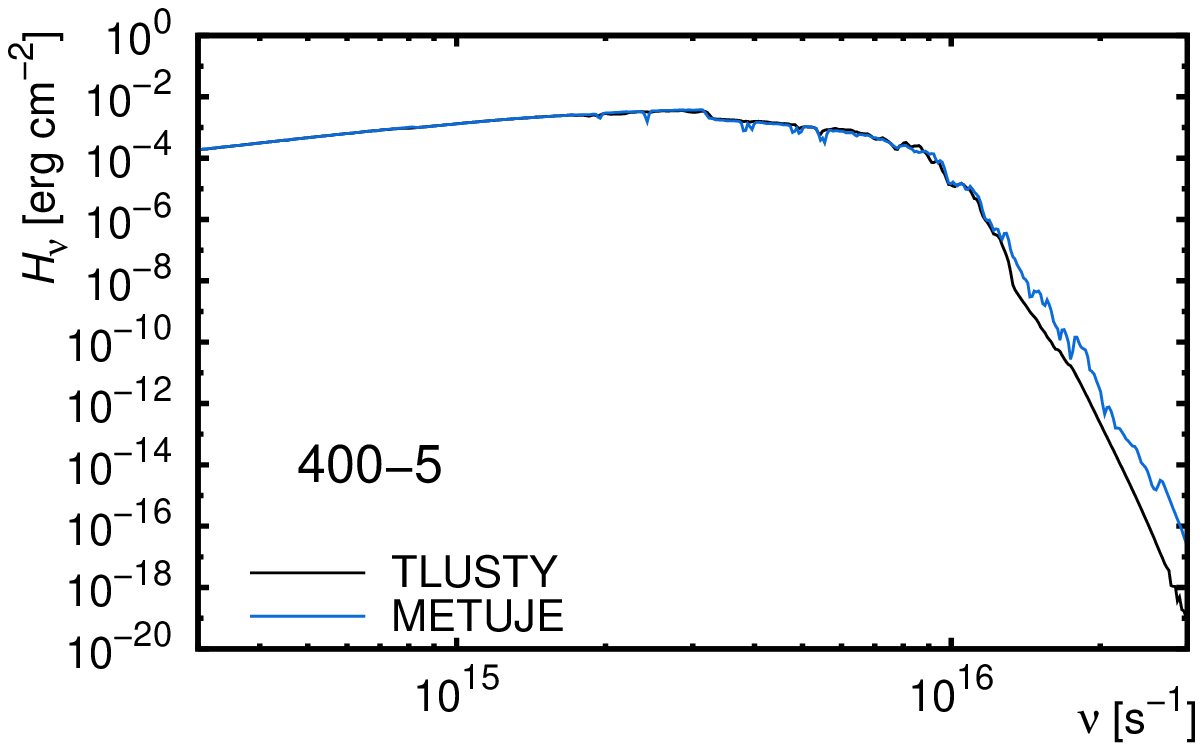}}\\
\resizebox{0.310\hsize}{!}{\includegraphics{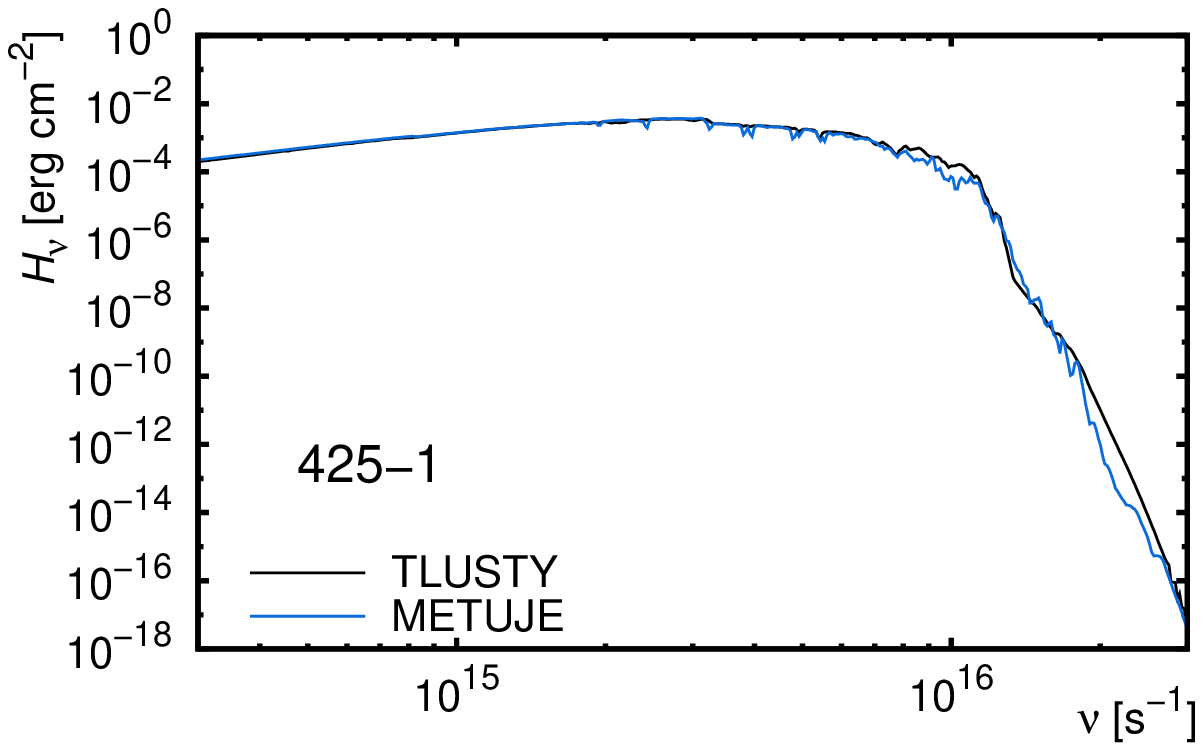}}
\resizebox{0.310\hsize}{!}{\includegraphics{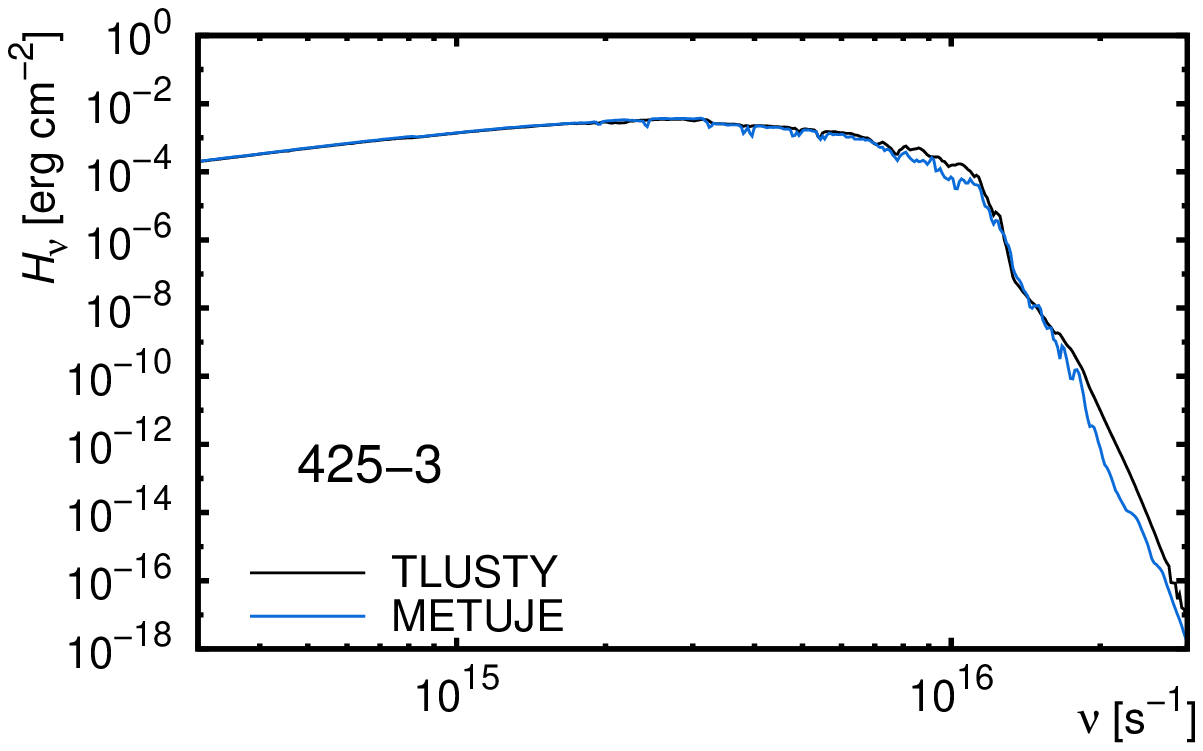}}
\resizebox{0.310\hsize}{!}{\includegraphics{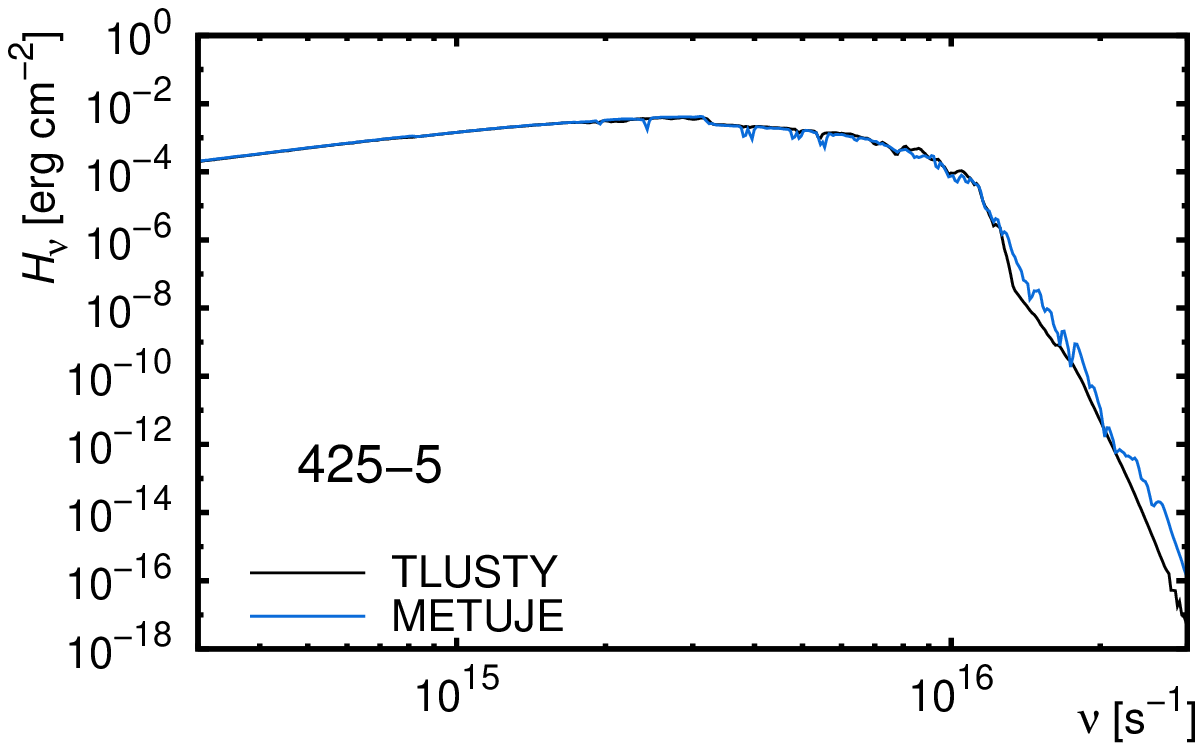}}\\
\resizebox{0.310\hsize}{!}{\includegraphics{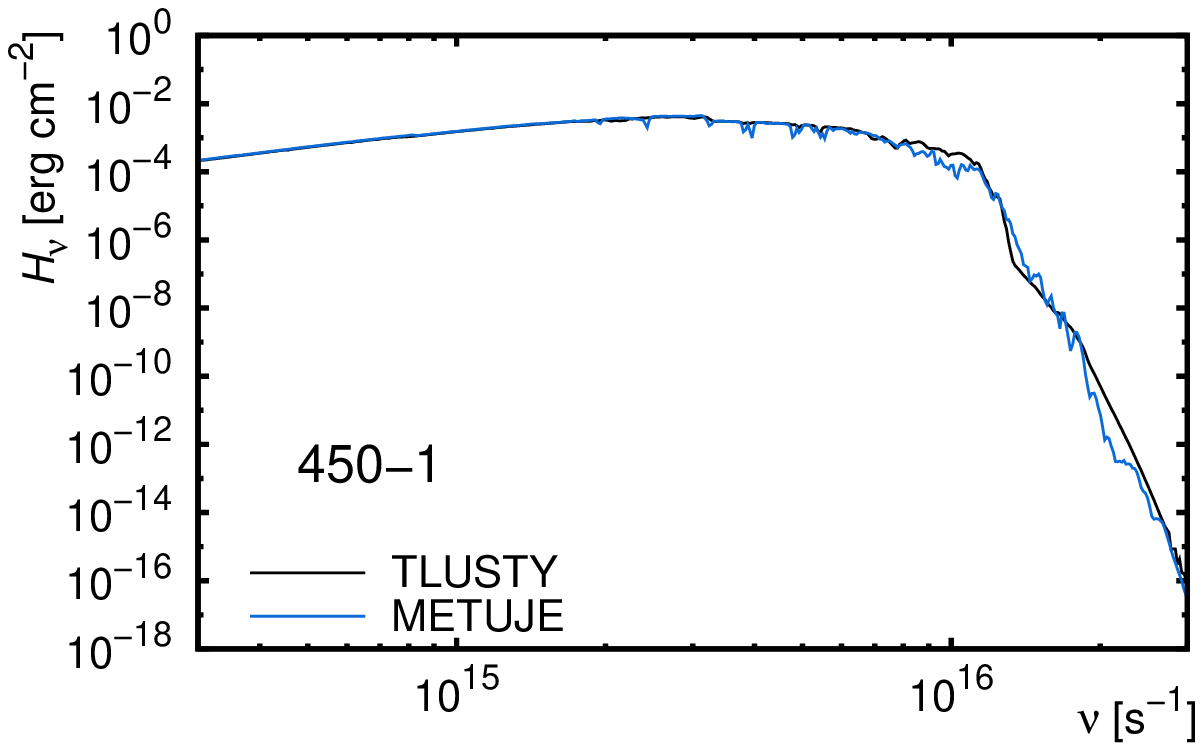}}
\resizebox{0.310\hsize}{!}{\includegraphics{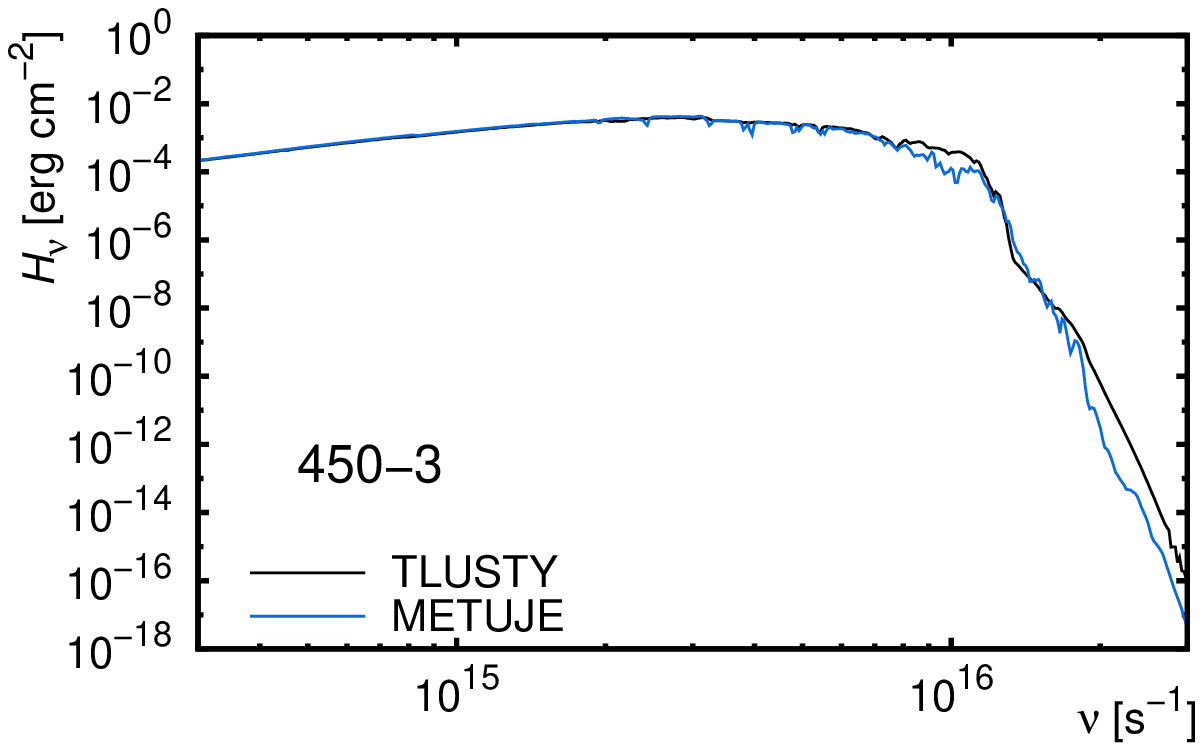}}
\resizebox{0.310\hsize}{!}{\includegraphics{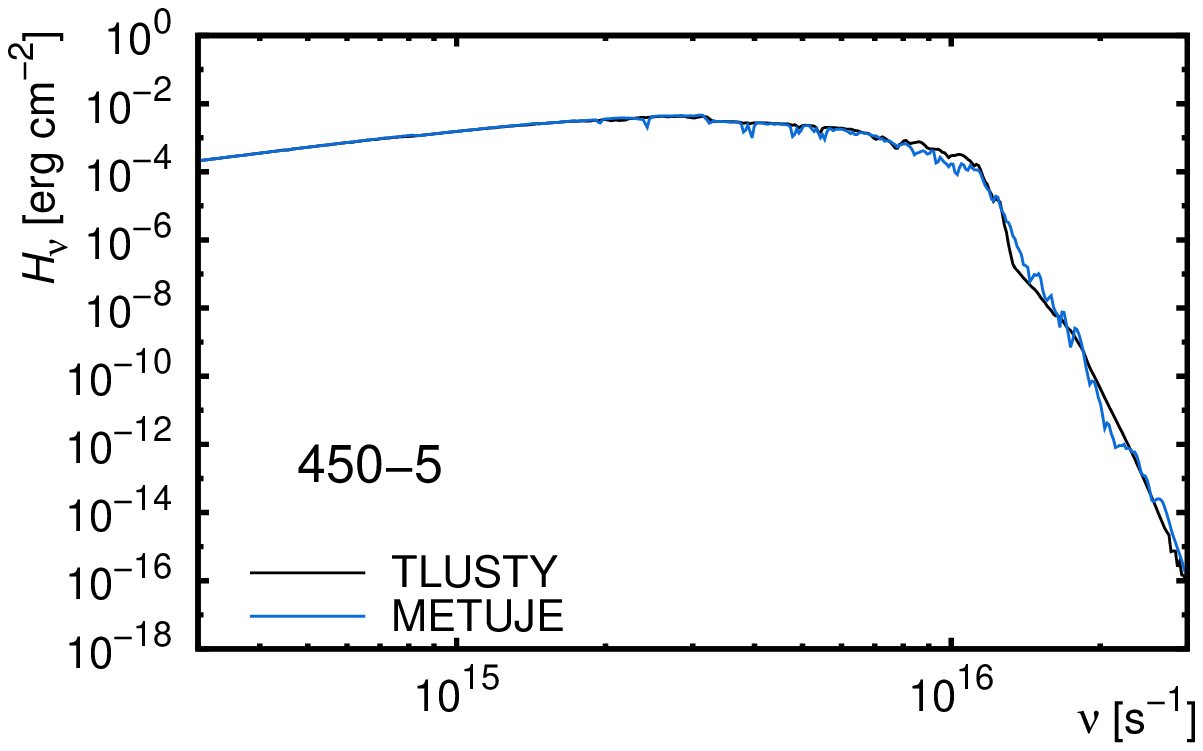}}
\caption{Comparison of the emergent flux from TLUSTY and METUJE models for LMC
stars. The graphs are plotted for individual model stars from
Table~\ref{ohvezpar} (denoted in the graphs).}
\label{tokmetluv}
\end{figure*}

The photospheric structure of METUJE global models in Figs.~\ref{tepmetlum} and
\ref{tepmetluv} nicely agrees with results of the hydrostatic planparallel
TLUSTY code. The agreement is even slightly better than for Galactic stars,
which is likely caused by lower iron abundance, consequently by lower influence
of iron lines on temperature. The METUJE and TLUSTY emergent fluxes in
Figs.~\ref{tokmetlum} and \ref{tokmetluv} also reasonably agree, with the exception
of the frequency region above roughly $7\times10^{15}\,\text{s}^{-1}$ where the
flux is blocked by the wind. This blocking is weaker for winds with lower
mass-loss rates. Larger differences appear in \ion{He}{ii} Lyman continuum (see
Figs.~\ref{tokmetlum} and \ref{tokmetluv}), because the continuum originates in
the wind. The differences in \ion{He}{ii} Lyman continuum are lower for the
hottest stars, where helium is nearly completely ionized.

The derived mass-loss rates are given in Table~\ref{ohvezpar}. Together with
mass-loss rates calculated for Galactic O stars \citep{cmfkont}, their
metallicity dependence can be fitted using the formula \begin{multline}
\label{dmdtmm} \log\zav{\frac{\dot M}{1\, \msr }}= - 5.70+ 0.50
\log\zav{\frac{Z}{Z_\odot}}\\ + \hzav{1.61- 0.12 \log\zav{\frac{Z}{Z_\odot}}}
\log\zav{\frac{L}{10^6L_\odot}}. \end{multline} The fit provides relatively
accurate approximation of our results with a typical error of about 10 -- 30 \%.
Because the $\Teff$ range of calculated Galactic O star wind models is narrower,
the formula is valid for $\Teff=30\,000-42\,500\,$K for $Z/Z_\odot=0.2-1$, while
at Magellanic Cloud metallicities can be used up to $\Teff=45\,000\,$K.
Equation~\eqref{dmdtmm} shows that with decreasing metallicity not only does the mass-loss
rate decrease due to weaker line force, but also the luminosity dependence
becomes steeper. The steeper luminosity dependence at low metallicity is caused
by weaker blocking of the flux by the wind, especially for frequencies
$\nu\gtrsim7\times10^{15}\,\text{s}^{-1}$ (compare plots in
Figs.~\ref{tokmetlum} and \ref{tokmetluv}), and by variation of the slope of the
line-strength distribution with metallicity \citep{pusle}. On the other hand,
the average metallicity dependence $\dot M\sim Z^{0.59}$ is less steep than that
derived by \citet[$\dot M\sim Z^{0.69}$]{vikolamet} and \citet[$\dot M\sim
Z^{0.67}$]{nlteii}, because the wind blanketing effect is weaker at low
metallicity.

In Eq.~\eqref{dmdtmm} we neglected a mild additional temperature dependence of
mass-loss rates, which is non-monotonic. Most intriguing is a missing clear
dependence of mass-loss rates in Eq.~\eqref{dmdtmm} on stellar luminosity class or
mass. \citet{cak} predict the dependence $\dot M\sim L^{1/\alpha}
M_\text{eff}^{1-1/\alpha}$, where $M_\text{eff}=M(1-\Gamma)$ and $\Gamma$ is the
Eddington factor, which for a canonical value of $\alpha=2/3$ \citep{pusle}
gives $\dot M \sim L^{3/2} M_\text{eff}^{-1/2}$. However, in our approach, all
stellar parameters depend on $\Teff$ and spectral type, which may lead to
simplification of the relationship. CMF models with base flux taken from
hydrostatic atmosphere models \citep{fosfor} predict dependence of mass-loss
rates on the luminosity class. This dependence disappears in the present models
as a result  of a stronger reduction of mass-loss rates connected with
abandoning of core-hale approximation in spectroscopically less evolved stars.

\section{Comparison with observations}

\subsection{Mass-loss rates}

\begin{figure}[t]
\centering
\resizebox{\hsize}{!}{\includegraphics{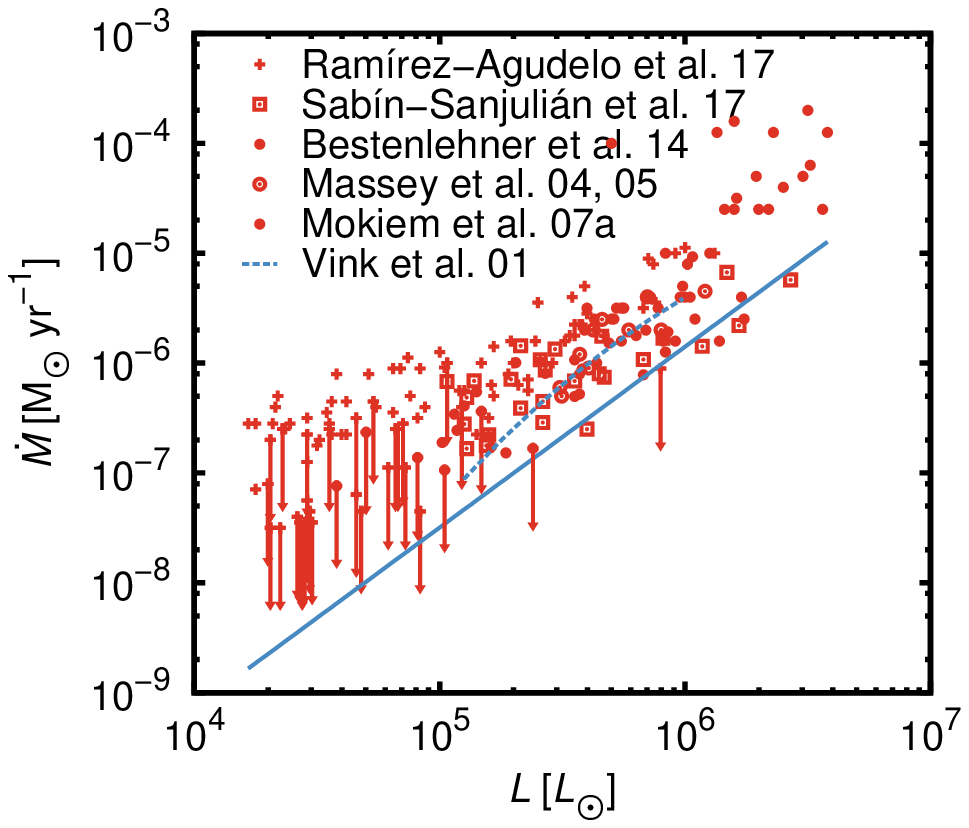}}
\resizebox{\hsize}{!}{\includegraphics{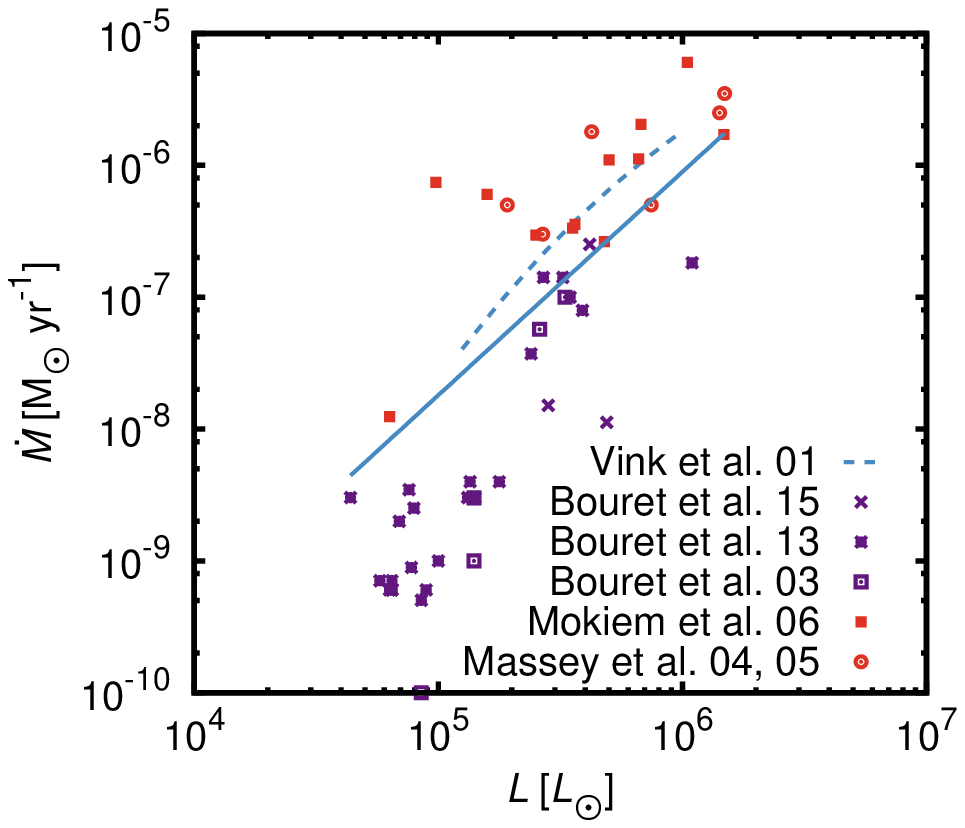}}
\caption{Predicted dependence of the mass-loss rate on luminosity
Eq.~\eqref{dmdtmm} (solid line) for LMC ({\em upper panel}) and SMC ({\em lower
panel}) in comparison with data derived from optical spectroscopy \citep [red
symbols,][] {mamrac1,mamrac2,mokm,mokv,nejlepsi,ramirez,sabina} and ultraviolet
spectroscopy \citep[violet symbols,][]{bourak,boulam,nemracna}. Vertical arrows
denote upper limits. The results of \citet{nemracna}
were derived for stars from IC 1613 and WLM dwarf galaxies, which have a similar
metallicity as SMC. Dashed lines denote predictions of \citet{vikolamet} for
giants from Table~\ref{ohvezpar}.}
\label{dmdtl}
\end{figure}

The mass-loss rate determinations for Magellanic Cloud stars are typically based
either on the optical spectroscopy (mostly H$\alpha$ line) or on the ultraviolet
spectroscopy (P~Cygni lines). We collected available recent observational
mass-loss rate estimates derived for Magellanic Cloud stars from the optical and
ultraviolet spectroscopy and plotted them as a function of stellar luminosity in
a comparison with our predicted mass-loss rates in Fig.~\ref{dmdtl}. The
observational mass-loss rate estimates determined from H$\alpha$ line (assuming
the $\beta$-velocity law) are higher possibly as a result of neglected clumping.
Consequently, the H$\alpha$ line alone is not sufficient to determine a precise
value of the mass-loss rate \citep[e.g.,][]{pulamko}. On the other hand,
although the differences between ultraviolet line profiles calculated with and
without clumping are rather tiny \citep{prvnifosfor,bezchuch}, precise
spectroscopy is able to reveal the influence of clumping on spectra. In the case
of mass-loss rates derived from ultraviolet spectroscopy, we therefore used only
mass-loss rate estimates corrected for clumping. However, these estimates
account only for optically thin clumps (microclumping), while a more general
approach allows also for optically thick clumps
\citep[macroclumping,][]{osporcar, chuchcar, sund, clres1, clres2, shendelori,
brawrp}.

The mass-loss rates derived from optical spectroscopy are higher on average by a
factor
of 6.6 than the theoretical predictions for LMC and by a factor of 3.3
for SMC. We also note that the ratio between optical mass-loss rates and
theoretical prediction for LMC stars with $L\lesssim10^5\,L_\odot$ is higher
than for more luminous stars. This is the main reason why the average ratio is
higher for LMC stars than for SMC stars, because the SMC sample contains mostly
more luminous stars. On the other hand, the predicted mass-loss rates are
consistent with upper limits derived by \citet{mokv}, \citet{ramirez}, and
\citet{sabina}. The latter upper limits are not plotted in Fig.~\ref{dmdtl}. As
a result of the density squared dependence of recombination rates, the H$\alpha$
mass-loss rate estimates are sensitive to small-scale inhomogeneities
\citep[clumping, e.g.,][]{pulchuch}. In the presence of clumping, the H$\alpha$
mass-loss rates are overestimated by a factor $C_\text{c}^{1/2}$, where
$C_\text{c}$ is the clumping factor. The difference between the observational
and theoretical mass-loss rates would imply a clumping factor (averaging results
for LMC and SMC) $C_\text{c}= 5.0^2=25 $ if the difference is purely the result
of the influence of clumping on observations. However, clumping also affects the
theoretical predictions \citep{muij} in the case when clumping starts close to
the star. Assuming optically thin inhomogeneities (microclumping) throughout the
whole wind, the mass-loss rate scales on average as $\dot M\sim
C_\text{c}^\alpha$ with $\alpha\approx1/4$ \citep[Table~3]{muij}. Consequently,
if the ratio 5.0 between the mass-loss rates is both due to influence of
clumping on observations and predictions, then the required clumping factor is only
moderate; from $C_\text{c}^{1/2}C_\text{c}^{1/4}= 5$ we derive
$C_\text{c}=9$. This would also imply that the true mass-loss rates (i.e., with
clumping included in the hydrodynamical models) are by a factor of
$C_\text{c}^{1/4}= 1.7$ higher than those predicted here. However, if clumping
starts above the critical point, where the wind mass-loss rate is estimated,
then the difference between observation and theory would imply a relatively large
clumping factor, $C_\text{c}= 25.$ These results are comparable with that derived
for Galactic stars \citep{cmfkont}, who found that $C_\text{c}\geq8$. On
the other hand, porosity either in the spatial or velocity space may lead to a
decrease of the mass-loss rate \citep{muij,supujo}.

We have demonstrated that the clumping factor is nearly the same in LMC and SMC
as in our Galaxy. This agrees with observational studies that also found that
the level of clumping does not depend on metallicity \citep{march,mokz}. The
observational dependence of mass-loss rate on luminosity in Fig.~\ref{dmdtl} is
wider for low-luminosity ($L\lesssim10^5\,L_\odot$) giants from the sample of \citet{ramirez}. This may indicate higher levels of clumping in some of these stars.

The mass-loss rates predicted by our hydrodynamic models either agree with or are
slightly higher than mass-loss rates derived by \cite{bourak, boulam, nemracna}
from UV analysis for high-luminosity ($L\gtrsim2\times10^{5}\,L_\odot$) SMC
stars (see lower panel of Fig.~\ref{dmdtl}). The UV analysis typically accounts
for optically thin clumps (microclumping). A more general analysis that accounts
also for optically thick clumps \citep{osporcar, sund,clres2} may explain why
some of the predicted mass-loss rates are higher than those derived from
observations.

For stars with low luminosity ($L\lesssim2\times10^{5}\,L_\odot$), the mass-loss
rates derived from UV wind line profiles are significantly lower than the
predicted mass-loss rates (Fig.~\ref{dmdtl}, see also \citealt{martin}). This
discrepancy between theory and observations is termed "weak wind problem". This
problem is possibly caused by an overly long shock cooling time, which becomes
comparable to the characteristic flow time in low-density environment
\citep{cobecru,nlteiii,lucyjakomy,huslab}. Most of the wind remains hot in such
a case and does not leave any imprints in the ultraviolet spectra.

\citet{massamm} derived mass-loss rates of Magellanic Cloud O stars from
mid-infrared excesses. Comparison with our model mass-loss rate shows that our
estimates are a factor of 50 lower for SMC stars and a factor of 24 lower
for LMC stars. We have not detected any significant luminosity dependence of the
difference for SMC stars and increase of the difference with decreasing
luminosity for LMC stars. Repeating the same analysis as that done for H$\alpha$
mass-loss rates, this would imply a clumping factor of at least 120, which is
significantly higher than that derived in the H$\alpha$ formation region. The
possibility that the difference is caused by small velocity gradients
\citep[suggested by][]{massamm} is not supported by the results of our models.

From the absence of frequency dependence of the observed pulse-arrival times in
SMC binary pulsar PSR J0045-7319, \citet{kaspik} derived an upper limit of the
B1V star wind mass-loss rate $\dot
M<1.1\times10^{-10}\,M_\odot\,\text{year}^{-1}$ \citep[taking into account the
wind terminal velocity derived in][]{metuje}. With luminosity of a B1-type star
\citep{har} we derive from Eq.~\eqref{dmdtmm} $\dot M=3. 5
\times10^{-10}\,M_\odot\,\text{year}^{-1}$, not far from the observational
constraint.

In Fig.~\ref{dmdtl} we also compared our model results with theoretical
predictions of \citet{vikolamet}  calculated for giants from
Table~\ref{ohvezpar} and metallicities corresponding to LMC and SMC. Because
\citeauthor{vikolamet} predictions assume \citet{angre} reference solar
abundances, which are higher than those derived by \citet{asp09}, we scaled the
SMC and LMC abundances by a ratio of the mass-fraction of heavier elements
derived by \citet{asp09} and mass-fraction derived by \citet{angre}. Our
predicted mass-loss rates are typically a factor of 2 -- 3 lower than the
predictions of \citeauthor{vikolamet} The difference is lower for SMC stars than
for LMC stars and could possibly be attributed to a different method of the
calculation of the line force \citep{cmfkont}. While \citeauthor{vikolamet} use
the Sobolev method, our models are based on the CMF line force. We derived
similar results for Galactic stars \citep{cmfkont}, for which the empirical
mass-loss rate estimates corrected for clumping may be reconciled with
theoretical predictions in such a way that the average ratio between individual
mass-loss rate estimates is not higher than about $ 1.6 $.

\subsection{Terminal velocities}

According to the predictions of hot-star wind theory, the wind terminal velocity
$v_\infty$ is proportional to the escape speed $v_\text{esc}$ \citep{cak}.
Although the wind terminal velocity can be readily derived from the observed
spectra, the predictions are sensitive to many approximations involved in the
calculation of the radiative force, including the treatment of X-rays and
clumping. For SMC stars we found the average ratio $v_\infty/v_\text{esc}=2.1$
and for LMC stars we found the same value as for Galactic stars \citep{cmfkont}
$v_\infty/v_\text{esc}=1.9$. This is at odds with the findings of \citet{leir},
who predict $v_\infty\sim Z^{0.13}$ from their
theoretical models. Moreover, our predicted ratio $v_\infty/v_\text{esc}$ shows
a large scatter and is around 2 for stars with
$T_\text{eff}\gtrsim32\,000\,\text{K}$ and increases up to 3 for the coolest
stars considered here.


The observational results show large scatter and do not provide a clear clue to
the metallicity variations of $v_\infty$. However, contrary to the Galactic O
stars \citep{cmfkont}, the theoretical results do not disagree with terminal
velocities derived from observations, which give
$v_\infty/v_\text{esc}=2.2\pm0.3$ for SMC and $v_\infty/v_\text{esc}= 2.6\pm0.7$
for LMC stars (averaging the results of \citealt{pulmoc,prvnifosfor,mamrac2},
and \citealt{boulam} for $ 30\,000\,\text{K}\leq \Teff\leq45\,000\,$K).

\subsection{Ionization fractions}

Ionization fractions of some ions can be derived from ultraviolet wind line
profiles. However, for such an analysis the wind mass-loss rates are required,
because the depth of the line profile depends also on the wind density.
Moreover, the strength of the line profiles is affected by optically thick
inhomogeneities \citep[macroclumping, porosity,][]{osporcar, sund, clres2}.
Ionization fractions are furthermore influenced by clumping and X-ray ionization
\citep{macown,pahole,pulchuch, lojza}.

\begin{figure}[t]
\centering
\resizebox{\hsize}{!}{\includegraphics{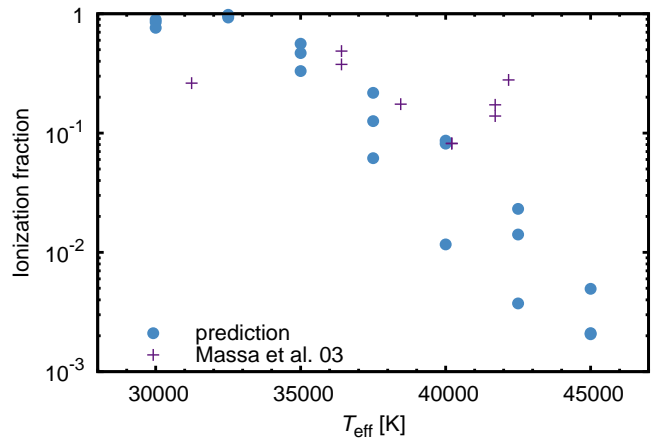}}
\caption{Comparison of predicted \ion{P}{v} ionization fractions (blue dots) with
ionization fractions derived from observations \citep[crosses]{maso} at the
point where the wind velocity is equal to half of its terminal value for LMC
stars as a function of stellar effective temperature. The ionization fractions
derived from observations were scaled by the ratio of mass-loss rate used to
derive the observational fractions and our predicted mass-loss rate for LMC
metallicity.}
\label{fosfor}
\end{figure}

Therefore, we selected just the \ion{P}{v} ion for the comparison of predicted
ionization fractions and ionization fractions derived from observations. The
ultraviolet lines of this ion are not saturated and this ion is not
significantly affected by X-ray ionization \citep{fosfor}. \ion{P}{v} gained
considerable attention due to its unexpectedly weak line profiles \citep{full}.
The ionization fractions of \ion{P}{v} from LMC stars as a function of wind
velocity were derived by \citet{maso} using the approximate SEI (Sobolev with
exact integration) method. To account for the mass-loss rate dependence of
observational indicators, we scaled \citet{maso} ionization fractions by the
ratio of the mass-loss rates used to derive these fractions and our predicted
mass-loss rates for LMC stars. The comparison of ionization fraction derived
from observations and theory is given in Fig~\ref{fosfor}. The comparison shows
that our models are able to reliably reproduce the \ion{P}{v} ionization
fraction in the winds of LMC stars for stars with $\Teff\leq40\,000\,\text{K}$.
For hotter stars, \ion{P}{vi} dominates in the models and the predicted
\ion{P}{v} ionization fractions are lower than those derived from observations. 

\subsection{Observed spectra}

The comparison with observed spectra provides the most detailed test of any
model atmosphere. Such comparison is not the aim of the current paper, because
some effects that are important for comparison with observations (e.g., clumping
and X-rays) are missing in the discussed models. In addition, we calculated a
grid of typical models and did not attempt to fit any specific star using our
models. However, the emergent spectrum from our models should not be too far
away from the observed one.

\begin{figure*}[t]
\centering
\resizebox{\hsize}{!}{\includegraphics{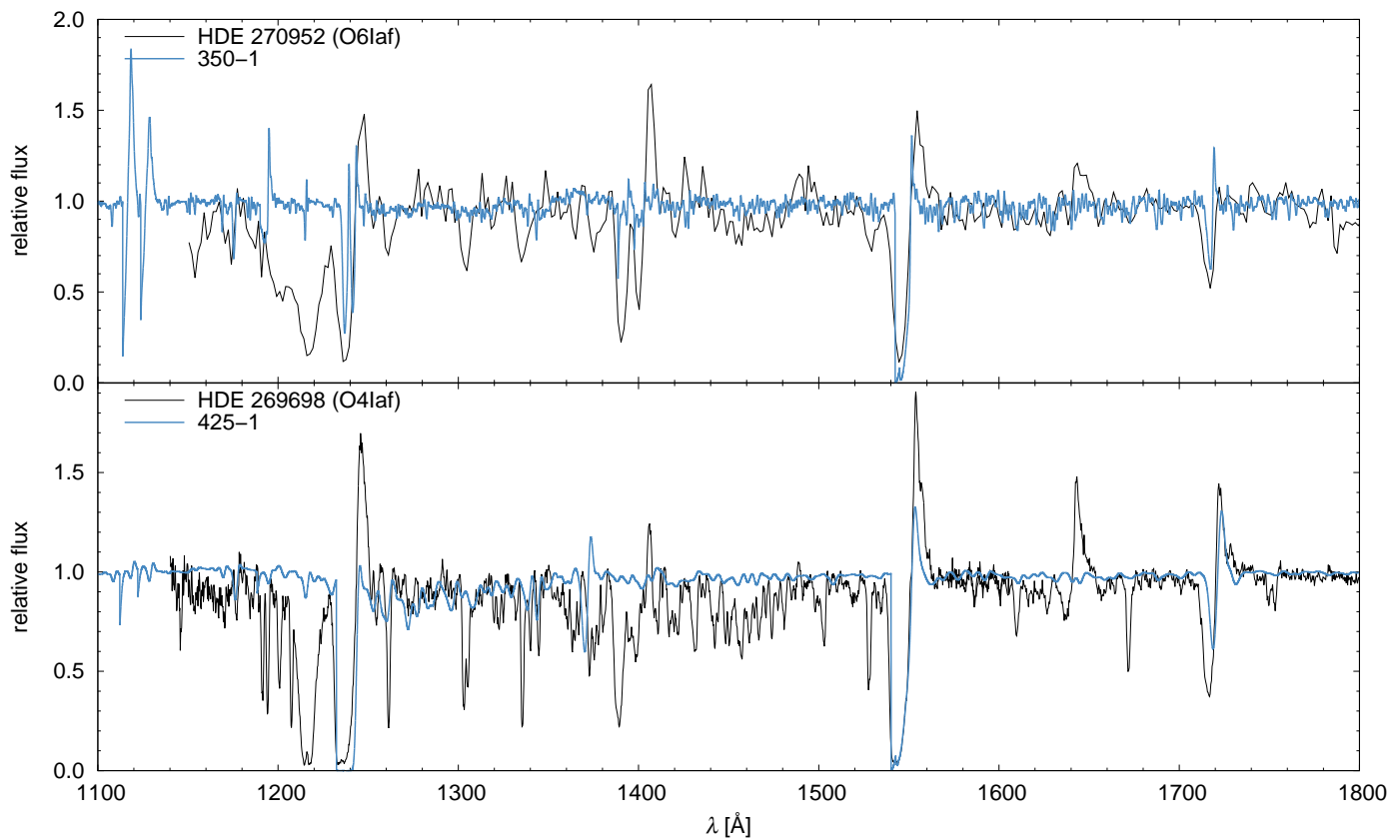}}
\caption{Comparison of predicted LMC 425-1 and 450-1 star spectra with HST/FOS
and IUE spectra of stars HDE~269698 and HDE~270952 with parameters close to that
of the grid.}
\label{spektravmm}
\end{figure*}

To demonstrate this, we compared the observed spectra of two stars with one of
the models from the grid with parameters close to the studied stars. We selected
LMC stars HDE~269698 (Sk~-67~166) and HDE~270952 (Sk~-65~22) studied by
\citet{prvnifosfor}. We compared the predicted spectra with HST/FOS (HDE~269698,
y14m0c05t and y14m0c06t) and IUE (HDE~270952, SWP~1628) observations in
Fig.~\ref{spektravmm}. The observations were downloaded and processed using the
SPLAT package \citep{splat,pitr}. The comparison shows that our models are able
to predict reasonable wind profiles. However, the photospheric line profiles are
not well reproduced, because we focus on wind dynamics and do not include
photosphere fully self-consistently. For example, we use the Sobolev
approximation to calculate the line source function in kinetic equilibrium
equations, which is an  oversimplification that may affect the population numbers.

\section{Conclusions}

We calculated global (unified, hydrodynamic) model atmospheres for a set of
model O stars with metallicities corresponding to those in Large and Small
Magellanic Clouds. Our models solve CMF radiative transfer, kinetic equilibrium
(NLTE), and hydrodynamical equations from (quasi-)hydrostatic atmosphere
outwards to expanding stellar wind. Therefore, the models predict the radial
variations of density, velocity, and temperature simply from basic global stellar
parameters.

As a result of lower metallicity of Magellanic Clouds, the line radiative force
is weaker, which leads to lower wind mass-loss rates with respect to the
Galactic stars. On average, for Galactic and Magellanic Cloud metallicities the
mass-loss rate scales with metallicity as $\dot M\sim Z^{0.59}$. We provide a
more accurate formula that fits the wind mass-loss rate predicted by our models
as a function of stellar luminosity and metallicity.

In comparison with observations, the mass-loss rates following from our
hydrodynamic models are lower than mass-loss rates derived from H$\alpha$
emission line profiles and can be reconciled with H$\alpha$ diagnostics assuming
a clumping factor of $C_\text{c}=9$. On the other hand, the model mass-loss rates
either agree with or are higher than the mass-loss rates derived from ultraviolet
line profiles. The model \ion{P}{v} ionization fractions agree with results
derived from observations for LMC stars with $\Teff\leq40\,000\,\text{K}$. Taken
together, our theoretical predictions provide reasonable models with consistent
mass-loss rate determination and can be used to study stars from Magellanic
Clouds.

\begin{acknowledgements}
This work was supported by grant GA \v{C}R 13-10589S. Access to computing and
storage facilities owned by parties and projects contributing to the National
Grid Infrastructure MetaCentrum provided under the programme "Projects of Large
Research, Development, and Innovations Infrastructures" (CESNET LM2015042) is
greatly appreciated. The Astronomical Institute Ond\v{r}ejov is supported by the
project RVO:67985815.
This research made use of the IUE data derived from
the INES database and of the HST data derived from the MAST database using the
SPLAT package.
\end{acknowledgements}

\Online

\end{document}